\documentclass[11pt,a4paper]{article}
\usepackage{amssymb,latexsym,amsmath,amscd,amsthm}
\usepackage[latin1]{inputenc}
\usepackage{epsfig}
\usepackage[active]{srcltx}
\usepackage[dvipsnames,table]{xcolor}
\usepackage{xspace}
\usepackage{hhline}
\usepackage{color}
\usepackage[hyphens]{url}
\usepackage[breaklinks,colorlinks=true,linkcolor=black,
citecolor=black, urlcolor=black]{hyperref}
\usepackage{multirow}
\usepackage{caption}
\usepackage{float}
\usepackage{appendix}
\usepackage[active]{srcltx}
\usepackage[english]{babel}
\usepackage{pgf,tikz,pgfplots}
\usepackage{mathrsfs}
\usepackage[round,authoryear]{natbib}
\usetikzlibrary{arrows}
\usepackage{geometry}
\usepackage{lscape}
\usepackage{authblk}

\definecolor{pink}{rgb}{1,0,1}
 
\newcommand{\eri}{\texttt{Erik+2}\xspace}
\newcommand{\saq}{\texttt{SAQ}\xspace}

\newcommand{\pl}{\texttt{PL}\xspace}
\newcommand{\asaq}{\texttt{ASAQ}\xspace}

\newcommand{\nj}{\texttt{NJ}\xspace}
\newcommand{\ml}{\texttt{ML}\xspace}
\newcommand{\fp}{\texttt{4P}\xspace}
\newcommand{\mll}{\texttt{ML}}

\renewcommand{\gg}[1]{{p\ell}_{#1}}

\newcommand{\CC}{\textit{CC}\xspace}
\renewcommand{\CD}{\textit{CD}\xspace}
\newcommand{\DD}{\textit{DD}\xspace}

\theoremstyle{definition}
\newtheorem{defi}{Definition}[section]

\theoremstyle{plain}
\newtheorem{lema}[defi]{Lemma}
\newtheorem{thm}[defi]{Theorem}

\newtheorem{rk}[defi]{Remark}

\newtheorem{nota}[defi]{Notation}
\newtheorem{teo-def}[defi]{Theorem/Definition}

\newif\ifprivate
\privatetrue

\def\???{\ifprivate {\bf {???}} \marginpar{{\Huge {\bf ?}}}
\else \fi}

 \title{Designing weights for quartet-based methods when data is heterogeneous across lineages}

 \newcommand*\samethanks[1][\value{footnote}]{\footnotemark[#1]}
 \author{Marta Casanellas \thanks{Institut de Matematiques de la UPC-BarcelonaTech (IMTech), Universitat Polit\`ecnica de Catalunya, Centre de Recerca Matem\`atica, Barcelona, Spain}, Jesús Fernández-Sánchez\samethanks, Marina Garrote-López\thanks{Max Planck Institute for Mathematics in the Sciences, Leipzig, Germany}, Marc Sabat\'{e}-Vidales\thanks{The University of Edinburgh, Edinburgh, United Kingdom}}
 
 \date{}

\begin{document}
\maketitle






\begin{abstract}
	Homogeneity across lineages is a common assumption in phylogenetics according to which nucleotide substitution rates remain constant in time and do not depend on lineages. This is a simplifying hypothesis which is often adopted to make the process of sequence evolution more tractable. However, its validity has been explored and put into question in several papers. On the other hand, dealing successfully with the general case  (heterogeneity across lineages) is one of the key features of phylogenetic reconstruction methods based on algebraic tools. 
	
	The goal of this paper is twofold. First, we present a new weighting system for quartets (\asaq) based on algebraic and semi-algebraic tools, thus specially indicated to deal with data evolving under heterogeneus rates.  This method combines the weights two previous methods by means of a test based on the positivity of the branch length estimated with the paralinear distance. 
	\asaq is statistically consistent when applied to GM data, considers rate and base composition heterogeneity among lineages and does not assume stationarity nor time-reversibility. Second, we test and compare the performance of several quartet-based methods for phylogenetic tree reconstruction (namely, Quartet Puzzling, Weight Optimization and Wilson's method) in combination with \asaq weights and other weights based on algebraic and semi-algebraic methods or on the paralinear distance. These tests are applied to both simulated and real data and support Weight Optimization with \asaq weights as a reliable and successful reconstruction method. 
\end{abstract}

\section{Introduction}



Molecular phylogenetic reconstruction faces several problems, still nowadays. Even if one restricts to gene tree reconstruction, one has to take into account the amount of  available data (which might be low with respect to the number of taxa), and depending on the method applied, the selection of a suitable evolutionary model, the inherent difficulty of estimating parameters for the most complex models, and  the incorporation of heterogeneity across sites and/or across lineages, among others \citep[cf.][and the references therein]{jermiin2020,zou2019}.

Most of the available methods strongly depend on the evolutionary model assumed (this is the case of maximum likelihood, bayesian or distance-based approaches, to a more or less extent) and some have to estimate the substitution parameters for each possible tree topology.
Recent work minimizes the relevance of the selection of the substitution model if the tree topology is correctly inferred \citep[see][]{abadi} and in this case, a \emph{general Markov model} (GM for short) could be used \cite[see also][]{kaehler2015}.
Regarding topology reconstruction, some methods that avoid parameter estimation and allow complex substitution models such as GM are those based on the \emph{paralinear distance}  or on \emph{algebraic tools} \citep[phylogenetic invariants and related tools, see][]{AllmanRhodeschapter4}.

The paralinear distance \citep{lake1994} is a measure that attempts to estimate the evolutionary distance (in terms of expected number of substitutions per site) when sequences evolve under a GM model. It may overestimate the expected number of substitutions when the process is far from stationary, or branch lengths are long \cite[see][]{kaehler2015,Zou2012}, so more elaborate methods are provided in the quoted papers. However, it is a widely used distance due to its simple formula and its generality, and it has recently been proven to be consistent for the multispecies coalescent model as well, see \citet{allman2021,allman2019}.

Several algebraic methods for phylogenetic reconstruction have been proposed in the last years, see for example  SVDQuartets \citep{chifmankubatko2014}, \eri  \citep{fercas2016} --both already implemented in PAUP* \citep{paup}--, \texttt{Splitscores} \citep{allmankubatkorhodes}, or \saq \citep{casfergar2020}. The methods \eri, \texttt{Splitscores} and \saq are based on the GM model  and, in particular, they are not subject to stationarity or time-reversibility, and account for different rates of substitution at different lineages (heterogeneity across lineages). The three of them consider \emph{algebraic} conditions in the form of rank constraints of a \emph{flattening} matrix obtained from the observed distribution of characters on a sequence alignment. Only \saq considers also the stochastic description of the evolutionary model \citep[see][]{AllmanSemialg}, which translates into \emph{semi-algebraic} conditions. On the other hand, \eri also allows across-sites heterogeneity as it is able to deal with data from a mixture of distributions on the same tree. Despite the potential of algebraic-based methods for topology reconstruction (already pointed out in the book by \citet{felsenstein2004}), some of these methods may not work well for short alignments 
specially in presence of the long branch attraction phenomenon. For instance, \eri is highly successful on different types of quartet data (also on the Felsenstein zone) but requires at least one thousand sites to outperform maximum likelihood or neighbor-joining (briefly \nj)
\citep{fercas2016}. By taking into account the stochasticity of the substitution parameters,  \saq overcomes this problem at the expense of performing slightly worse than \eri for very large amounts of data (10,000 sites or more) \citep{casfergar2020}.

Algebraic methods are mainly aimed at recovering \emph{quartet} topologies (or splits in some cases) and some account for the possibility of being implemented into \emph{quartet-based methods}.
Quartet-based methods (\emph{Q-methods} for short) have been questioned in the literature. For instance, \citet{Ranwez2001} evaluated two quartet-based methods, their weight optimization method (briefly WO) and the quartet-puzzling method (QP) by \citet{strimmer1996}, and they weighted quartets using a maximum likelihood approach for the Kimura 2-parameter model. Their main conclusion was that both QP and WO give worse results than neighbor-joining   or than a maximum likelihood approach applied directly to the whole set of taxa. As pointed out by \citet{Ranwez2001}, the
weaknesses of Q-methods are very likely due to the method of weighting the quartets rather than to the method of combining them.
As far as we are aware, the only work that evaluates the use of algebraic methods as input of quartet-based methods is \cite{rusinko2012}, which is restricted to QP with the Jukes-Cantor model.
On the other hand, the correct management of long-branch attraction is crucial for obtaining a successful quartet-based method \citep{stjohn2003}.
%
These claims and remarks motivate part of the present work.
We expect that, as \saq, \eri and the paralinear method handle successfully the long-branch attraction problem, Q-methods with these weighting systems  improve their performance.

Recently, \citet{zou2019} proved that if a machine learning approach is applied to weightening the quartets, QP can have a similar performance than \nj specially under substitution processes that are heterogeneous across lineages. Precisely, handling heterogeneity of substitution rates across lineages is one of the key features  of algebraic methods based on the GM model of nucleotide substitution.

The goal of this work is to test the performance of several quartet-based methods  for  phylogenetic tree topology reconstruction when applied with input weights from different methods consistent with the general Markov model. We test QP, WO and the method WIL \citep[proposed in][]{willson} with different systems of input weights: two systems, \pl and \fp, derived from paralinear distances and the \emph{four-point condition} \citep{lake1994,gascuel94,wnjw}, and three systems based on algebraic and semi-algebraic methods, namely, \saq, \eri and the new proposed method \asaq (see below) that combines the paralinear distances and algebraic methods. We test exhaustively all these combinations on simulated data evolving either under the GM model or the homogeneous general (continuous) time-reversible model (GTR) on twelve taxon trees,  and provide a comparison when input weights  from a  maximum-likelihood approach are used. We also compare the performance to a global \nj,  and test some of the methods on real data, specifically the eight species of yeast studied in \cite{Rokas2003} and the Ratite mitochondrial DNA data studied in \cite{phillips2010}.

The new method \asaq  (standing for Algebraic and Semi-Algebraic Quartet reconstruction) is a topology reconstruction method for four taxa which combines \eri and \saq by means of the \emph{paralinear method}. This is a quartet-reconstruction method based on a statistic (see (1) in the Methods section) that assesses the positivity of the estimated length (using the paralinear distance) of the interior branch of the quartet. When data are unmistakably generated by a quartet, the topology output by any reconstruction method should be consistent with the positivity of this statistic. \asaq uses  the paralinear method  to either ratify the results of \eri, or to rely on \saq when there is an inconsistency between the outputs of \eri and the paralinear method.
By proceeding like this, \asaq ensures an overall better performance on quartets than \eri, \saq, and the paralinear method itself.

As \asaq is statistically consistent with the GM model, it accounts for rate and base composition heterogeneity  among lineages and does not assume stationarity nor time-reversibility. 
We test the performance of \asaq on a wide scenario of simulated data: on the tree space proposed by \citet{huelsenbeck1995}, on quartets with random branch lengths and on mixture data on the same topology with 2 categories. All simulations used data either generated from the GM model or from a GTR model, with different sequence lengths.
One can use \asaq with mixtures of distributions on the same tree topology with two or three categories (or partitions) as this was already implemented by \eri.
%
Although \asaq is based on \saq and the paralinear method, which are not guaranteed to be consistent on mixtures, it is highly successful on this kind of data, see the Results section.

\section{Methods}


In this section, we describe the new method \asaq for quartet reconstruction, the quartet-based methods applied, and the simulation studies performed.

\subsection{Description of the new method \asaq}
\asaq is a quartet reconstruction method based on a pair of previous methods by the authors \citep{fercas2016, casfergar2020}.

For a phylogenetic tree $T$ with an interior node $r$ as root,  we consider a Markov process on $T$ by assigning transition matrices $M_e$ at the edges of $T$ and a distribution $\pi$ at $r$. As no restrictions are imposed on the transition matrices or $\pi$, this is usually called a \emph{general Markov model} (GM) on $T$.
One can compute the theoretical joint distribution $p$ of patterns at the leaves of $T$ in terms of the entries of $M_e$ and $\pi$
and we say that $p$ \emph{has arisen on} $T$  with certain \emph{substitution parameters}.

We consider fully-resolved (unrooted) trees on a set of four taxa $[4]=\{1,2,3,4\}$: the three quartet trees shall be denoted as $12|34$, $13|24$, $14|23$, according to the bipartition induced by the interior edge.

In \cite{fercas2016}, the first two authors introduced \eri, a reconstruction method essentially based  on the rank of flattening matrices obtained from a distribution of nucleotides at the leaves of a tree.
The method allows the possibility of dealing with data from mixtures of distributions with up to 3 categories (this upper bound is a theoretical restriction, not computational).
%
%
In the more recent method \saq, \citet{casfergar2020} create a more sophisticated method combining the rank of the flattening matrices with the stochastic information available in the data via a result of \citet{AllmanSemialg}.
%
Both methods \eri and \saq, as well as their associated weighting system, are brefly described in Appendix A.1.
%
%
%
They have been widely studied and the reader is referred to the cited publications for their performance on different scenarios.

As already explained in the introduction, while \saq usually outperforms \eri for short
alignments (length $\leq$ 1000), \eri obtains better results as the length of the alignment increases. This consideration leads us to introduce \asaq, a new combined method of \eri and \saq that tries to apply one or the other according to whether the input pattern distribution is consistent with
the positivity of the estimated length of the interior branch. To this end we use the paralinear distance and the paralinear method (see Appendix A.1, Lemma A.1 in the Appendix, and \citet{lake1994}).
%

For a distribution of patterns at the set of leaves, $p\in \mathbb{R}^{256}$, we compute all paralinear distances $d_{x,y}$ between pairs $x,y \in [4]$.
%
Then, given a bipartition $A|B$ of the set $[4]$, $A=\{i,j\}$, $B=\{k,l\}$, we define the following quantity 
\begin{eqnarray}\label{eq:defgamma}
	\gg{A|B}(p)=\min\{d_{i,k}+d_{j,l} \, , \, d_{i,d}+d_{j,l}\}-d_{i,j}-d_{k,l}.
\end{eqnarray}
{The quantity above is the ``neighborliness'' measure used in \citet{gascuel94}, the ``paralinear method'' used in \cite{lake1994}, and was presented by  \cite{Buneman} as a measure of twice the length at the interior edge (when $d$ is a tree metric).
	It is worth pointing out that at most one of the three values $\gg{12|34}(p)$, $\gg{13|24}(p)$, and $\gg{14|23}(p)$ is strictly positive, see Lemma \ref{lemma:def_gamma}. 
	We denote by $\texttt{PL}(p)$ the collection of these values:
	$$\texttt{PL}(p)=(\gg{12|34}(p), \gg{13|24}(p), \gg{14|23}(p)).$$}
If $p$ has arisen on a quartet $A|B$ and
the entries of the Markov matrix at the interior edge were strictly positive, both quantities inside the minimum in \eqref{eq:defgamma} are equal (as can be deduced from the 4-point condition). Moreover, $\gg{A|B}(p)$ is the unique positive quantity in the triplet $\texttt{PL}(p)$ and its value coincides with twice the paralinear distance of the interior edge (see Theorem \ref{main:app}).
The \emph{paralinear method} is the quartet reconstruction method that outputs the tree $T$ with highest $\gg{T}$ value. It is a statistically consistent quartet-inference method that satisfies the ``strong property II" as stated in the quoted paper \cite{sumner2017} (see Theorem \ref{main:app} of the Appendix).

Therefore, we propose the following quartet reconstruction method \asaq: it checks whether both methods \eri and the paralinear method \pl  output the same quartet, and
\begin{itemize}
\item [(1)] if \eri and \pl agree, then \asaq outputs the topology and weights of \eri;
\item [(2)] if they do not agree, then \asaq outputs the topology and weights of \saq
\end{itemize}
Note that an inconsistency between \eri and the paralinear method implies that the algebraic conditions used by \eri are not in concordance with the fact that the substitution parameters must be \emph{stochastic} (note the role of the positiveness of the entries of the Markov matrix to prove that $d_{x,y}\geq 0$ in Lemma \ref{paralinearMarkov}). The positiviness of transition matrices implies semi-algebraic conditions on the joint distributions at the leaves \citep{AllmanSemialg}. In this case we rely on \saq, as it is the unique method that takes into account both the semi-algebraic and the algebraic constraints. In  Theorem \ref{main:app} we prove  that \asaq is as well a statistically consistent quartet reconstruction method for the general Markov model.

We cannot claim that \asaq is statistically consistent for mixtures because consistency is not known to hold for the paralinear method or \saq  in this scenario. However,  the simulation studies in \citet{casfergarrote2020} show a good performance of \saq in mixture data from the same tree and this will lead to a good performance of \asaq as well (see the Results section). We denote by \asaq ($m=k$) the use of \asaq with \eri estimating mixtures on the same tree with $k$ categories.   The limit on the number of categories $m=3$ for quartets comes from the theoretical foundations of \eri, as a larger amount of categories would make unfeasible the identifiability of the tree topology by this method.

As a topology reconstruction method, the paralinear method is highly successful (see Results section, Figure \ref{treespaces}). Nevertheless, we found that using the paralinear method in order to ratify or not the results of \eri gives a better performance on topology reconstruction.


\subsection{Quartet-based methods (Q-methods)}
\label{sec:methods_quartets}
We have implemented different quartet-based methods (Q-methods) with different input weights. Quartet Puzzling (QP), Weight Optimization (WO) and Willson's (WIL) methods have been programmed in C\verb!++!.
%

QP amalgamates quartets on a randomized order and seeks to maximize the total sum of weights. Weight Optimization uses quartet weights to dynamically define the taxon addition order, seeking to maximize the total weight at each step.  
WO is known to reconstruct the correct tree if the input quartets are correctly weighted. Instead of constructing a tree that maximizes the total weight at each step, the essential idea of Willson's method is attaching new taxa in such a way that the new tree at each step is highly consistent with the input quartets. 
All these algorithms are initialized at a random 4-tuple.

Since the output of Q-methods strongly depends on the choice of the initial quartet, each one of them has been applied 100 times to each alignment and then the majority rule consensus tree (briefly MRCT) of these $100$ replicates has been computed. In order to evaluate the difference between two trees we use the Robinson-Foulds distance (RF for short), \citep{RobinsonFoulds81}. For the computation of the majority rule consensus tree and the RF distance the available functions in the Python Library \textit{DendroPy} have been used, see \citet{dendropy}.

\paragraph{Input Weights}

We require the weights of the quartet reconstruction methods to be positive and normalized.
Details about the input weights obtained from \asaq are provided in the previous section.
Further
details about the weighting system for all the considered methods are moved to Appendix A.1.
For the paralinear method the weights are denoted as \pl and are obtained after normalizing the exponentials of the scores given by \eqref{eq:defgamma}. We consider already published weighting systems for \saq, \eri and maximum likelihood (\ml): see \citet{fercas2016} for \eri, \cite{casfergar2020} for \saq, and the posterior probabilities used in \citet{strimmer_bayesian1997} for \ml.  
%



Maximum likelihood computations have been performed assuming the most general continuous-time homogeneous model (same instantaneous rate matrix throughout the tree but no constraints on the entries of the rate matrix or assumption of stationarity), which will be denoted as \mll(homGMc).
To this aim we used the \textit{baseml} program from the \textit{PAML} library \citep{yang1997}
with the UNREST model and let it infer the instantaneous rate matrix (common to all lineages) and the distribution at the root.


\subsection{Description of the simulated data}\label{sec_simul}


We consider two different scenarios of simulated data: one for testing \asaq as a quartet reconstruction method (described in Section \ref{sec:simdata_quartet}) and another for testing different Q-methods with several weighting systems (see Section \ref{sec:simdata_large}). For the first we use the simulated data introduced by \citet{fercas2016} whereas for the second we follow the approach of \citet{Ranwez2001} and consider $12$-leaf trees.

%
%


\paragraph{Evolutionary models} For the trees described below, we generate data evolving either on a general Markov model (GM, see section 2.1) or on a homogeneous general time-reversible model (GTR). By a homogeneous GTR model we mean a continuous-time GTR model that shares the same instantaneous mutation rate matrix $Q$ across all branches of the tree. GTR data have been generated using \texttt{Seq-gen} \citep{Rambaut1997} while GM data have been generated using the software \texttt{GenNon-h} \citep{GenNonh}.


\subsubsection{Simulated data for quartet reconstruction}\label{sec:simdata_quartet}

\paragraph{Tree space}  
The first data set we use to test \asaq corresponds to the tree space suggested by \citet{huelsenbeck1995}. We consider quartets as in Figure \ref{tree}.a in the Appendix, with branch lengths given by a pair of parameters $a$ and $b$ which vary between 0 and 1.5 in steps of 0.02. Branch lengths are always measured as the expected number of elapsed substitutions per site. The resulting \textit{tree space} is shown in Figure \ref{tree}.b in the Appendix. The upper left region of this tree space corresponds to the ``Felsenstein zone'', which contains trees subject to the long branch attraction phenomenon.
For each of the two nucleotide substitution models considered in the paper, GM and GTR, and for each pair $(a,b)$ of branch lengths, we have simulated one hundred alignments. The considered alignment lengths are of 500, 1 000 and 10 000 sites.

\paragraph{Random branch lengths} 
Following \citet{casfergar2020}, we test \asaq 
on 10 000 alignments generated from quartets whose branch lengths are randomly generated according to a uniform distribution in the intervals $(0,1)$ or $(0,3)$. These alignments are obtained according to both substitution models, GM and GTR, and are either 1 000 or 10 000 sites long. We represent the weights output by \asaq in a ternary plot (also called a simplex plot) as in \cite{strimmer1997}. 

\paragraph{Mixture data} 
The performance of \asaq is also tested in the scenario of data sampled from a mixture of distributions.
According to the approach of \citet{Kolaczkowski2004}, we consider the mixture of distributions as follows.
%
We partition the alignment into two categories of the same sample size both evolving under the GM model on the same quartet topology as Figure \ref{tree}.a but the first category corresponds to branch lengths $a = 0.05$, $b = 0.75$, while the second corresponds to $a = 0.75$ and $b = 0.05$ (see Figure \ref{kolaczkowski} in the Appendix). The internal branch length takes the same value in both categories and varies from 0.01 to 0.4 in steps of 0.05. The total length of the alignments considered is 1 000 or 10 000 sites.

\begin{figure}
	\begin{center}
		\includegraphics[width=15cm, height=6cm]{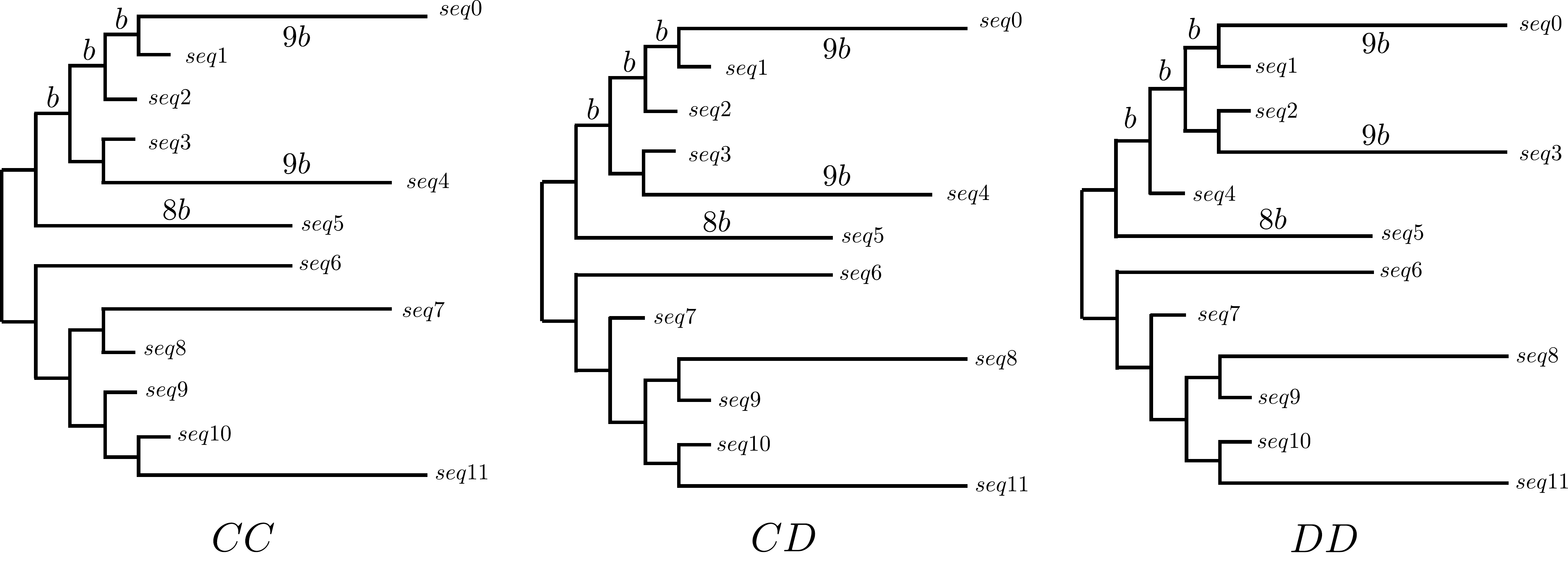}
	\end{center}
	\caption{\label{fig:TreeTopologies}  \footnotesize The three different tree topologies $CC$, $CD$ and $DD$ on 12 taxa considered to test the Q-methods with different weighting systems. They are obtained by glueing a combination of two trees ($C$ and $D$) by the root. Here, the parameter $b$ represents the length of internal branches. These tree topologies have been taken from \citep{Ranwez2001}.}
\end{figure}

\subsubsection{Simulated data for larger trees }\label{sec:simdata_large}


We followed \citet{Ranwez2001} to test the performance of different quartet-based methods (QP, WO and WIL) with different weighting systems. To this end, we considered the three 12-taxon topologies depicted in Figure \ref{fig:TreeTopologies} denoted as \textit{CC}, \textit{CD} and \textit{DD}, and fixed the ratio among branch lengths as in \cite{Ranwez2001}, depending on a parameter $b$ denoting the internal branch length, which
is varied in the set $\{0.005,0.015,0.05,0.1,0.25,0.5\}$. 

For each tree topology and for each $b$ we have considered 100 alignments generated under the GM model with lengths 600 (in order to match the alignment length considered in \cite{Ranwez2001}), 5 000, and 10 000. In addition, we have considered collections of 100 alignments of lengths 600 and 5 000 bp. generated on the tree \textit{DD} under the GTR model with $b=0.005,\ldots, 0.1$ (we chose this tree topology because it is arguably the hardest to reconstruct, see the Results section). 

\paragraph{Mixture data} We have considered a 2-category mixture model, 
that is, we generated alignments evolving on a the topology CD but whose sites evolve following two systems of substitution parameters: the first system corresponds to the branch lengths described in the tree \textit{CD}, while the second system corresponds to \textit{CD} after exchanging the branch lengths of $seq3$ and $seq4$, and the lengths of $seq7$ and $seq8$ (see Figure \ref{mixtureCD} for details). 
The parameters that were varied in this framework were the proportion $p$ of sites in the first category (which was varied in $0.25,0.50,$ and $0.75$) and  the internal edge length $b$ which was varied as above in $0.005$, $0.015$, $0.05$, and $0.1$. The lengths of the alignments considered were 600, 1 000 and 10~000~bp.

\subsection{Real data}\label{sec_real}

\paragraph{Yeast data} \label{sec_realdata}
We analyse the performance of \asaq on real data on the eight species of yeast studied in \citet{Rokas2003} with the concatenated alignment provided by \citet{jayaswal2014}. We investigate whether the quartets output by \asaq support the tree $T$ of \citet{Rokas2003}, the alternative tree $T'$ of \citet{Phillips2004} (see Figure \ref{fig:trees_simul}), or the mixture model proposed by \citet{jayaswal2014}. Although the tree $T$ is widely accepted by the community of biologists, its correct inference is known to depend on the correct management of heterogeneity across lineages, as an inaccurate underlying model usually reconstructs $T'$  \citep{Rokas2003, Phillips2004, jayaswal2014}. According to \citet{jayaswal2014} these data are best modeled by considering, apart from heterogeneity across lineages, two different rate categories (discrete $\Gamma$ distribution) plus invariable sites. In our setting, this is translated into a mixture distribution with 3 categories ($m=3$ in \asaq).

\paragraph{Ratites and tinamous mitochondrial data}
The phylogeny of ratites and tinamous has been debated largely \citep[see][and the references therein]{phillips2010}. There is evidence of a higher rate of evolution among the tinamous relative to the ratites   \citep[see e.g.]{paton2002}, so these data are likely to be analysed by the methods proposed here. Moreover, the recoding used in \citet{phillips2010} to sort out this problem is questioned in \citet{veraruiz2021}. We do our analyses using the third codon position in the mitochondrial alignment for 24 DNA sequences provided in this last paper.  We run WO and QP with weights obtained by \asaq, \saq, \eri, \fp, \pl. For each combination of methods we randomly selected as starting quartets either one hundred 2125 (approximately 20\% of all possible quartets), or 5313 quartets (equal to 50\%), and then we performed the MRCTs.

We analyse the results obtained in comparison to the following trees: (A) the tree that groups Tinamous and Moas as proposed by \citet{phillips2010}  and displayed in \citet[Fig. 4]{veraruiz2021},  (B) the mt consensus tree of  Figure 1a of \citet{phillips2010} and (C) the Ratite paraphyly tree of Figure 1b of \citet{phillips2010}. If we call CEK to the largely established clade CEK=((cassowary, emu), kiwis), then these trees can be summarized as:
\begin{itemize}
	\item[A:]  (outgroup, neognathus, (ostrich, (rheas, ((moas,  tinamous), CEK))));
	\item[B:] (outgroup, neognathus, (tinamous, (moas, rheas, (ostrich, CEK))));
	\item[C:] (outgroup, neognathus, ((tinamous, moas), (rheas, (ostrich, CEK)))).
\end{itemize}
As unrooted trees, A and C have 21 interior edges and B has 20.

\section{Results}


In this section, we describe the results obtained and benchmark them with published results
for the sake of completeness. The interested reader is referred to the corresponding papers for details of the methods therein.

\subsection{Results on quartets}

\subsubsection{Tree space}

The performance of \asaq and \pl on data generated on the tree space of section 2.3.1 is represented in Figure \ref{treespaces} (for GM data) and in Figure \ref{treespaces_GTR} in the Appendix (for GTR data).
In black we represent 100\% success, in white 0\% success, and gray tones correspond to regions of intermediate success accordingly. The 95 \% and 33 \% isoclines are represented with a white and black line, respectively.
These simulation studies show a consistent performance according to the results by \cite{huelsenbeck1995} and \cite{fercas2016}, with the usual decreasing performance at the Felsenstein zone and an improvement of both methods with sample size.
%
Figures \ref{treespaces} and \ref{treespaces_GTR} show  that the performance of \asaq is better than that of \pl, both for GM and GTR data.
%

For completeness, the average performance of \asaq and \pl on this tree space for different alignment lengths and underlying models is compared to other methods in Table \ref{tab:mean_sd}: we include the average results of \saq \citep[as shown in]{casfergar2020} and \eri and \ml as published in  \cite{fercas2016}. 
This comparison shows that for GM data the best results are achieved by \asaq, while for GTR data \ml obtains the best performance for alignments of length 500 and 1 000 bp. and \eri does so for  long alignments (10 000 bp.).

\begin{table}[]
\begin{center}
	\textbf{Average success of different methods on the tree space.}
	\vspace*{5mm}
	
	\begin{tabular}{cc|ccccc}
		\hline
		simulations & base pairs & \asaq & \saq & 
		\eri & \pl & \ml \\ \hline \hline
		\multirow{3}{*}{GM} & 500 & \textbf{85.3} & {84.6}  & 72.4   &  82.1  & 72.1\\
		& 1~000  & \textbf{90} & {88.8}  & 80.3  & 87.8  & 73.6 \\
		& 10~000  & \textbf{98.4} & 96.8 & 97.1 & 97.7 & 75.4\\
		\hline
		\multirow{3}{*}{GTR}  & 500 & 79.9 & 78.4 & 74.8  & 78.7  & \textbf{88.0}  \\
		& 1~000 & 86.9 & 83.5  & 84.3 & 85.8  & \textbf{93.4} \\
		& 10~000  & 97.9 & 94.5 & \textbf{99.2} & 96.9 & 98\\
		\hline
	\end{tabular}
\end{center}

\caption{\label{tab:mean_sd} \footnotesize
	Average success of several methods applied to data simulated on the tree space of Figure \ref{tree} b). \asaq and \pl are compared to the results for \saq obtained in \citep{casfergar2020}, and for \eri and maximum likelihood \ml  in \citep{fercas2016}.  \mll(homGMc)
	is applied when data are generated under a GM model, while \ml estimates a homogeneous GTR model when data are generated under GTR, see \cite[Table 1]{fercas2016}. In each row of the table, the highest success is indicated in bold font. }
\end{table}

\begin{figure}[]
\centering
\addtolength{\leftskip} {-2cm}
\addtolength{\rightskip}{-2cm}
\begin{center}
	\hspace{0.4cm} Performance on GM data for the treespace \\
	\vspace{0.1cm}
	\hspace{0.3cm} \asaq \hspace{6.1cm} \pl \\
	\vspace{-0.5cm}
	\includegraphics[scale=0.45]{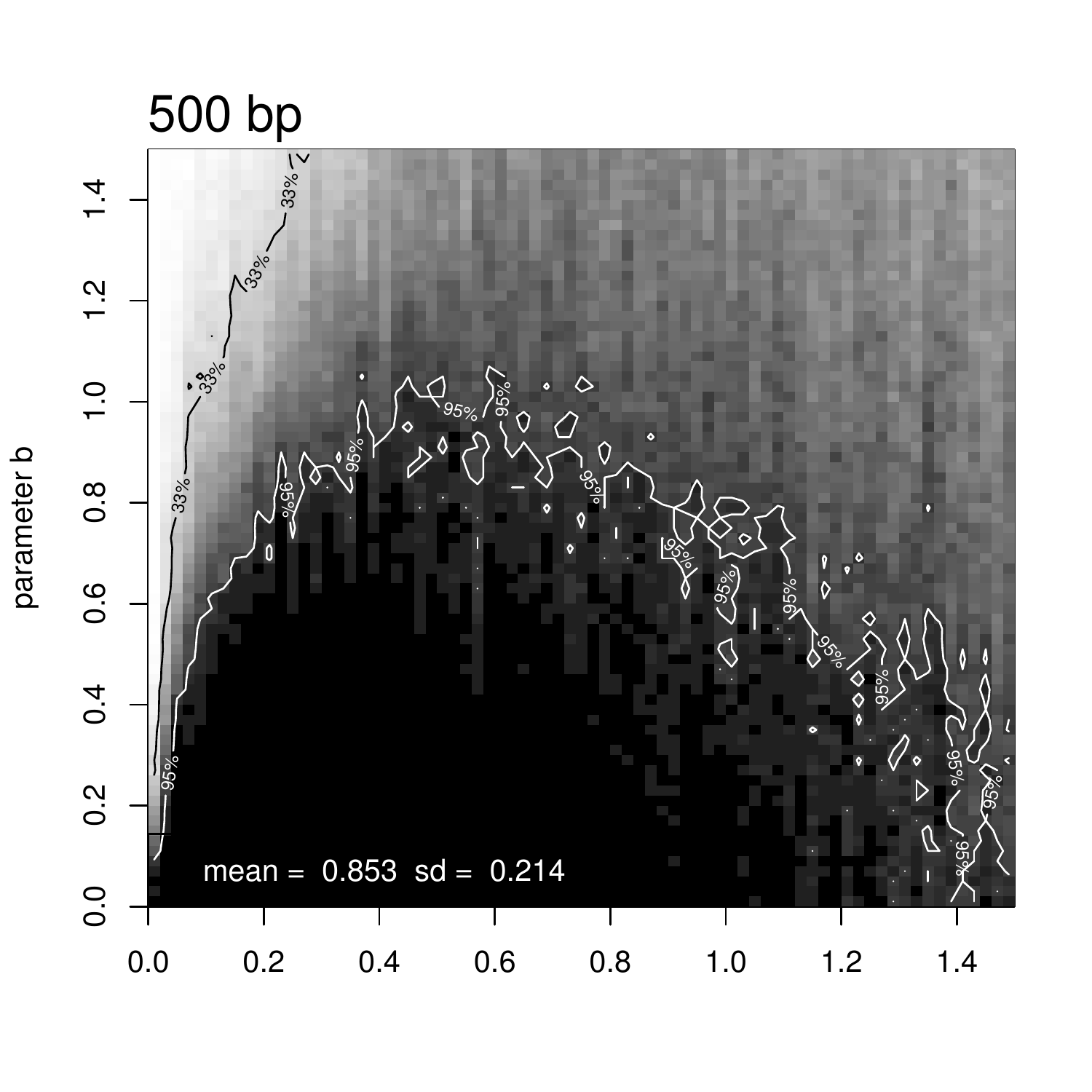}
	\includegraphics[scale=0.45]{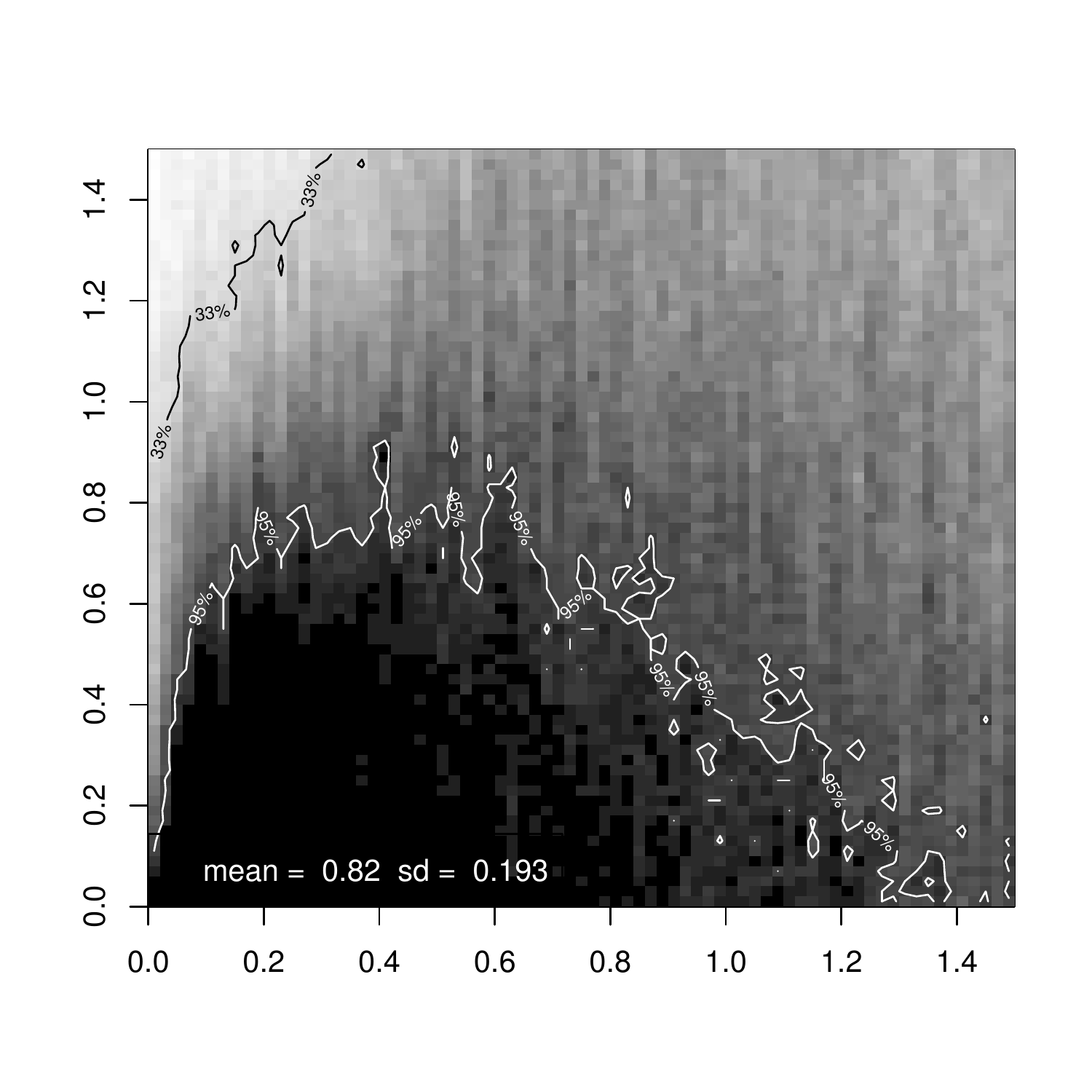}\\
	\includegraphics[scale=0.45]{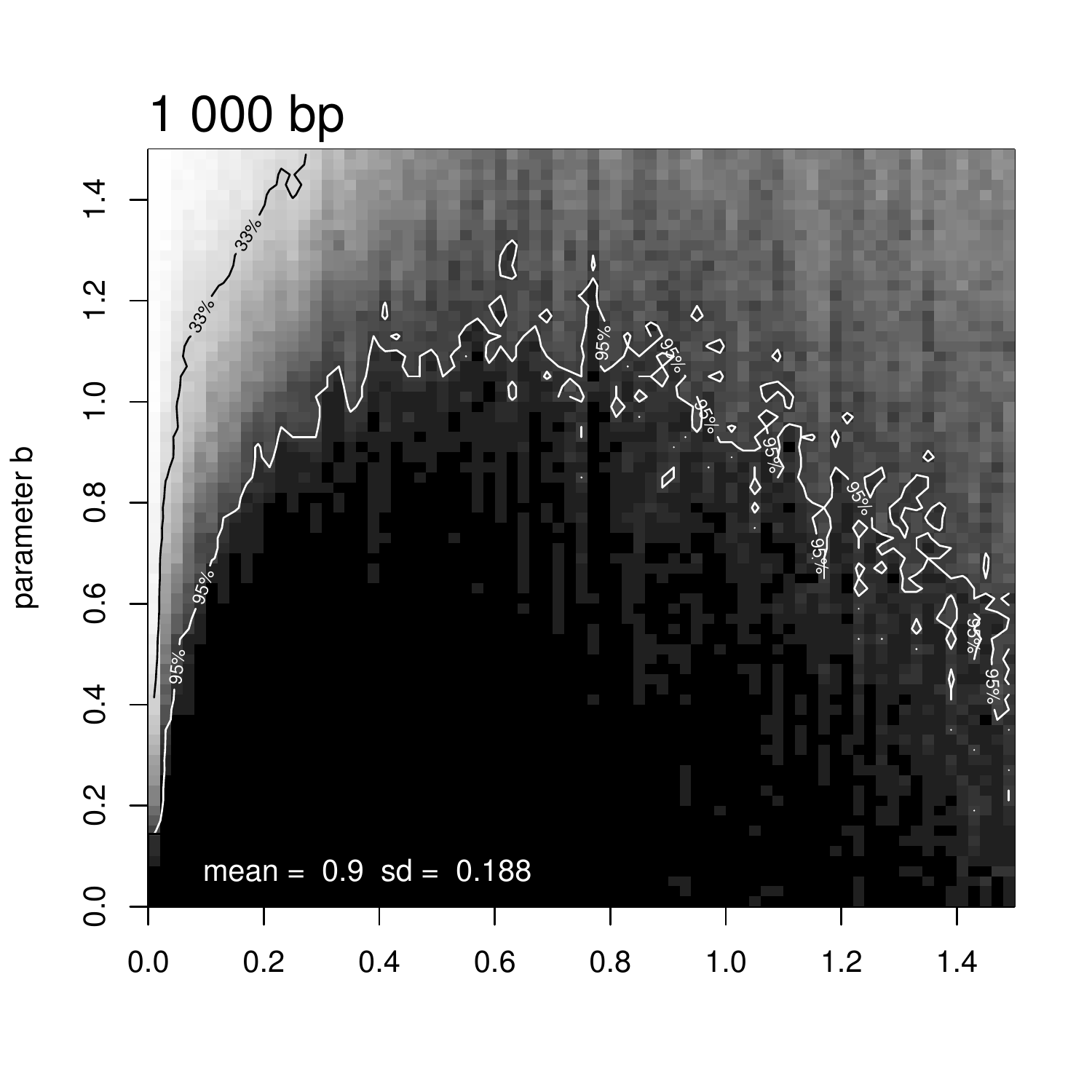}
	\includegraphics[scale=0.45]{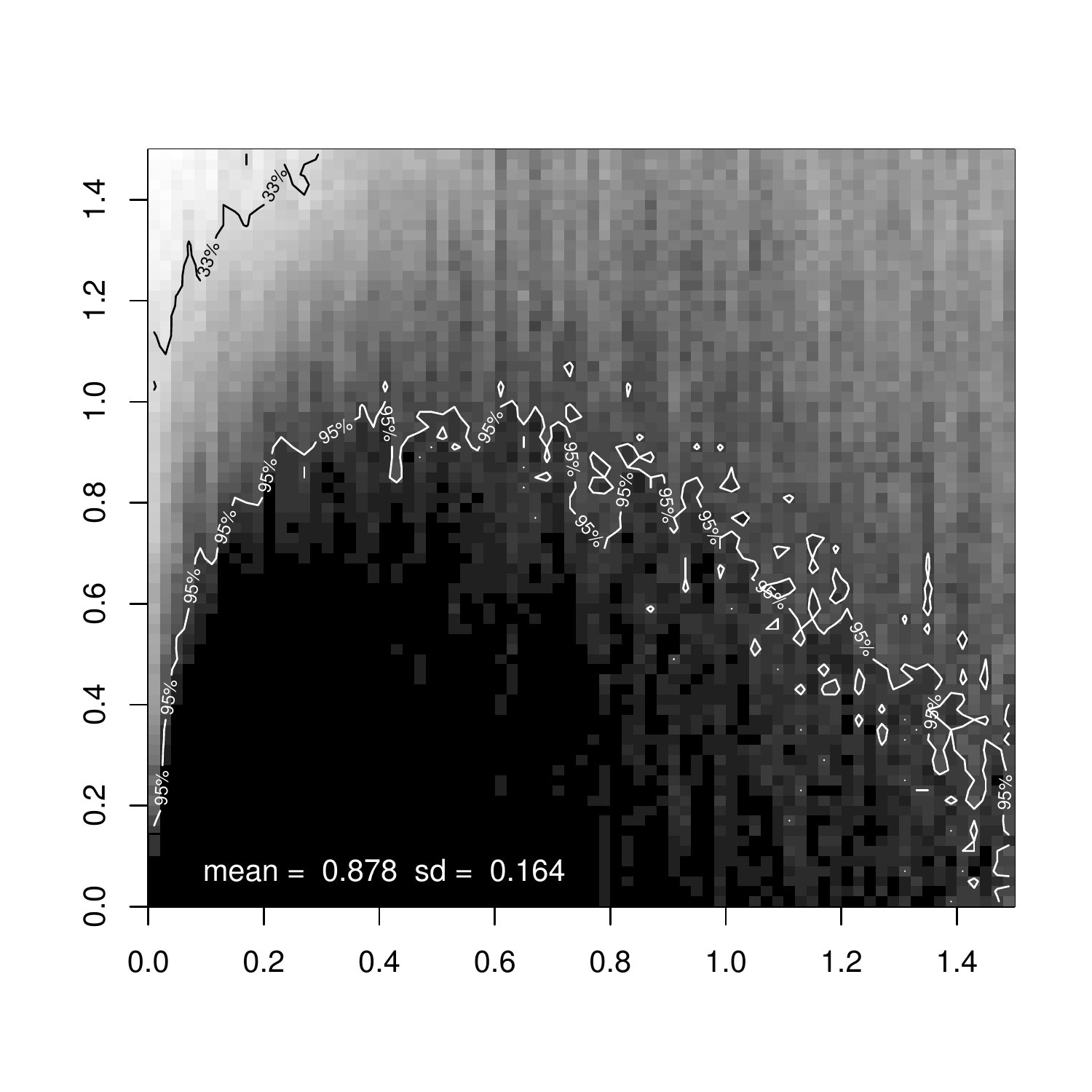}\\
	\includegraphics[scale=0.45]{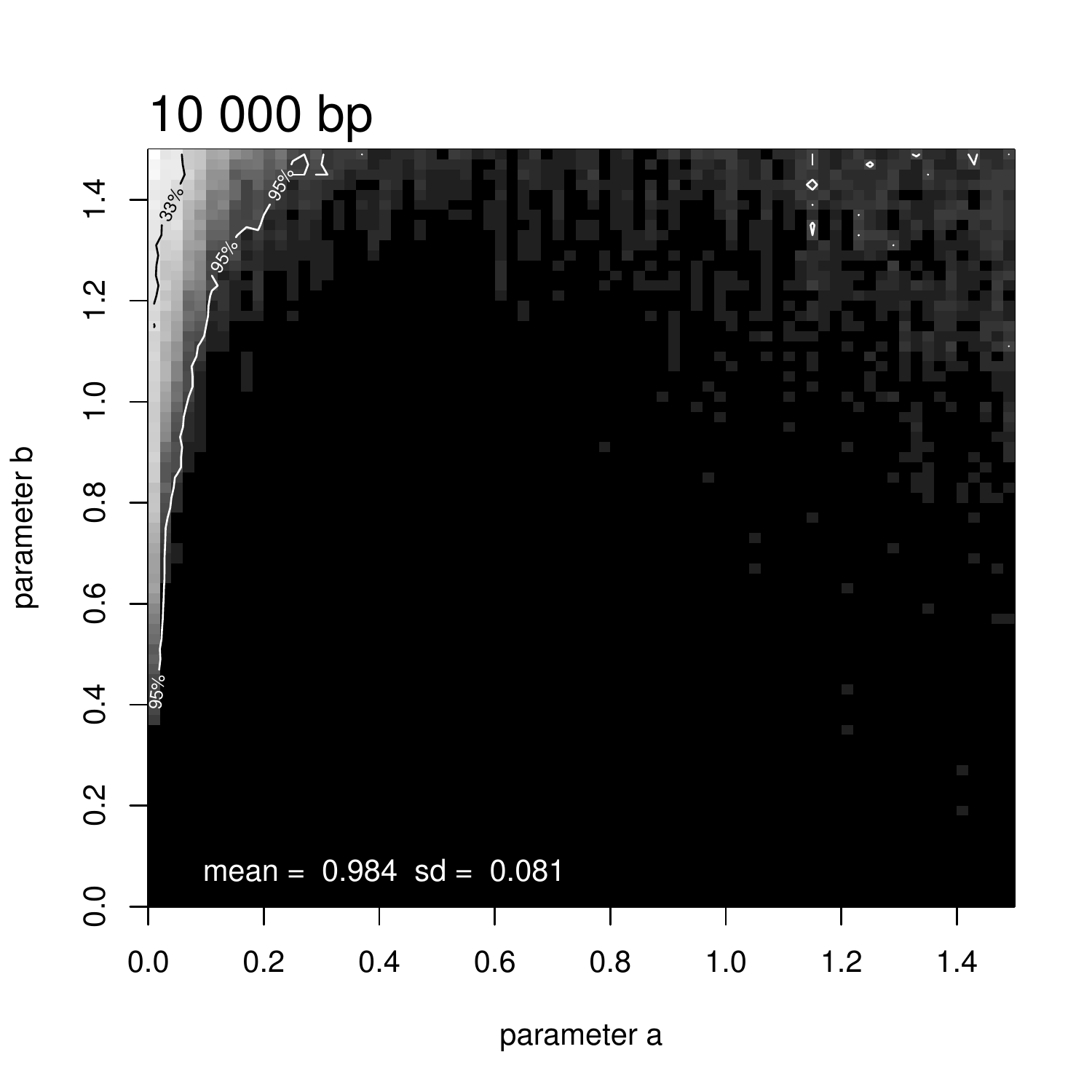}
	\includegraphics[scale=0.45]{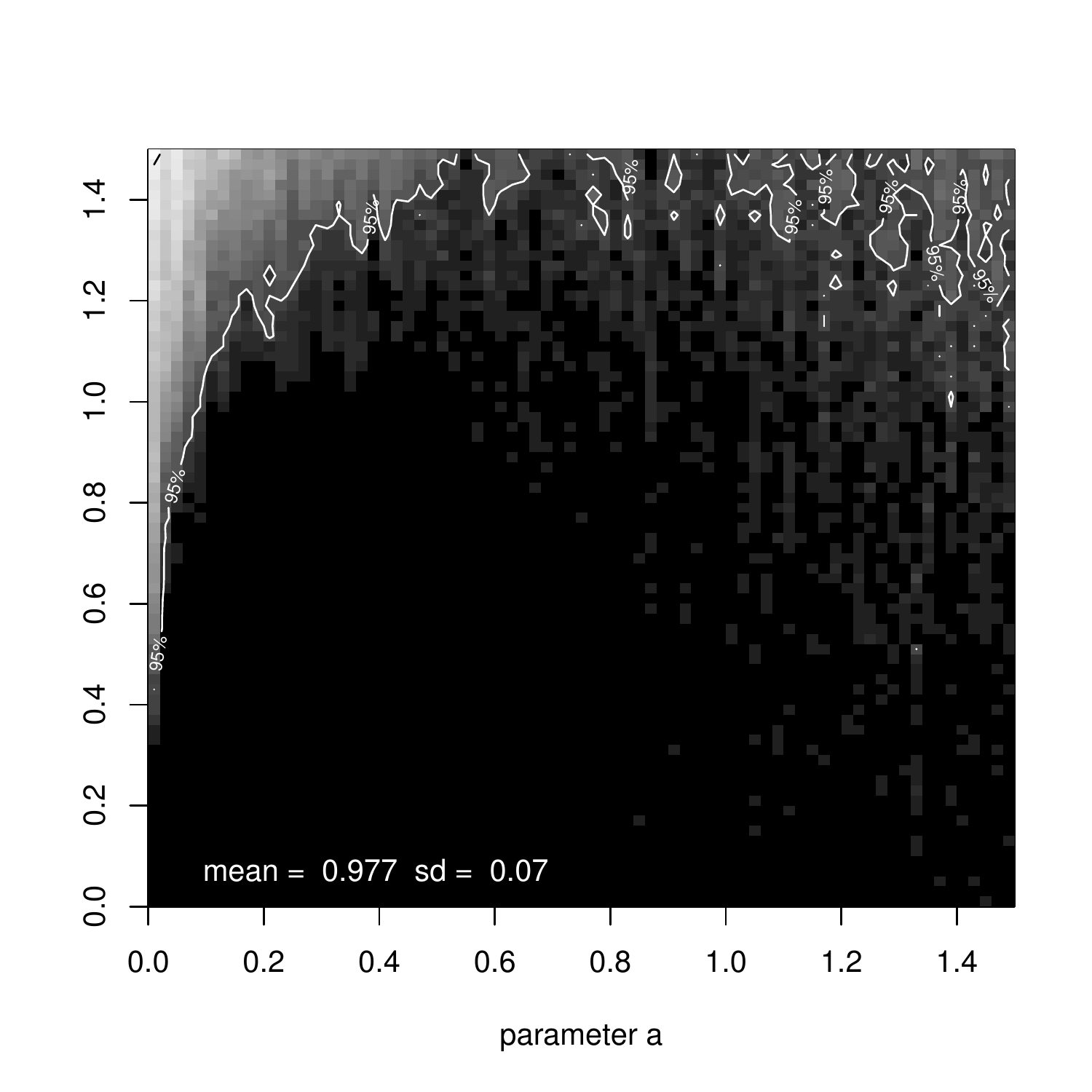}\\
	
\end{center}
\caption{\label{treespaces} \footnotesize
	Performance of \asaq (left) and \pl (right) in the tree space of Figure \ref{tree} b) on alignments of length  500 bp (top), 1 000 bp (middle) and 10 000 bp (bottom) generated under the GM model. Black is used to represent 100\% of successful quartet reconstruction, white to represent 0\%, and different tones of gray the intermediate frequencies. The 95\% contour line
	is drawn in white, whereas the 33\% contour line is drawn in black. 
}
\end{figure}

\subsubsection{Random branch lengths}
\begin{figure}[]
\centering
Performance on GM data for random branches\\
\includegraphics[scale=0.25]{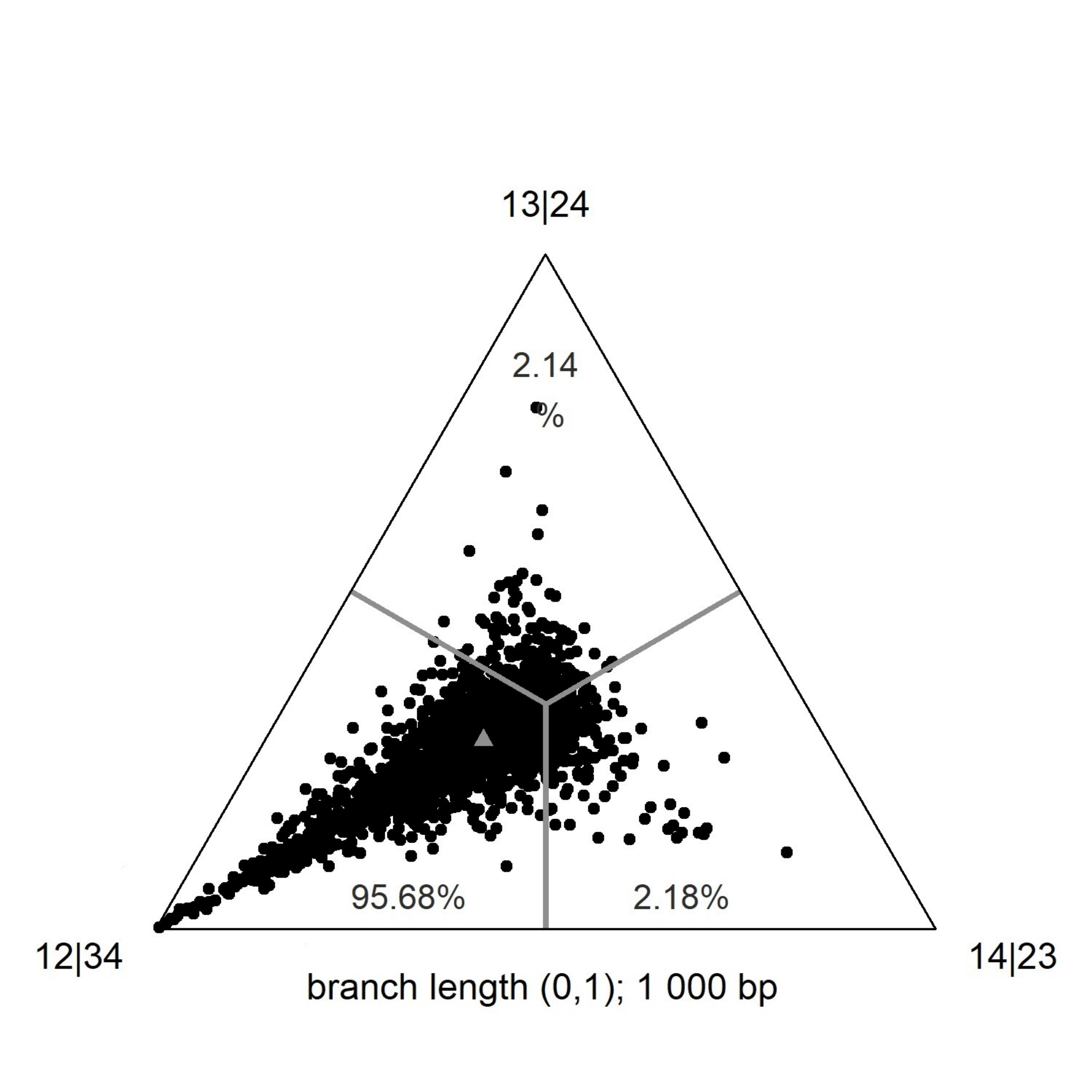}
\includegraphics[scale=0.25]{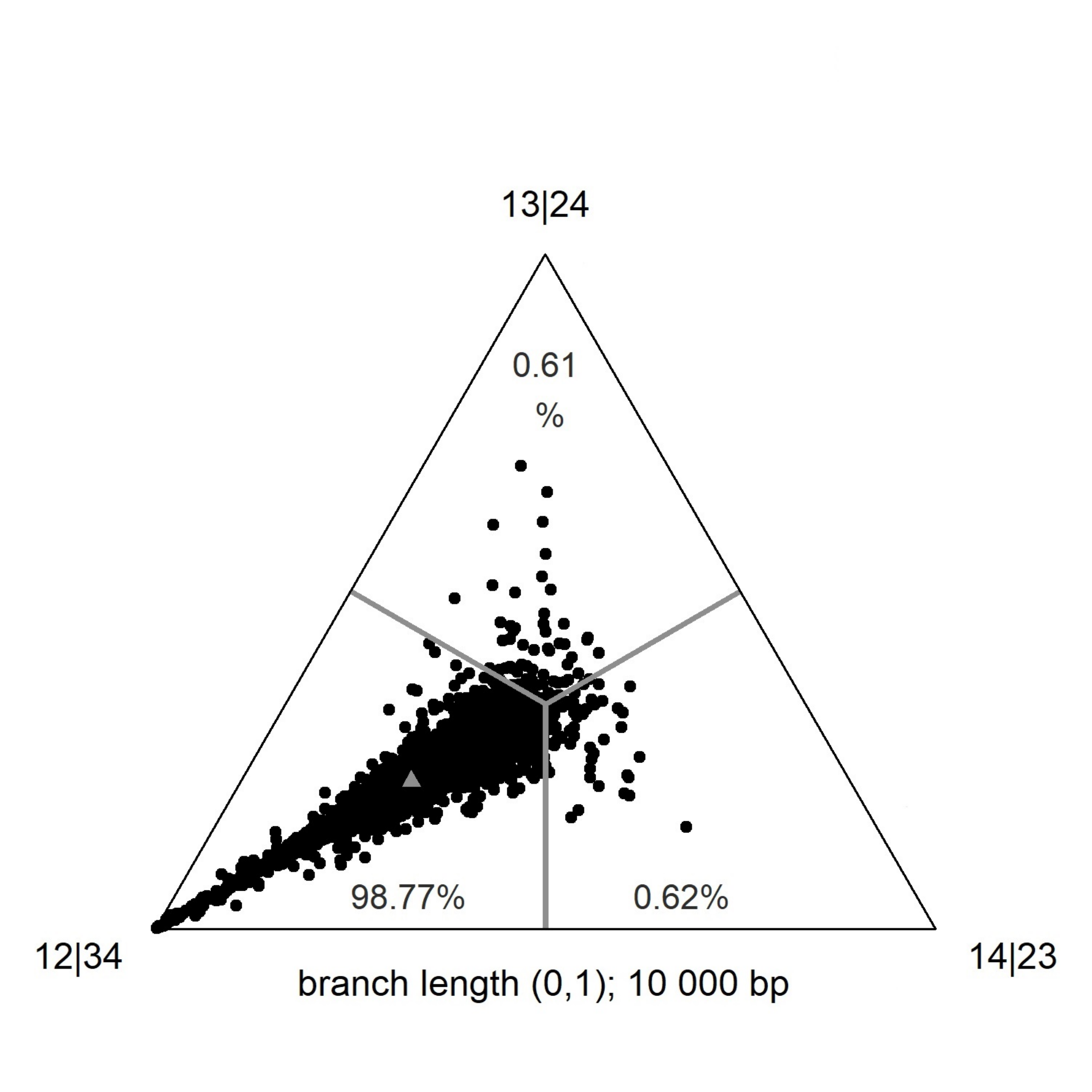} \\
\includegraphics[scale=0.25]{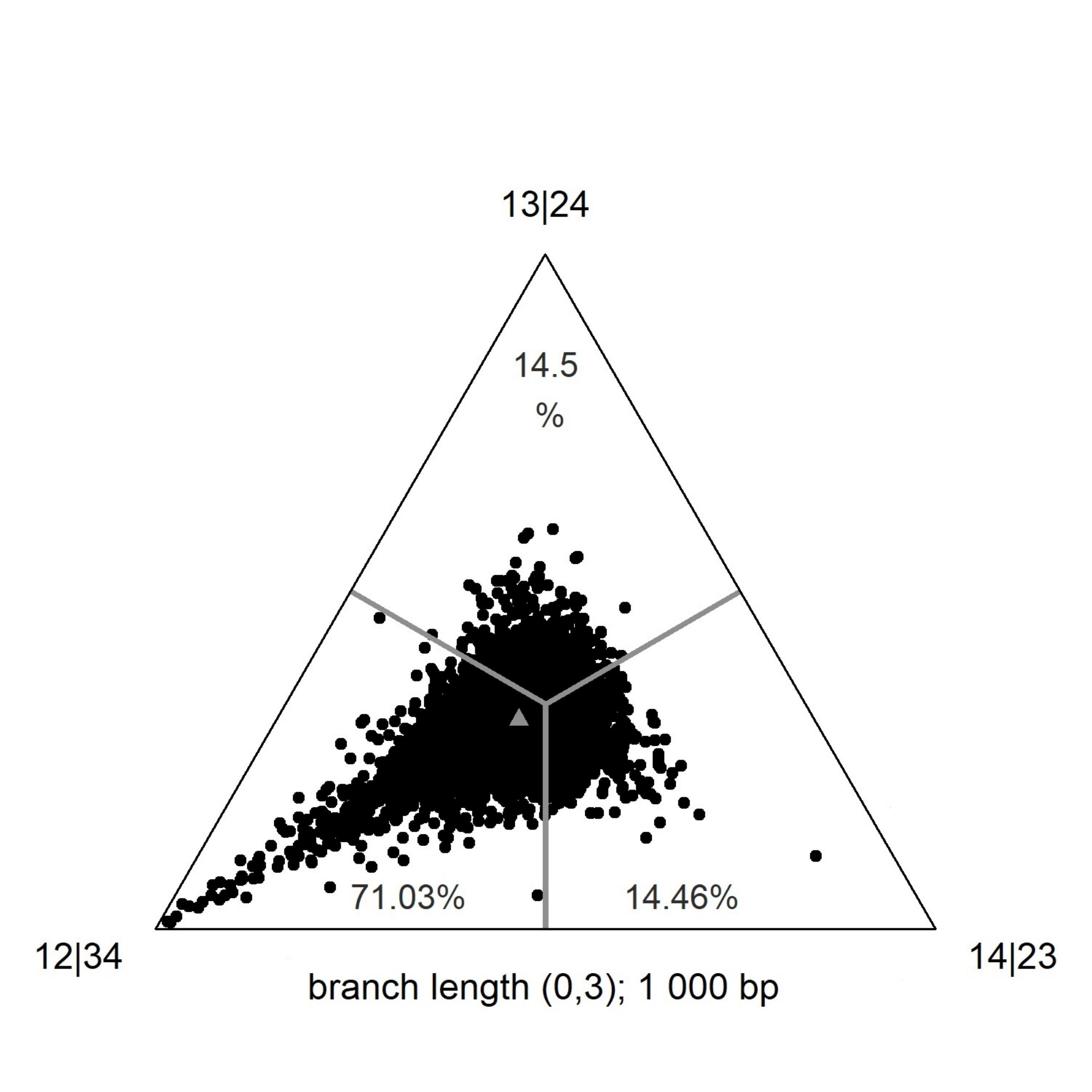}
\includegraphics[scale=0.25]{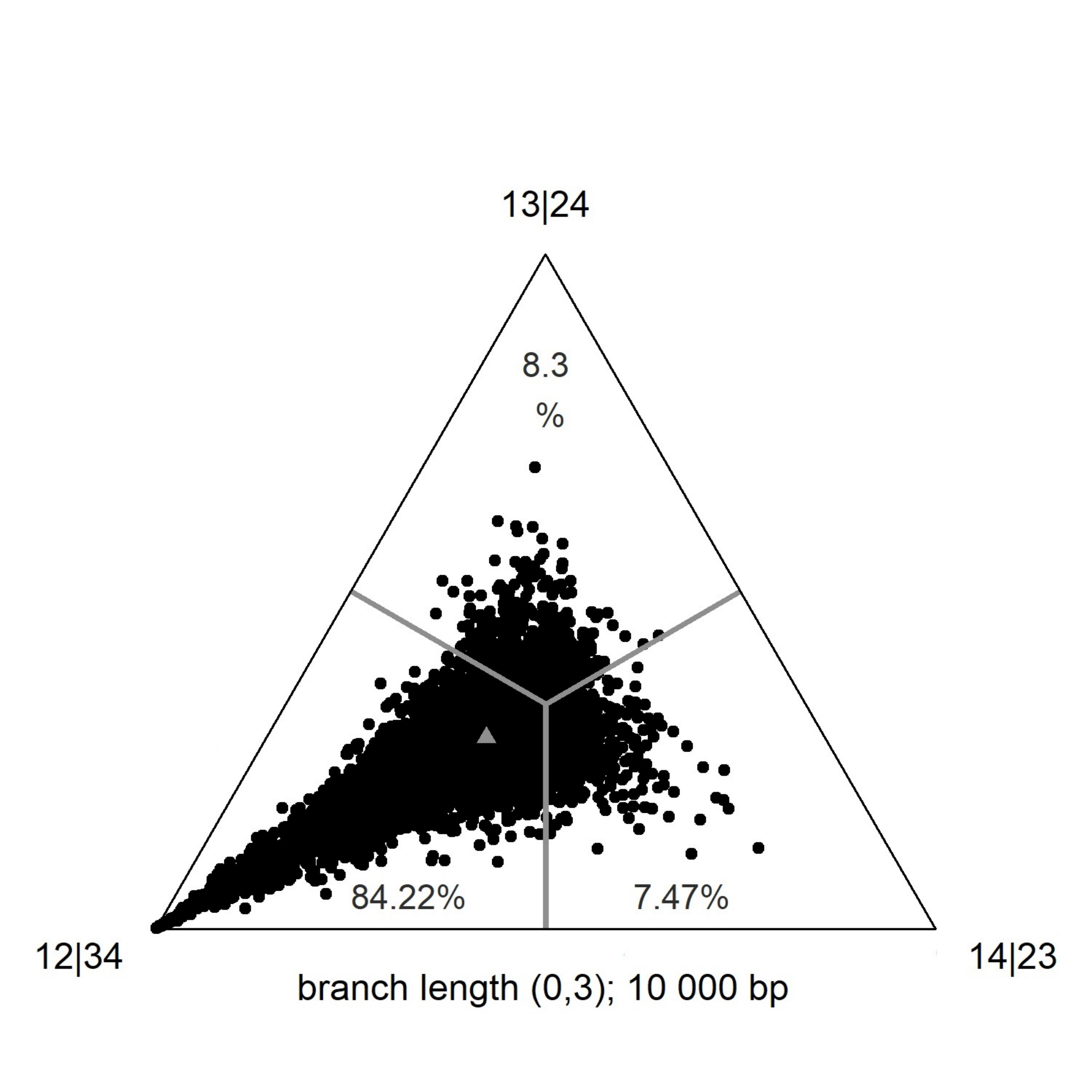}

\caption{\label{ternary_GM} \footnotesize
	Ternary plots corresponding to the weights of \asaq applied to 10000 alignments generated under the GM model on the  $12|34$ tree. On each triangle the bottom-left vertex represents the underlying tree $12|34$, the bottom-right vertex is the tree $13|24$ and the top vertex is $14|23$.
	The small grey triangle depicted represents the average point of all the dots in the figure.
	Top: correspond to trees with random branch lengths uniformly distributed between $0$ and $1$; bottom: random branch lengths uniformly distributed between $0$ and $3$. Left : 1 000 bp; Right: 10 000 bp.}
\end{figure}

To visualize the overall distribution of the weights of \asaq  applied to trees with random branch lengths, in Figure \ref{ternary_GM} we show ternary plots corresponding to the GM alignments described in section \ref{sec:simdata_quartet}.
The ternary plots of the performance of \asaq when applied to the same setting with data generated under the GTR model are shown in Figure \ref{ternary_GTR}. In Table~\ref{tab:random_branches} we display the summary of average success of \asaq on these data.

Note that Table~\ref{tab:random_branches} shows a high performance of the method and the ternary plots show a clear distribution of points towards the bottom left corner (which represents the correct quartet) and weights symmetrically  distributed on the other corners. In particular, the method is not biased towards any of the incorrect topologies. We note that the level of success exhibited is quite sensitive to the branch length, being much higher for branch lengths in (0,1) than in (0,3).
We do not appreciate a remarkable difference between the performance of \asaq when applied to GTR data.

\vspace{3mm}
\begin{table}[h]
\textbf{Average success of \asaq applied to data generated on $12|34$ with random branch lengths}
\begin{center}
	\def\arraystretch{1}		
	\begin{tabular}{c|cc|cc}
		\hline
		\multirow{2}{*}{branch length} & \multicolumn{2}{c|}{GM} & \multicolumn{2}{c}{GTR} \\ \cline{2-5}
		& 1 000 bp.  & 10 000 bp. & 1 000 bp.   & 10 000 bp.   \\ \hline \hline
		(0,1)                 & 95.68      & 98.77      & 94.65       & 98.42        \\
		(0,3)                 & 71.03      & 84.22      & 69.37       & 85.12 \\
		\hline
	\end{tabular}	
\end{center}

\caption{\label{tab:random_branches} \footnotesize Average success of \asaq on alignments of lengths 1 000 and 10 000 bp. generated on the tree $12|34$ under the GM and GTR models with random branch lenghts uniformely distributed in (0,1) (first row) and (0,3) (second row). The plots corresponding to these data are shown in Figures \ref{ternary_GM} and \ref{ternary_GTR}.}
\end{table}

\subsubsection{Mixture data}

\begin{figure}[h]
\centering
Performance on mixture data \\
\includegraphics[scale=0.65]{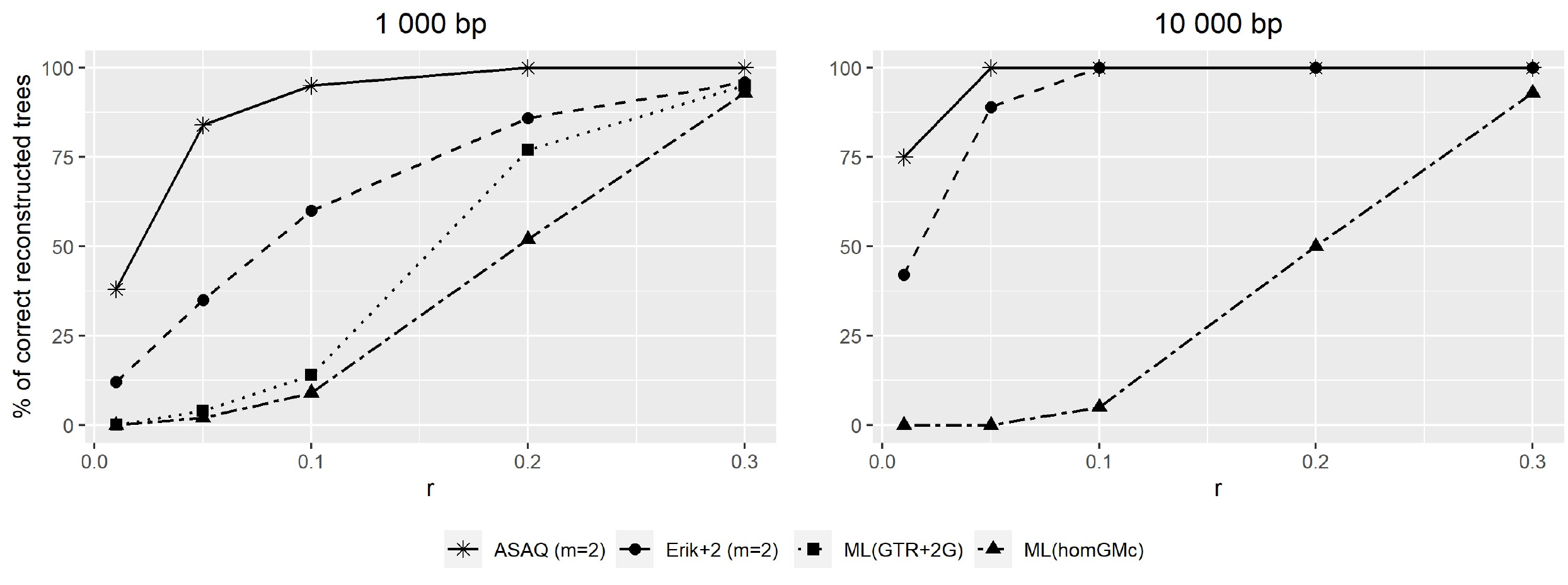}
\captionsetup{width=1.1\textwidth}
\caption{\label{fig:barplot_mixtures} \footnotesize These plots represent the percentage of correctly reconstructed trees by several methods applied to the mixture data described in the Method section; the value $r$ refers to the branch length of the interior edge of the trees (see Figure \ref{kolaczkowski}).
	We compare the results of \asaq ($m=2$) and the results presented in \cite[Figure 5]{fercas2016} for
	\eri ($m=2$), \mll(homGMc) and a \ml estimating a heterogeneous across lineages GTR with two categories of discrete $\gamma$ rates across sites, \mll(GTR+2$\Gamma$) (only for length 1 000 bp.).
}
\end{figure}

In Figure \ref{fig:barplot_mixtures} and Table \ref{tab_mixt_p50} we show the performance of the method \asaq with $m=2$ categories when applied to data from mixtures as described in section \ref{sec:simdata_quartet} and in Figure \ref{kolaczkowski}.
Based on the results of Figure 5 in \cite{fercas2016}, we also provide a comparison to the success of \eri $(m=2)$ and two versions of maximum likelihood on the same data.
We did not include the performance of \saq, as it is very similar to that of \asaq.
%
In Table \ref{tab_mixt_p50}, the interested reader will also find the results obtained by applying \asaq with $m=1$ category to the same data. The performance of \asaq ($m=1$) is similar to \pl in general, and to \asaq (m=2) for short alignments. As the length of the alignment increases, we note that \asaq (m=2) outperforms both the unmixed \asaq and \pl.

Figure \ref{fig:barplot_mixtures} shows an increasing accuracy of all methods when the value of the parameter $r$ (the branch length of the interior edge of the two trees involved) is increased. This was expected as larger values of $r$ represent larger divergence between sequences at the left and the right of the trees.
We note that  \asaq (m=2) outperforms the other methods, with a high level of success in all cases, even when the length of the alignments is 1 000 bp.
The average success of \asaq $(m=2)$ applied to the simulated data is $96.64\%$ for 1 000 bp., $99\%$ for 10 000 bp. and $99.92\%$ for 100~000 bp (results not shown).

\subsection{Results of Q-methods}

We proceed to describe the performance of WO, WIL and QP (see section \ref{sec:methods_quartets}) applied to the input weights from \asaq, \saq, \eri, \pl and \fp on the trees $CC$, $CD$ and $DD$ presented in section \ref{sec:simdata_large}. The weights from \pl and \fp are both defined in terms of the paralinear distance (see Appendix A.1) and they produce similar results. Because of this, we omit the performance of \fp in some cases.
We also add the results obtained of a global \nj applied to the same data.
To this end we used the \nj algorithm implemented in the R package \texttt{APE} \citep{ape} with paralinear distance.

\subsubsection{Unmixed data}

\begin{figure}
\centering
%
\includegraphics[scale=0.5]{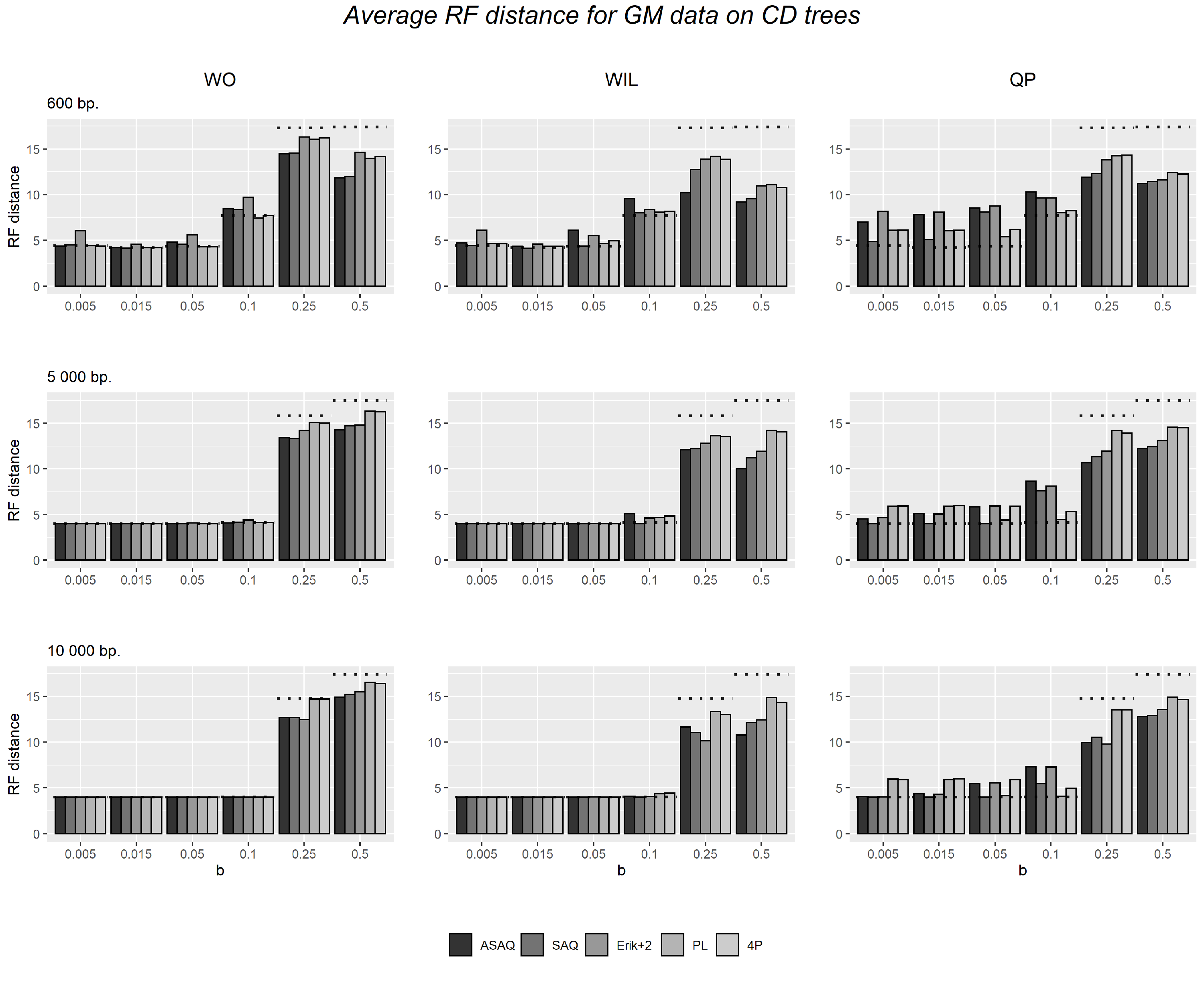}
\captionsetup{width=1\textwidth}
\caption{\label{fig:CDgennonh}  \footnotesize Average Robinson-Foulds distance for GM data simulated on the tree \CD with alignment length 600 bp. (above), 5 000 bp. (middle) and 10 000 bp. (below).
	The Q-methods WO (left), WIL (center) and QP (right) are applied with different systems of weights, namely \asaq, \eri, \pl and \fp.
	The dotted horizontal lines represent the average RF distance of the tree reconstructed using a \emph{global} \nj with paralinear distance. Concrete values of these results are detailed in Table \ref{tab_CDgennonh}.
}
\end{figure}

The results on simulated GM data obtained by the combination of methods and weights presented in this paper
are summarized in Figure \ref{fig:CDgennonh} for  \CD, in Figure \ref{fig:CCgennonh} for \CC, and in Figure \ref{fig:DDgennonh} for \DD. The results for GTR data (for the tree \DD) are shown in Figure \ref{fig:DDgtr}.
The height of the bars shows the average of the RF distance from the original tree to the consensus tree of 100 replicates for each of the 100 generated alignments.
We also plot in the figures the result of a global \nj with paralinear distance.
The values of these results are detailed in Tables \ref{tab_CCgennonh} (for \CC), \ref{tab_CDgennonh} (for \CD) and \ref{tab_DDgennonh} (for \DD) for GM data and in \ref{tab_DDgtr} for GTR data on \DD. In these tables we also include the results obtained when applying \ml weights. Since \ml did not converge for some quartets (specially in the presence of long branches and when trying to maximize the  likelihood for GM data generated on another quartet), we write between parentheses the number of alignments considered for \ml (in the computation of the average RF distance we neglected the alignments where \ml did not converge).

By comparing the performance of the three Q-methods, we observe a slightly better performance of WIL and WO compared to QP. In all the methods the accuracy drops for long branches, but  WO seems to perform slightly better for short and medium branches ($b\leq 0.1$) while WIL does better for longer branches $b\geq 0.25$. Among the three Q-methods, QP is the most sensitive to the election of the system of weights.

Another general remark is that all Q-methods with weights from \asaq, \saq, \eri and \pl outperform \nj for $b>0.1$ on GM data, while none of them (with any system of weights) beats a global NJ for  $b$ smaller than or equal to 0.1. Figure \ref{fig:DDgtr} depicts that a global NJ is the best option for GTR data on \DD trees.


The reconstruction algorithms have had considerably more success when reconstructing the tree \CC in constrast to \CD or \DD for GM data, in concordance with the results obtained by \cite{Ranwez2001}.
In the tree topologies \CD and \DD, the distance between $seq9$ and $seq10$ is $4b$, while the distance between $seq8$ and $seq11$ is $20b$. The same happens between species $seq1$ and $seq2$, and $seq0$ and $seq3$ of the \DD topology. Thus, for an alignment generated from these trees, there is a  high probability that two separate lineages evolve in a convergent manner to the same nucleotide at the same site, creating a \textit{long branch attraction} situation and making the reconstruction methods to infer the wrong topology.
We note specially good results of WO and WIL when dealing with  \CC and \CD trees (see Figures \ref{fig:CCgennonh} and \ref{fig:CDgennonh}) and short branches, probably because these Q-methods succeed in reconstructing the $C$ subtree.

Weights from \asaq, \saq, \eri, \pl and \fp produce overall comparable results, although \asaq and \saq do better when dealing with long branches. We also note an improvement in the results when the length of the alignment increase, specially for the \CC case or for GTR data (Fig \ref{fig:DDgtr}) and hardly noticeable for the other two trees on GM data.

The simulation study shows a big difference between the general performance of the Q-methods with input weights from \asaq, \eri and \pl in contrast to weights obtained by \ml, specially when reconstructing the \CC and \CD trees (see Tables  \ref{tab_CCgennonh}, \ref{tab_CDgennonh} and \ref{tab_DDgennonh}).
This is probably due to the inconsistence beetween the \ml weights computed and the general Markov model used to generate the data (see subsections 2.2 and 2.3). However, even when the model used for ML estimation  matches the model that generated the data (as for GTR, see Table \ref{tab_DDgtr}), \ml seems to be the worse weighting system.
When data are generated under the GTR model on \DD trees (Figure \ref{fig:DDgtr} and Table \ref{tab_DDgtr}), all Q-methods except \ml have a remarkable high performance when alignments are long enough ($\geq$ 5 000 bp.), specially WO.
For short alignments, \pl seems a good choice for the input weights with these data.


All in all, we observe that both \asaq and \pl weights (combined with one or another Q-method) give rise  to good results, better than \ml and comparable with the results obtained by a global \nj and even better when dealing with long branches.



%
%
%

\subsubsection{Mixture data}

\begin{figure}
\centering
\includegraphics[scale=0.6]{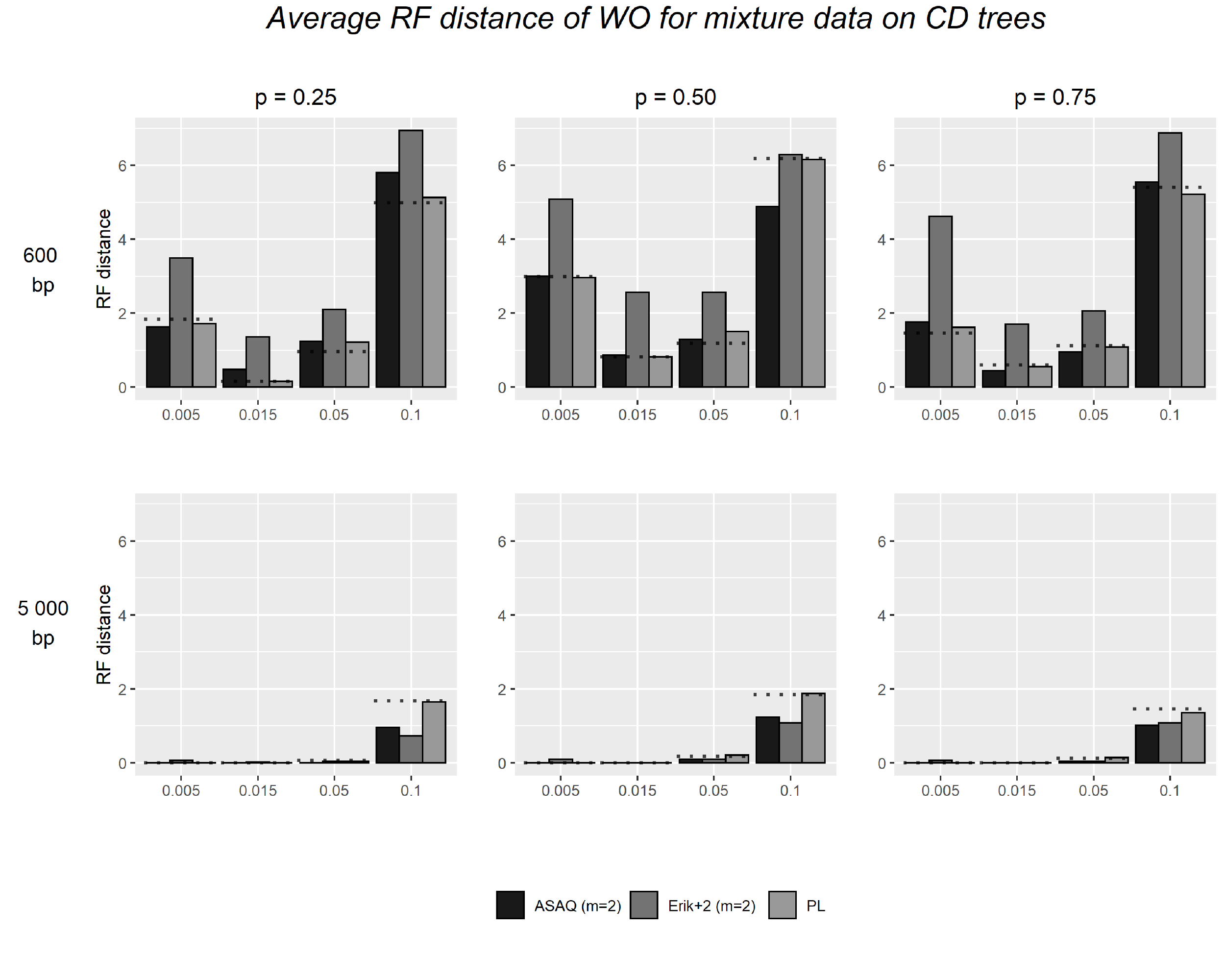}
\captionsetup{width=1\textwidth}
\caption{\label{fig:mixt_p}  \footnotesize 	Average Robinson-Foulds distance on mixture data simulated on the tree \CD for $p=0.25$ (left), $p=0.5$ (middle) and $p=0.75$ (right) with alignment length 600 bp. (above) and 5 000 bp. (below).
	We omit the results for 10 000 bp. as they are very similar to those obtained for 5 000 bp.
	The Q-method WO is applied with different systems of weights, namely  \asaq with 2 categories, \eri with 2 categories, and \pl.
	The dotted horizontal lines are the result of a global \nj with the paralinear distance.
	Results obtained by WO, QP, WIL for the different proportions between the two categories, $p=0.25$, $p=0.5$ and $p=0.75$, can be found in Tables \ref{tab_mixt_p25}, \ref{tab_mixt_p50} and \ref{tab_mixt_p75}, respectively.
	%
}
\end{figure}

The results on mixture data obtained by  a global \nj  and by QP, WO and WIL with input weights from \pl, \asaq and \eri adapted to 2-category data (as described in section \ref{sec:simdata_large})  are summarized in Figure \ref{fig:mixt_p} for WO and in Tables \ref{tab_mixt_p25} to \ref{tab_mixt_p75} for all methods. 
%
%
Tables \ref{tab_mixt_p25} to \ref{tab_mixt_p75} have a similar structure as Tables \ref{tab_CCgennonh}-\ref{tab_DDgtr} and correspond to the results obtained for different proportions between the two categories; $p=0.25$, $p=0.5$ and $p=0.75$, respectively (we recall that the proportion $p$ of sites of the first category of the alignment were generated assuming the branch lengths of the \CD tree in Fig. \ref{fig:TreeTopologies} ).
%
%
It is worth pointing out that the reconstruction of mixture data from \CD trees is more accurate in average than when applied to unmixed data (Figure \ref{fig:CDgennonh}).
This is probably due to the fact that when the branch lengths of species $seq3$, $seq4$ and $seq7$, $seq8$ are exchanged (see Figure \ref{fig:TreeTopologies}), reconstruction methods have an easier job to make the appropriate splits, since the long branch attraction situation that was provoked by the quartet of species $\{seq8,seq9,seq10,seq11\}$ does no longer exist with the new branch lengths. This is consistent with the observation that the reconstruction results are more accurate for low values of $b$.
For short alignments (600 bp.), we note that \asaq and \pl weights provide better results than \eri. 
For 5 000 bp. \asaq and \eri outperfom \pl. Moreover, if $b=0.1$, the combination WO+\asaq or WO+\eri beats a global \nj. For 10 000 bp. the difference between the performance results is even larger (see Tables \ref{tab_mixt_p25}, \ref{tab_mixt_p50} and \ref{tab_mixt_p75}).

%
%
As expected, the length of the alignment improves the performance of these methods, reducing the impact of the long branch attraction effect for high values of $b$.

\subsection{Results on real data}

%
\paragraph{Yeast data}
As shown in Table \ref{table_real}, the claim by \citet{jayaswal2014} is fully corroborated by our analysis: WO+\asaq correctly reconstructs $T$ when 3 categories are considered, but it reconstructs $T'$ otherwise.
Similarly the RF distance of the consensus tree obtained by WIL+\asaq to the tree $T$ is smaller than the RF distance to $T'$ only when 3 categories are considered, in concordance with \citet{jayaswal2014}.
When applied to these data, \nj with paralinear distances reconstructs the tree $T$.

\begin{table}[H]
\centering
\def\arraystretch{1}
\begin{tabular}{c|cc|cc}
	& \multicolumn{2}{c|}{RF distance to $T$} & \multicolumn{2}{c}{RF distance to $T'$} \\ \cline{2-5}
	& WIL  & WO & WIL & WO         \\ \hline \hline
	\asaq (m=1)
	& 1.974    & 2 & 0.026         & 0        \\
	\asaq (m=2)
	& 1.922    & 2 & 0.078         & 0        \\
	\asaq (m=3)
	& {0.004} & \textbf{0}  & 1.996         & 2 \\
	\hline
\end{tabular}
\caption{\label{table_real} \footnotesize Robinson-Foulds distance of the consensus tree obtained by WIL and WO with \asaq weights (with different number of categories) applied to the trees $T$ and $T'$ suggested in \citet{Rokas2003} and \citet{Phillips2004}, respectively. }
\end{table}

\paragraph{Ratites and tinamous}

First we analyse the results obtained by 50 majority rule consensus trees obtained from WO and QP on 100 initial quartets weighted with different methods. In Table \ref{table_ratites} we show the average results in comparison to the trees $A$, $B$ and $C$ detailed in section \ref{sec_realdata}. The number of interior edges obtained by each method shows that WO produces more resolved trees than QP (independently of the weighting method) and that \asaq, \saq and \eri (m=1) give rise to more resolved trees that \pl and \fp with both WO and QP.

\begin{table} \footnotesize

\begin{tabular}{l||cc||cc|cc|cc|cc|cc|cc|}
	& \multicolumn{2}{c||}{}  & \multicolumn{4}{c|}{A} & \multicolumn{4}{c|}{B} & \multicolumn{4}{c|}{C}   \\
	\hline
	&  \multicolumn{2}{c||}{interior edges} & \multicolumn{2}{c|}{ RF dist} & \multicolumn{2}{c|}{norm. RF} & \multicolumn{2}{c|}{ RF dist} & \multicolumn{2}{c|}{norm. RF} & \multicolumn{2}{c|}{ RF dist}  & \multicolumn{2}{c|}{norm. RF}  \\
	\hline
	weights & QP & WO & QP & WO & QP & WO & QP & WO & QP & WO &  QP & WO & QP & WO \\
	\hline
	\saq & 14,3 & 20,5 &
	9,7 & 10,0 & 0,3 & 0,2 &
	6,7 & 7,0 & 0,2 & 0,2 &
	9,7 & 10,0 & 0,3 & 0,2
	\\
	\hline
	\asaq $_1$ & 13,8 & 19,5 &
	9,7 & 11,9 & 0,3 & 0,3 &
	6,7 & 8,9 & 0,2 & 0,2 &
	9,7 & 11,9 & 0,3 & 0,3
	\\
	\hline
	\asaq $_2$ & 14,0 & 19,8 &
	9,0 & 9,9 & 0,3 & 0,2 &
	6,2 & 6,9 & 0,2 & 0,2 &
	9,0 & 9,9 & 0,3 & 0,2
	\\
	\hline
	\eri $_1$  & 11,0 & 19,0 &
	12,3 & 14,4 & 0,4 & 0,4 &
	11,3 & 12,0 & 0,4 & 0,3 &
	12,3 & 14,4 & 0,4 & 0,4
	\\
	\hline
	\eri $_2$ & 10,1 & 15,8 &
	12,0 & 11,8 & 0,4 & 0,3 &
	11,0 & 9,4 & 0,4 & 0,3 &
	12,0 & 11,8 & 0,4 & 0,3
	\\
	\hline
	\pl & 8,5 & 12,4 &
	15,3 & 17,4 & 0,5 & 0,5 &
	14,3 & 16,4 & 0,5 & 0,5 &
	15,3 & 17,4 & 0,5 & 0,5
	\\
	\hline
	\fp & 8,3	&12,7 &
	15,3 & 17,7 & 0,5 & 0,5 &
	14,3 & 16,7 & 0,5 & 0,5 &
	15,3 & 17,7 & 0,5 & 0,5
	\\
	\hline \hline
	\nj & \multicolumn{2}{c||}{21}  & \multicolumn{2}{c|}{8}   & \multicolumn{2}{c|}{0.19}  & \multicolumn{2}{c|}{5}  & \multicolumn{2}{c|}{0.12}  & \multicolumn{2}{c|}{8}  &
	\multicolumn{2}{c|}{0.19} \\
	\hline
\end{tabular}

\caption{\label{table_ratites}  \footnotesize
	%
	For each weighting system indicated in the left column and combined with either QP or WO, the first column depicts the average number of interior edges after 50 trials. 
	The Robinson-Foulds distance and normalized RF distance (that is, dividing by the number of interior edges) of the resulting consensus tree relative to the trees $A$, $B$ and $C$ (see section \ref{sec_realdata}) is shown in the remaining columns.
	The last row shows these values when a global \nj is applied.
}
\end{table}


Then we study the MRCTs obtained for these real data with WO (trees displayed in Figure \ref{fig:5_consensustrees}). When we use \asaq and we do the MRCT for 2125 or 5313 initial quartets (with both $m=1$ and $m=2$), we obtain the phylogeny
\begin{itemize}
\item[D:] (outgroup, neognathus, tinamous, (rheas, (ostrich, (moas, CEK)))).
\end{itemize}
This phylogeny is also supported by a global \nj, although \nj splits tinamous and outgroup against the others. The same tree as \nj is obtained with \saq (both for 2125 or 5313 initial quartets). Although the deepest interior branch lengths within ratites obtained by \nj are very small: (rheas,(ostrich, (moas, CEK):0.0025):0.003), \saq and \asaq give full support to these splits. Indeed, even when  we use \saq or \asaq (with both $m=1$ or $2$) with only 100 initial quartets, \emph{all} 50 MRCTs share these splits.

Unexpected trees are obtained by \eri, \pl or \fp (either not solving correctly the neognathus clade, or not giving a CEK clade nor putting all tinamous in a clade).

\subsection{Execution time}
The computations on this paper have been performed on a computer with 6 Dual Core Intel(R) Xeon(R) E5-2430 Processor (2.20 GHz) equipped with 25 GB RAM running Debian GNU/Linux 8. We have used the g++ (Debian 4.9.2-10+deb8u2) 4.9.2 compiler and the C++ library for linear algebra \& scientific computing \textit{Armadillo} version 3.2.3 (Creamfields).

The average time required to compute \asaq and \pl weights for $100$ alignments of $4$ taxa and length $10~000$ bp is $8.7$ and $7.8$ seconds respectively. The average time to reconstruct a CD tree given the weights for its 495 quartet subtrees is $4$ seconds for WO, $89.5$ seconds for WIL and $1.8$ seconds for QP.

\section{Discussion and conclusions}

Via experiments on the simulation framework proposed by \citet{Ranwez2001} but considering more general models of nucleotide substitution (including GM model and mixtures of distributions), we observe a huge improvement on the performance of Q-based methods when weights from \asaq, \saq, \eri and \pl methods are considered. In general, the highest success is obtained by WO with the weighting system of \asaq or \pl. The success of these methods is compatible with a global \nj,
and outperforms it in the presence of mixtures or long branches when using \asaq as weighting system. Moreover these weights also outperform weights obtained by \ml, even when data is generated under a GTR model and \ml is estimating the same model. 


It is worth noting that in the previous studies of Q-methods by \citet{Ranwez2001,stjohn2003}, only weights from \ml and \nj were considered. The results in this paper validate WO and WIL as successful phylogenetic reconstruction methods if their input is a system of reliable weights. Moreover, as \asaq assumes the most general Markov model and can deal with mixtures, this opens the door to use Q-methods under complex models.

We need to mention that the comparison performed against the \ml weights (\ml as input of Q-methods) may not be totally fair because the model used in parameter estimation (time-continuous unrestricted model) does not fit the model used to simulate the GM data.  It would certainly be interesting to develop a quartet maximum likelihood estimadion based on the GM model and perform new tests. In any case, the results of \ml weights are much worse than those of \asaq, \eri or \pl also for GTR data, see Figure \ref{fig:DDgtr} and Table \ref{tab_DDgtr} in the Appendix.

We have not considered a comparison to a global maximum likelihood estimate because  in \cite{Ranwez2001} it had a similar performance than \nj (subject to the assumption of a correct substitution method), in \citet{stjohn2003} it is claimed that \ml usually gives worse results than \nj), and there is no available software that implements \ml under a general Markov model.

We have observed a good performance of input weights from \asaq, \eri and \pl on simulated mixture data on 12-taxon trees, specially for 5 000 bp or more. \asaq and \pl are not likely to be statistically consistent for general mixtures on the same tree. Nevertheless, as \eri is consistent on mixture data and \pl is known to be consistent on some type of mixtures \citep[see][]{allman2019}, this suggests that \asaq (being based on the accordance of \eri and \saq) might also be consistent on some types of mixture data, which would explain the good results obtained. We would like to point out that the mixture model allowed in \eri  (and hence in \asaq) is actually more flexible than we mentioned. Indeed, the rank conditions considered in \eri for quartets still hold  if we let one cherry or one leaf evolve under a mixture with any number of categories (while the other cherry evolves under a single partition), see \citet{CasFerBirkhauser}. This makes the mixture model underlying \eri and \asaq more general, with implications in the mixture model that can be considered for Q-methods with input weights from \asaq.

Our results on real data, validate WO + \asaq as the most reliable method (both for yeast or ratites/tinamous data). Moreover, it is relevant to note that for the ratites/tinamou data, the topology obtained by WO + \asaq or \saq, or by \nj with paralinear distances does not agree with any of the topologies proposed in \citet{phillips2010}.

Note that the Q-methods considered here have higher order of computational complexity than \nj, but we have not yet explored the possibility of considering a subset of the possible 4-tuples as suggested in \cite{snir2010} or \cite{davidson2015}. It would be interesting to further explore these other versions of Q-methods with weights from \asaq and \pl, or even to restrict to quartets with highest weights as starting point for Q-methods.

On another direction, our results show that \asaq is a powerful reconstruction method for quartet topology reconstruction. It assumes the most general model of nucleotide substitution (a general Markov model) of independently and identically distributed sites but it can also account for mixtures of distributions (with up to three categories). As it is based on the algebraic and semi-algebraic description of the model, it does not need to estimate the substitution parameters. In this sense, \asaq could be easily adapted as a suitable method for dealing with amino acid substitutions as well. The incorporation of invariable sites seems also plausible via the results in \cite{jayaswal2007,steel2000}. We plan to incorporate these features in a forthcoming version of the software.

As mentioned in the introduction, \asaq is part of the set of phylogenetic reconstruction tools that are based on algebra. Most of these methods only reconstruct quartets because no statistically consistent (and computationally affordable) algebraic method for larger trees has been designed yet.

We are aware that we have not made a study of the method from the statistical point of view and the claim on the efficiency of the method is solely based on the results obtained on large sets of simulated data. A study on the statistical efficiency of the method would certainly be relevant.

\bibliographystyle{apalike}


\appendix
\newpage
\section{Appendix A. Weighting systems and technical results about \asaq}


\renewcommand\thefigure{\thesection.\arabic{figure}}
\renewcommand\thetable{\thesection.\arabic{table}}
\setcounter{figure}{0}
\setcounter{table}{0}

\subsection{Weighting systems}

The input weights we consider in this paper are obtained from different quartet reconstruction methods: from \saq, \eri and \asaq,
from methods that use the paralinear distance (\pl and \fp) and maximum likelihood (\ml). For any of these methods we describe here the weighting score used. To this end, let $f$ be the vector of relative frequencies obtained from a DNA alignment of four taxa.



\paragraph{Weights for methods based on the paralinear distance, \pl and \fp.}
Consider two (ordered) nucleotide sequences $S_x$ and $S_y$ of the same length corresponding to two taxa $x$ and $y$, respectively. Let $J$ be the underlying joint probability matrix of $S_x$ and $S_y$, this is, the entry $(i,j)$ of $J$ is the probability (either theoretical or estimated by relative frequencies) of observing nucleotides $i$ and $j$ at the same position of $S_x$ and $S_y$ (so that the sum of entries in $J$ is one).
If $\det J\neq 0$ and all nucleotides are observed in $S_x$ and $S_y$, then the \emph{paralinear distance} \citep{lake1994} between $x$ and $y$ is
\begin{eqnarray} \label{new_paralinear}
	d_{x,y}=-\log \frac{|\det J|}{\sqrt{\det D_x} \sqrt{\det D_y}},
\end{eqnarray}
where $D_x$ and $D_y$ are the diagonal matrices whose diagonal entries are given by $J \mathbf{1}$ and $\mathbf{1}^t J$, respectively; if $\det J=0$, we take $d_{x,y}$ as infinity.
%
%
Based on the work by \citet{lake1994} and \citet{Ranwez2001}, given the vector $f$ of relative frequencies from a quartet alignment, in Section 2 we defined
\begin{eqnarray*}
	\gg{12|34}(f)=\min\{d_{1,3}+d_{2,4},d_{1,4}+d_{2,3}\}-d_{1,2}-d_{3,4}
\end{eqnarray*}
which represents twice the length of the interior edge of the quartet (see Theorem \ref{main:app}(b) below).
We define $\gg{13|24}(f)$ and $\gg{14|23}(f)$ similarly.

In \cite{wnjw}, the authors propose a slightly different way of weighting the quartets using the paralinear distance:
\begin{eqnarray*}
	\ell_{12|34}(f)=\frac{1}{4}(d_{1,3}+d_{1,4}+d_{2,3}+d_{2,4})-\frac{1}{2}(d_{1,2}+d_{3,4}),
\end{eqnarray*}
($\ell_{13|24}(f)$ and $\ell_{14|23}(f)$ are defined analogously). These are the measures that we use for what we call the \fp method (standing for ``four-point condition'').
In the same paper quoted above, the authors establish the (theoretical) equivalence between \nj and a consistent quartet-based method that uses these weights.  Observe also that these weights are equivalent to $\gg{12|34}(f)$ for theoretical data.

In order to assign a non-negative score to each quartet, we define 
the normalized weights for \pl and \fp as
\begin{eqnarray*}
	w_{p\ell}(f)=
	\frac{1}{\sum_i \exp(\gg{T_i}(f))} \left ( \exp(\gg{T_1}(f)), \exp(\gg{T_2}(f)), \exp(\gg{T_3}(f)) \right ).
\end{eqnarray*}

\begin{eqnarray*}
	w_{4p}(f)=
	\frac{1}{\sum_i \exp(\ell_{T_i}(f))} \left ( \exp(\ell_{T_1}(f)), \exp(\ell_{T_2}(f)), \exp(\ell_{T_3}(f)) \right ).
\end{eqnarray*}

%
%
%

\paragraph{Weights for maximum likelihood.}

We weighted quartet trees for the \ml method using the likelihoods associated to the three trees $T_1, T_2, T_3$ as done by \citet{Ranwez2001}:
if $l_{T_i}(f)$ is the likelihood for the tree $T_i$ (with the MLE parameters), then 
the normalized weights for the three quartets are defined as
\begin{eqnarray*}
	w_{ml} (f)= \frac{1}{\sum_i l_{T_i}(p)} \left (l_{T_1}(f), l_{T_1}(f), l_{T_1}(f) \right ). 
\end{eqnarray*}

\paragraph{Weights for \eri.}

Given a bipartition $ij|kl$ and a distribution $f\in \mathbb{R}^{256}$, denote by $Flat_{ij|kl}(f)$ the  flattening of $f$ according to that bipartition, that is, $Flat_{ij|kl}(f)$ is the $16\times 16$ matrix whose $(x_ix_j,x_kx_l)$-entry is the coordinate of $f$ that matches $x_ix_jx_kx_l$ in the convenient order.
Then, $Flat_{ij|kl}(f)$ has rank $\leq 4$, while the flattenings corresponding to the other two bipartions have rank 16 if the parameters are generic \citep{AllmanRhodeschapter4}.
%
%
The method \eri computes for each quartet tree $T$ the value
\begin{eqnarray} \label{weights_Erik2}
	e_T(f)=\frac{\delta_4(Flat_T^r(f))+\delta_4(Flat_T^c(f))}{2},
\end{eqnarray}
where $Flat_T^r$ and $Flat_T^c$ stand for the (normalized) matrices obtained by dividing the rows and columns of the flattening matrix $Flat_T$ by the sum of its entries, respectively (as a variation we can consider the distance $\delta_{4m}(\cdot)$ to rank $4m$ matrices when we deal with mixed data with $m\in \{1,2,3\}$ categories). The right quartet tree should be the one that gives the minimal value; this is the output of the method.

In order to have a normalized scoring system that allows us to compare and represent the output of $\eri$ for each quartet tree $T$, we consider the inverse of the value $e_T(f)$ and then divide but the sum of these scores for all three quartet trees:

\begin{eqnarray*}
	\eri(f)= \frac{1}{e}\left (e_{T_1}(f)^{-1},e_{T_2}(f)^{-1},e_{T_3}(f)^{-1}\right )
\end{eqnarray*}
where $e := e_{T_1}(f)^{-1} + e_{T_2}(f)^{-1} + e_{T_3}(f)^{-1}$.

\paragraph{Weights for \saq.}

The method \saq suggested by \cite{casfergar2020}  associates the following score to the tree $T=12|34$:

\begin{eqnarray*}
	s_T(f):=\sum_i \frac{\min \left \{\delta_4(psd(Flat_{13|24}(\tilde{f}_i))),
		\delta_4(psd(Flat_{13|23}(\tilde{f}_i)))\right \}}{\delta_4(psd(Flat_{12|34}(\tilde{f}_i)))}
\end{eqnarray*}
where the sum runs over the sixteen $12|34$ leaf-transformed distributions $\tilde{f}_i$, $\delta_4(\cdot)$ stands for the distance to the space of matrices with rank $\leq 4$ and $psd(M)$ is the closest symmetric and positive semi-definite matrix to the matrix $M$. If $T$ is any of the other two possible quartet trees, $s_T(f)$ is computed analogously by permuting the roles of the leaves accordingly. \saq  outputs the normalized three scores, that is, if $s := s_{T_1}(f) + s_{T_2}(f) + s_{T_3}(f)$, then
\begin{eqnarray*}
	\saq(f):=\frac{1}{s} \left (s_{T_1}(f),s_{T_2}(f),s_{T_3}(f)\right ).
\end{eqnarray*}

\subsection{The paralinear distance and theoretical foundations of \asaq}

The following lemma generalizes the nonnegativity and the additivity properties of this dissimilarity map to general Markov matrices, extending the results of \citet{lake1994} to general Markov matrices as far as they are non-singular.

\begin{lema} \label{paralinearMarkov}
	Let $\pi$ be the nucleotide distribution at $x$ and consider a  substitution process leading from $x$ to $y$ and ruled by a Markov matrix $M$. Then, the paralinear distance $d_{x,y}$ between $x$ and $y$ defined in \eqref{new_paralinear} coincides with
	\begin{eqnarray}\label{paralinear_markov}
		d_{x,y}=-\log \frac{|\det M| \, \sqrt{\det D(\pi)}}{\sqrt{\det D(M^t \pi)}},
	\end{eqnarray}
	(where $D(u)$ refers to the diagonal matrix whose diagonal entries are the coordinates of the vector $u$).  Moreover, we have that
	\begin{itemize}
		\item[a)] $d_{x,y}\geq 0$, and the equality holds if and only if $M=Id$ or is a permutation matrix;
		\item[b)] the dissimilarity measure $d_{x,y}$ is additive.
	\end{itemize}
\end{lema}

\begin{proof}
	We have that $J=D(\pi) M$ is the underlying joint probability matrix between $x$ and $y$ so that the sum of its entries is one. Note that $\pi=J\mathbf{1}$ and $\pi^t M=\mathbf{1}^t J$, so $D_x=D(\pi)$ and $D_y=D(M^t \pi)$.
	Therefore,
	\begin{eqnarray*}
		d_{x,y}=-\log \frac{|\det J|}{\sqrt{\det D_x}\sqrt{\det D_y}} =
		-\log \frac{|\det M| \, \det D_x}{\sqrt{\det D_x}\sqrt{\det D_y}} =
		-\log \frac{|\det M| \, \sqrt{\det D(\pi)}}{\sqrt{\det D(M^t \pi)}}.
	\end{eqnarray*}
	Now, we proceed to prove (a): $d_{x,y}\geq 0$ or equivalently, that
	\begin{eqnarray}\label{lessthanone}
		(\det M)^2 \; \frac{\det D(\pi)}{\det D(M^t\pi)} \leq 1.
	\end{eqnarray}
	%
	First of all, since the matrix $M$ is positive, we deduce that
	\begin{eqnarray*}
		\big |\det M  \big |=\left |\sum_{\{i_1,i_2,i_3,i_4\}=[4]}	\prod_j \;	 sgn(i_1,i_2,i_3,i_4) \, m_{i_j,j} \right |\leq \sum_{\{i_1,i_2,i_3,i_4\}=[4]}	\prod_j \;	m_{i_j,j}.
	\end{eqnarray*}	
	We have that
	\begin{eqnarray*}\label{aux_ineq}
		|\det M |\det D(\pi) & =  &
		|\det M | \,\prod_k \pi_k  \nonumber  \\
		& \leq & \prod_k \pi_k \; \left (\sum_{\{i_1,i_2,i_3,i_4\}=[4]}	\prod_j \;	m_{i_j,j}\right )  \\
		& = & \sum_{\{i_1,i_2,i_3,i_4\} = [4]}	\prod_j \; m_{i_j,j}\; \pi_j  \\
		& \leq & 	 \sum_{i_1,i_2,i_3,i_4 \in [4]}	\prod_j \;	 m_{i_j,j}\, \pi_{i_j} = \prod_{j} \sum_i m_{ji} \pi_j = \prod_i (M^t\, \pi)_i \\ & = & \det D(M^t\pi).
	\end{eqnarray*}
	By multiplying this inequality with $|\det M|\leq 1$ ($M$ is a Markov matrix), we obtain \eqref{lessthanone}. Note that a similar argument is used in the proof of the main theorem in \cite{Steel94}.

	Note that if $d_{x,y}=0$, then necessarily $\det M=1$. Hadamard's inequality  implies that a Markov matrix $M$ with determinant 1 is necessarily the identity matrix or a permutation matrix \citep{CGZ}. 
	Finally, the statement (b) follows easily from the expression \eqref{paralinear_markov}.
\end{proof}

Write $\Delta$ for the probability simplex in the space $\mathbb{R}^{256}$, that is, $\Delta$ is the set of all possible distribution vectors of patterns of nucleotides $p=(p_{x_1 x_2 x_3 x_4})_{x_1,x_2,x_3,x_4\in \{A,C,G,T\}}$.
%
%
From Lemma 8 of \cite{Buneman} applied to the paralinear distance, we get the following result (which can be also deduced almost directly from the definition \eqref{eq:defgamma}):
\begin{lema}\label{lemma:def_gamma}
	Assume that the determinant of every double marginalization of $p\in \Delta$ is non-zero (so that all values $d_{i,j}$, $i,j\in [4]$ can be computed). Then, $\gg{A|B}(p)$ is strictly positive for at most one bipartion~$A|B$.
\end{lema}

\begin{figure}
	\begin{center}
		\includegraphics[scale=0.5]{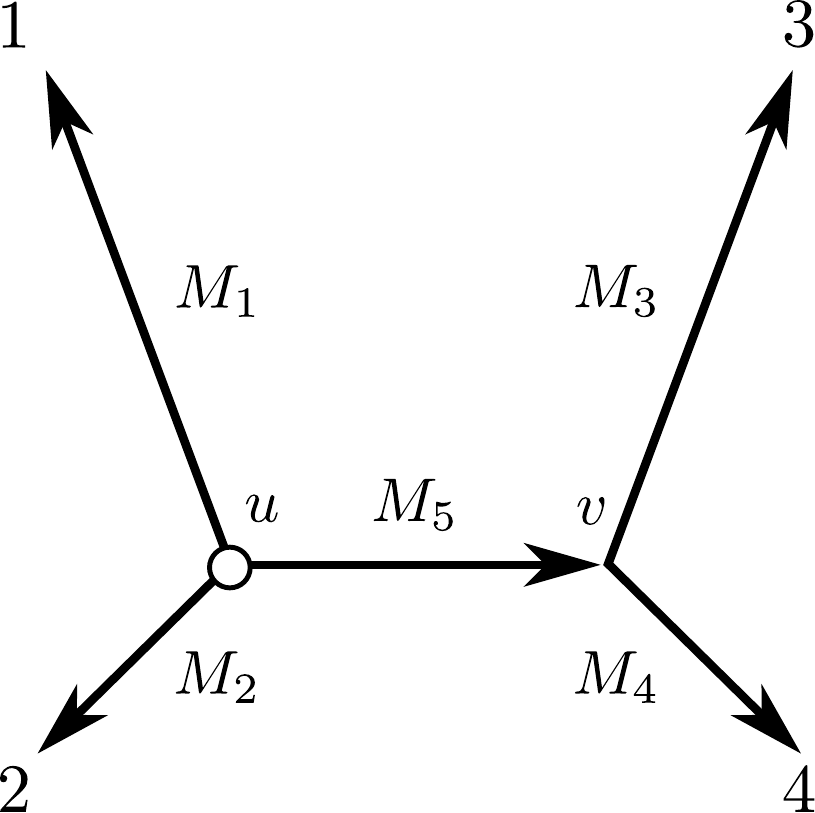}
	\end{center}
	\caption{\label{fig:tree} \footnotesize Markov process on the 12|34 quartet as in Theorem \ref{main:app} a). The root is located at the interior node $u$ close to the cherry $(1,2)$.
	}
\end{figure}

%
	%

\begin{thm}\label{main:app} Let $f$ be vector of relative frequencies of site patterns observed in an alignment of length $N$ generated according to a multinomial distribution $p$ (that is, $f$ is the vector of relative frequencies that arises as $N$ independent samples from  $p$). Then we have
	\begin{itemize}
		\item[a)] \[\pl(f)=(\gg{12|34}(f),\gg{13|24}(f),\gg{14|23}(f))\] is a quartet inference method  that satisfies the strong property II (in the sense of \citet{sumner2017}). Moreover, if $p$ arises from a Markov process on the tree $T=12|34$ of Figure \ref{fig:tree} with positive distribution $\pi$ at the root and invertible transition matrices, then the limit of the expectation of $\pl(f)$ as $N$ goes to infinity is
		\begin{eqnarray*}
			\lim_{N\rightarrow \infty} \mathbb{E}(\pl(f))=
			\left (\gg{12|34}(p),\gg{13|24}(p),\gg{14|23}(p) \right );
		\end{eqnarray*}
		\item[b)] if $p$ arises from a Markov process as in a), $\gg{12|34}(p)=2 d_{u,v}\geq 0$ and $\gg{C|D}(p)=-\gg{12|34}(p)$ for any other bipartition $C|D \neq 12|34$;
		\item[c)] the paralinear method (which associates to a frequency vector $f$ the tree $T_{A|B}$ with largest $\gg{A|B}(f)$) is statistically consistent for the general Markov model;
		\item[d)] \asaq is statistically consistent for the general Markov model.
	\end{itemize}
\end{thm}


\begin{proof}
	\begin{itemize}
		\item[(a)] The same proof of \citet[Theorem 3.1]{sumner2017} applied to the paralinear distance gives that $\pl(f)$ satisfies the strong property II with $\lambda=0$ in our case (see the quoted proof).  Indeed, it is straightforward to see that
		$\gg{A|B}(f)$ is invariant by the Markov action: $\gg{A|B}((X_1,\dots,X_4)*f)=\gg{A|B}(f)$ for any $4\times 4$ Markov matrices $X_i$ and any bipartition $A|B$. This can be immediately seen by observing that for any array $F$
		\begin{eqnarray*}
			\gg{A|B}(F)=\max \left \{\left | \frac{\det(N_{ac}) \det(N_{bd})}{\det(N_{ab}) \det(N_{cd})} \right |,
			\left | \frac{\det(N_{ad}) \det(N_{bc})}{\det(N_{ab}) \det(N_{cd})} \right |
			\right \},
		\end{eqnarray*}
		where $N_{ab}$ is the matrix obtained by the double marginalization of $F$ on components different from $a$ and $b$, with rows (respectively columns) labelled by the states at $a$ (resp. $b$) and $N_{ba} = N_{ab}^t.$
		%
		The claim about expectation follows by Taylor expansion of $\gg{A|B}(f)$ around the expectation of $f$ (as $\mathbb{E}(f)=p$ componentwise).
		
		\item[(b)] The first claim is a consequence of the additivity of the paralinear distance, see Lemma \ref{paralinearMarkov}). Indeed, under the assumption that $p$ arises from the tree $T=12|34$, every quantity $d_{i,j}$ ($i,j\in [4]$) can be written as the sum of the paralinear distances attached to the edges of $T$ between $i$ and $j$. For example, $d_{1,2}=d_{1,u}+d_{u,v}+d_{v,3}$. %
		Then the value $\gg{A|B}(p)$ results in $2 d_{u,v}$, which is nonnegative in virtue of (a) of Lemma \ref{paralinearMarkov}.
		
		Now, if $C|D$ is a bipartition other than $12|34$, then from \eqref{eq:defgamma} we immediately get $\gg{C|D}(p)=-\gg{12|34}(p)$ .

		\item[(c)]  Denote $\delta= 2 d_{u,v}$. Then, following $(a)$ we have that $$ \lim_{N \rightarrow \infty} \pl(f) =\lim_{N \rightarrow \infty} \left(\gg{12|34}(f),\gg{13|24}(f),\gg{14|23}(f)\right) = \pl(p) =(\delta, -\delta, -\delta),$$ where the last equality follows from $(b)$. Therefore, for any $\epsilon>0$ there exist $N_1$, $N_2$, $N_3$ such that for any $N>N_0=\max\left\{N_1,N_2,N_3\right\}$ we have
		$$\left|\gg{12|34}(f)-\delta\right|<\epsilon, \quad \left|\gg{13|24}(f)+\delta\right|<\epsilon, \quad \left|\gg{14|23}(f)+\delta\right|<\epsilon.$$
		Then, for any $\epsilon<\delta$ we have
		$$\gg{12|34}(f) > \delta-\epsilon > 0 > -\delta+\epsilon >\max \left\{\gg{13|24}(f), \gg{14|23}(f)\right\}.$$
		Then we can conclude that the method that chooses the tree $T$ with largest $PL_T(f)$ chooses $T_{12|34}$ with probability tending to 1 when $N$ tends to infinite.
		
		\item[d)] The statistical consistency of \asaq follows immediately from the statistical consistency of \eri \citep{fercas2016} and of the paralinear method obtained in c).
		Indeed, if $f\sim MultiNom(p,N)$ is sampled from a distribution $p$ that has arisen from $T=T_{12|34}$ with non-singular parameters (positive distribution at the root and invertible transition matrices), then
		the probability that both methods (when applied to $f$) select $T_{12|34}$ tends to 1 when $N$ tends to infinite.
	\end{itemize}

\end{proof}

Note that under the assumption that $M_5$ is \emph{diagonally largest in column}  (DLC for short, see \cite{chang1996}), we deduce that if $\gg{A|B}(p)=0$, then $M_5$ is necessarily the identity matrix.

\begin{rk}\rm In order to avoid numerical problems, we only compute $\gg{A|B}(f)$ if the condition number of the double marginal matrices (joint distributions at two leaves) involved in its computation are less than a certain tolerance. This is implemented using a parameter ``threshold'' set to $5000$, but it can be modified by the user and can be adapted to the alignment length. If $\gg{A|B}(f)$ cannot be computed, then this  is a sign of short sample size and \asaq outputs the topology and the weighting system given by \saq.
\end{rk}




\newpage
\section{Appendix B. Figures}
\setcounter{figure}{0}
\setcounter{table}{0}

\begin{figure}[H]
	\begin{center}
		\includegraphics[scale=0.5]{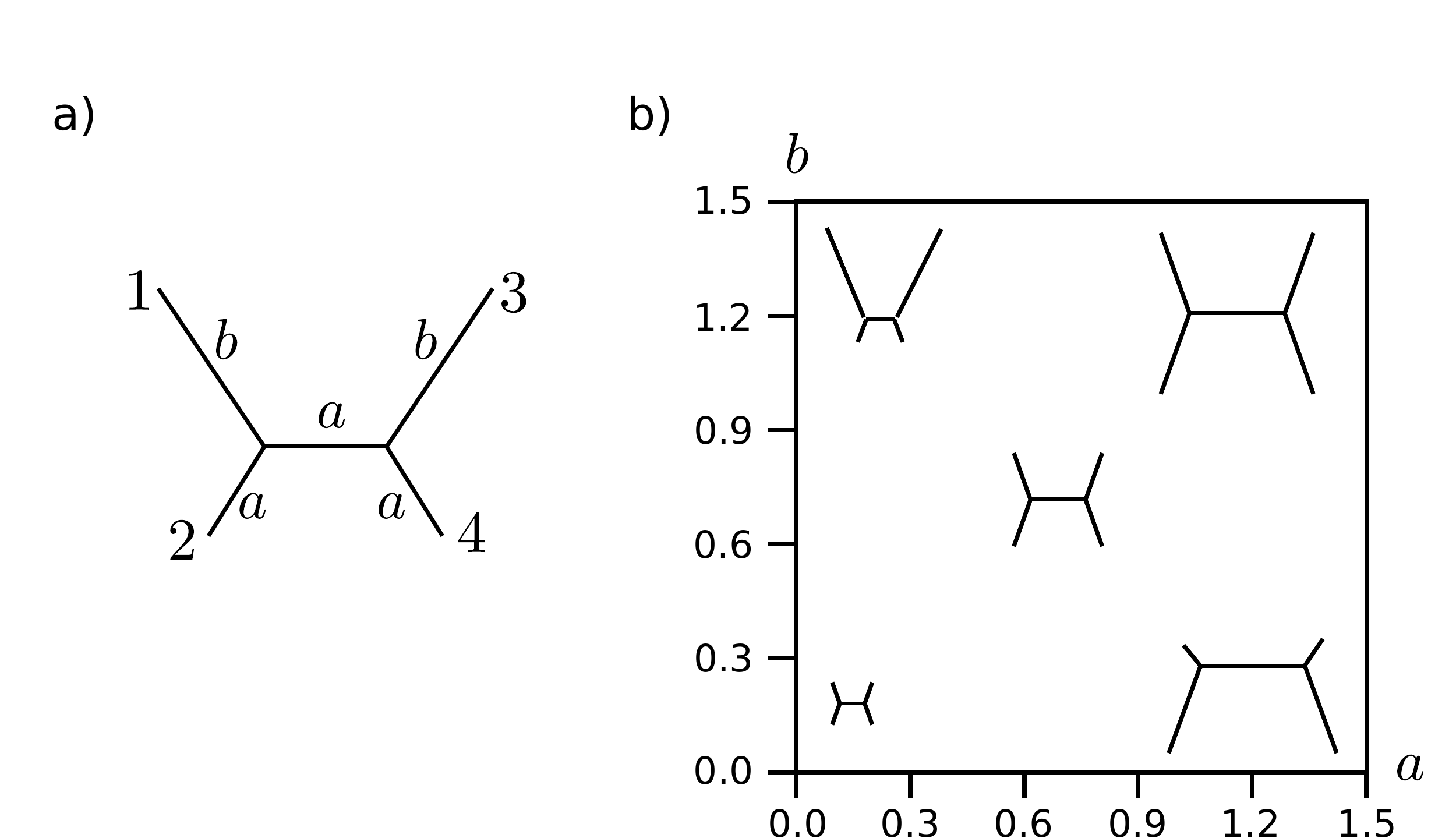}
	\end{center}
	\caption{\label{tree} \footnotesize
		a) 4-leaf tree where the length of two opposite branches and the interior branch are
		represented by $a$; the other two peripheral branches have length $b$.
		Branch lengths are measured as the expected number of substitutions per site. b) Tree space considered for the tree in a) where the branch lengths $a$ and $b$ are varied from 0.01 to 1.5 in steps of 0.02.
	}
\end{figure}

\begin{figure}[H]
	\begin{center}
		\includegraphics[scale=0.4]{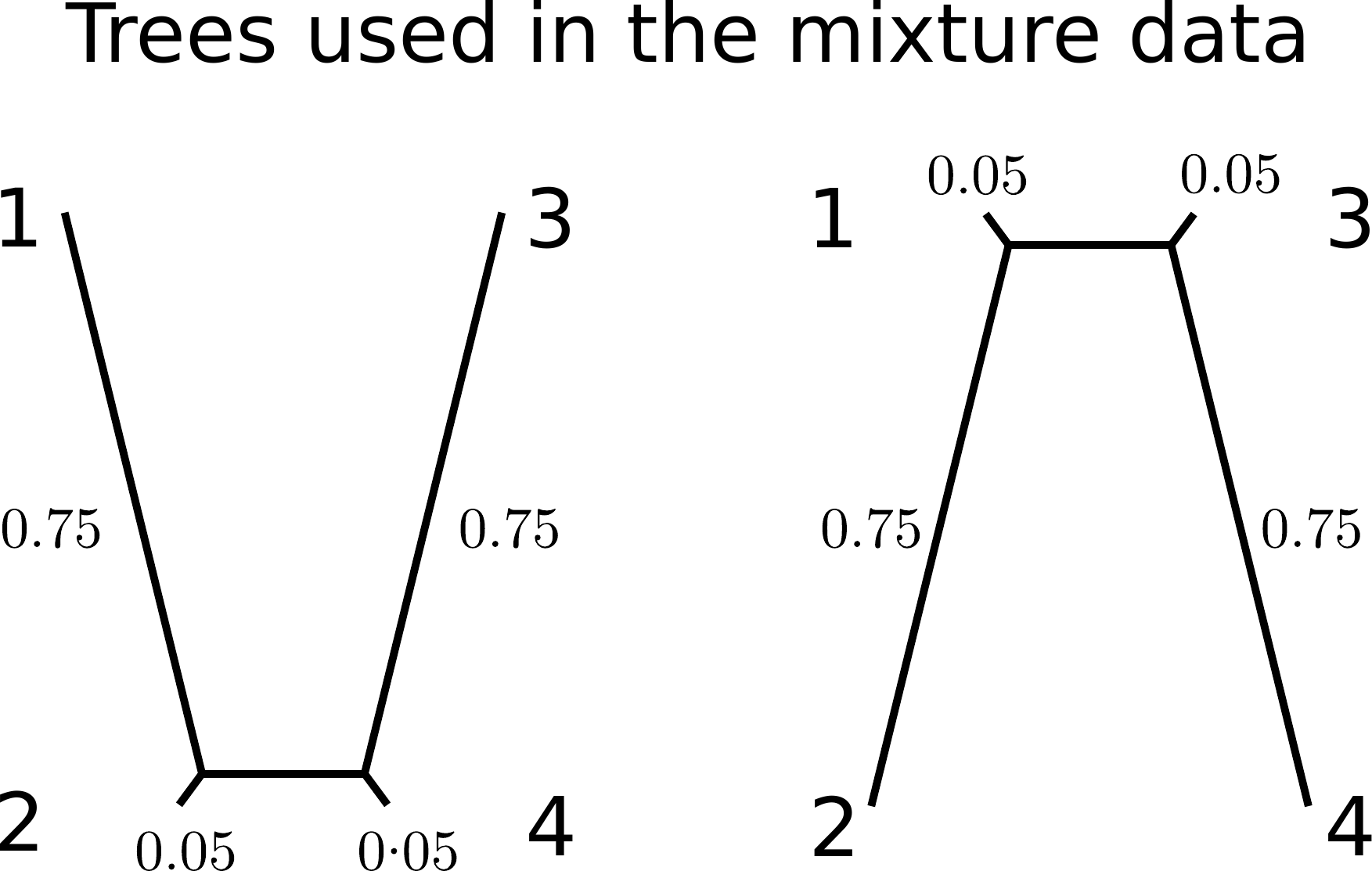}
	\end{center}
	\caption{\label{kolaczkowski} \footnotesize
		Trees considered 
		in the simulations of mixture data: the alignments have two  categories of the same size, each  evolving under the GM model on one of the trees depicted above with the systems of branch lengths indicated. The internal branch length takes the same value in both categories and is varied from 0.01 to 0.4 in steps of 0.05.
	}
\end{figure}

\begin{figure}[H]
	\begin{center}
		\includegraphics[scale=0.4]{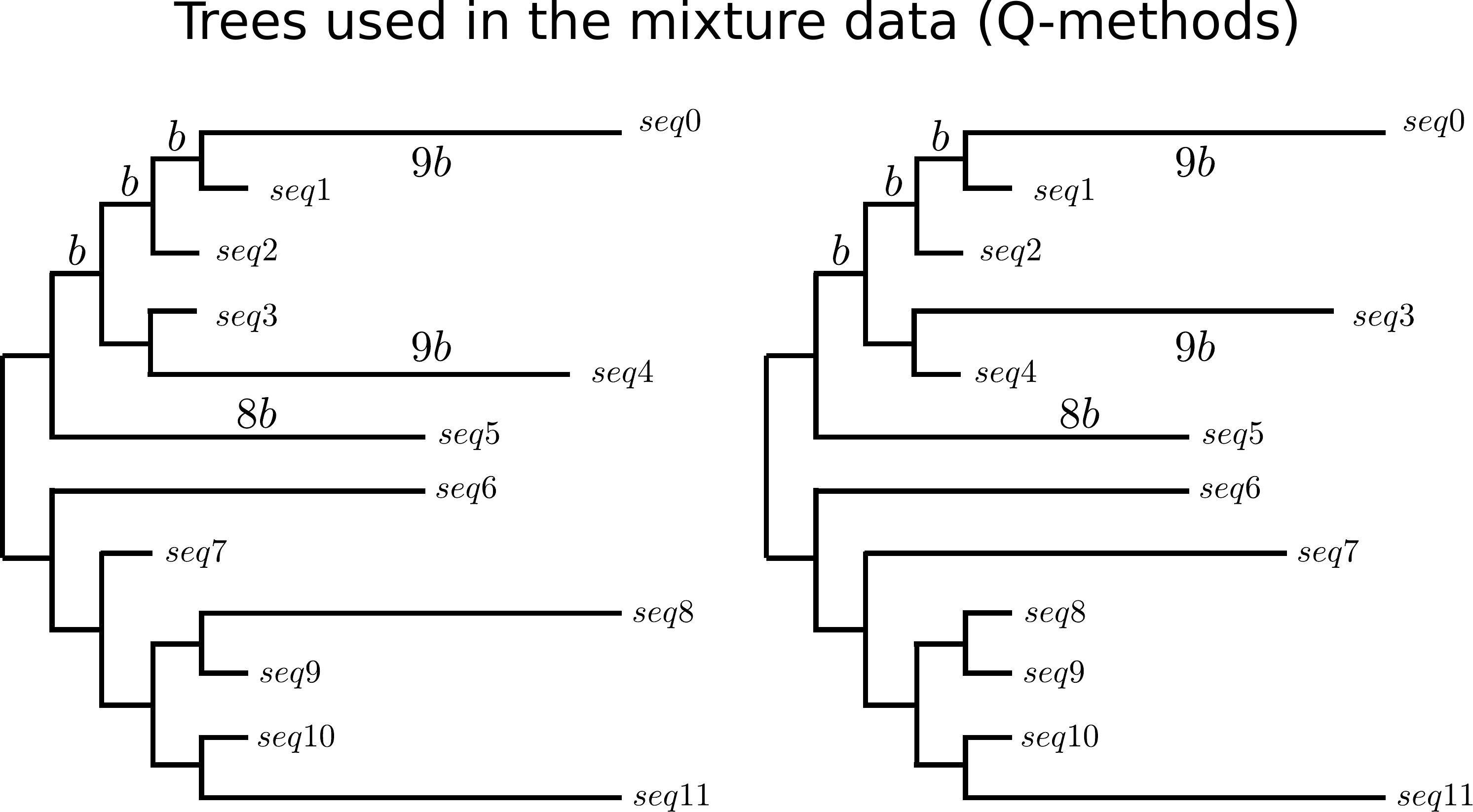}
	\end{center}
	\caption{\label{mixtureCD} \footnotesize
		Trees considered 
		in the simulations of mixture data for Q-methods: the alignment have two  categories and the parameter $p\in \{0.25,0.50,0.75\}$ represents the proportion of the first category relative to the second. Data are generated under the GM model on one of the trees depicted above with the systems of branch lengths indicated for different values of $b$.
	}
\end{figure}

\begin{figure}[H]
	\tikzstyle{vertex}=[circle,fill=black,minimum size=3pt,inner sep=0pt]
	\tikzstyle{hidden}=[circle,draw,minimum size=2pt,inner sep=0pt]
	\begin{tikzpicture}[scale=0.75]
		\node[hidden] (1) at (-3,7) [] {};
		\node[hidden] (2) at (-4,6) []{};
		\node[hidden] (3) at (-5,5) []{};
		\node[hidden] (4) at (-6,4) []{};
		\node[hidden] (5) at (-7,3) []{};
		\node[hidden] (6) at (-8,2) []{};
		\node[hidden] (7) at (-9,1) []{};
		
		\node[vertex] (8) at (-2,6) [label=right:\textit{\small \small  C. albicans}]{};
		\node[vertex] (9) at (-3,5) [label=right:\textit{\small \small S. kluyveri}]{};
		\node[vertex] (10) at (-4,4) [label=right:\textit{\small \small S. castelliii}]{};
		\node[vertex] (11) at (-5,3) [label=right:\textit{\small S. bayanus}]{};
		\node[vertex] (12) at (-6,2) [label=right:\textit{\small S. kudriavzevii}] {};
		\node[vertex] (13) at (-7,1) [label=right:\textit{\small S. mikatae}] {};
		\node[vertex] (14) at (-8,0) [label=below:\textit{\hspace{1cm}\small S. paradoxus}] {};
		\node[vertex] (15) at (-10,0) [label=below:\textit{\hspace{-0.3cm}\small S. cerevisiae}] {};

		\draw[line width=.4mm] (1) to (2);
		\draw[line width=.4mm] (2) to (3);
		\draw[line width=.4mm] (3) to (4);
		\draw[line width=.4mm] (4) to (5);
		\draw[line width=.4mm] (5) to (6);
		\draw[line width=.4mm] (6) to (7);
		
		\draw[line width=.4mm] (1) to (8);
		\draw[line width=.4mm] (2) to (9);
		\draw[line width=.4mm] (3) to (10);
		\draw[line width=.4mm] (4) to (11);
		\draw[line width=.4mm] (5) to (12);
		\draw[line width=.4mm] (6) to (13);
		\draw[line width=.4mm] (7) to (14);
		\draw[line width=.4mm] (7) to (15);

		\node[hidden] (21) at (7,7) [] {};
		\node[hidden] (22) at (6,6) []{};
		\node[hidden] (23) at (5,5) []{};
		\node[hidden] (24) at (4,4) []{};
		\node[hidden] (25) at (5,3) []{};
		\node[hidden] (26) at (2,2) []{};
		\node[hidden] (27) at (1,1) []{};
		
		\node[vertex] (28) at (8,6) [label=right:\textit{\small C. albicans}]{};
		\node[vertex] (29) at (7,5) [label=right:\textit{\small S. kluyveri}]{};
		\node[vertex] (30) at (6,4) [label=right:\textit{\small S. castelliii}]{};
		\node[vertex] (31) at (6,2) [label=right:\textit{\small S. bayanus}]{};
		\node[vertex] (32) at (4,2) [label=below:\textit{\hspace{0.5cm} \small S. kudriavzevii}] {};
		\node[vertex] (33) at (3,1) [label=below:\textit{\hspace{0.5cm} \small S. mikatae}] {};
		\node[vertex] (34) at (2,0) [label=below:\textit{\hspace{1cm}\small S. paradoxus}] {};
		\node[vertex] (35) at (0,0) [label=below:\textit{\hspace{-0.3cm}\small S. cerevisiae}] {};

		\draw[line width=.4mm] (21) to (22);
		\draw[line width=.4mm] (22) to (23);
		\draw[line width=.4mm] (23) to (24);
		\draw[line width=.4mm] (24) to (26);
		\draw[line width=.4mm] (24) to (25);
		\draw[line width=.4mm] (26) to (27);
		
		\draw[line width=.4mm] (21) to (28);
		\draw[line width=.4mm] (22) to (29);
		\draw[line width=.4mm] (23) to (30);
		\draw[line width=.4mm] (25) to (31);
		\draw[line width=.4mm] (25) to (32);
		\draw[line width=.4mm] (26) to (33);
		\draw[line width=.4mm] (27) to (34);
		\draw[line width=.4mm] (27) to (35);

	\end{tikzpicture}
	
	\caption{ \footnotesize The tree $T$ of \cite{Rokas2003} (left) and the alternative tree $T'$ of \cite{Phillips2004} (right) are constructed using the data provided by \cite{jayaswal2014} with 42 337 second codon positions of 106 orthologous genes of \emph{Saccharomyces cerevisiae}, \emph{S. paradoxus}, \emph{S. mikatae}, \emph{S. kudriavzevii}, \emph{S. castellii}, \emph{S. kluyveri}, \emph{S. bayanus}, and \emph{Candida albicans}.}\label{fig:trees_simul}
\end{figure}
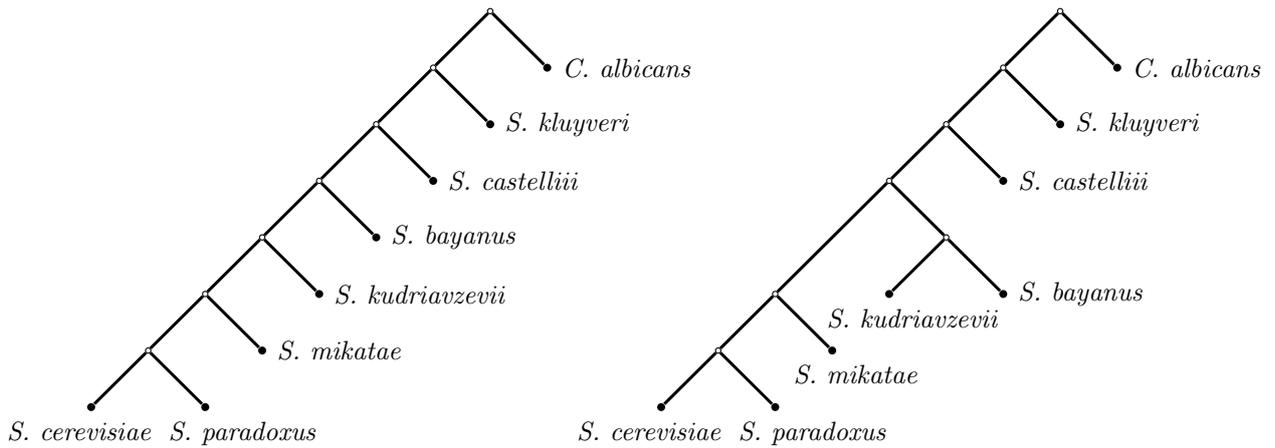


\begin{figure}[H]
	\begin{center}
		\includegraphics[scale=0.6]{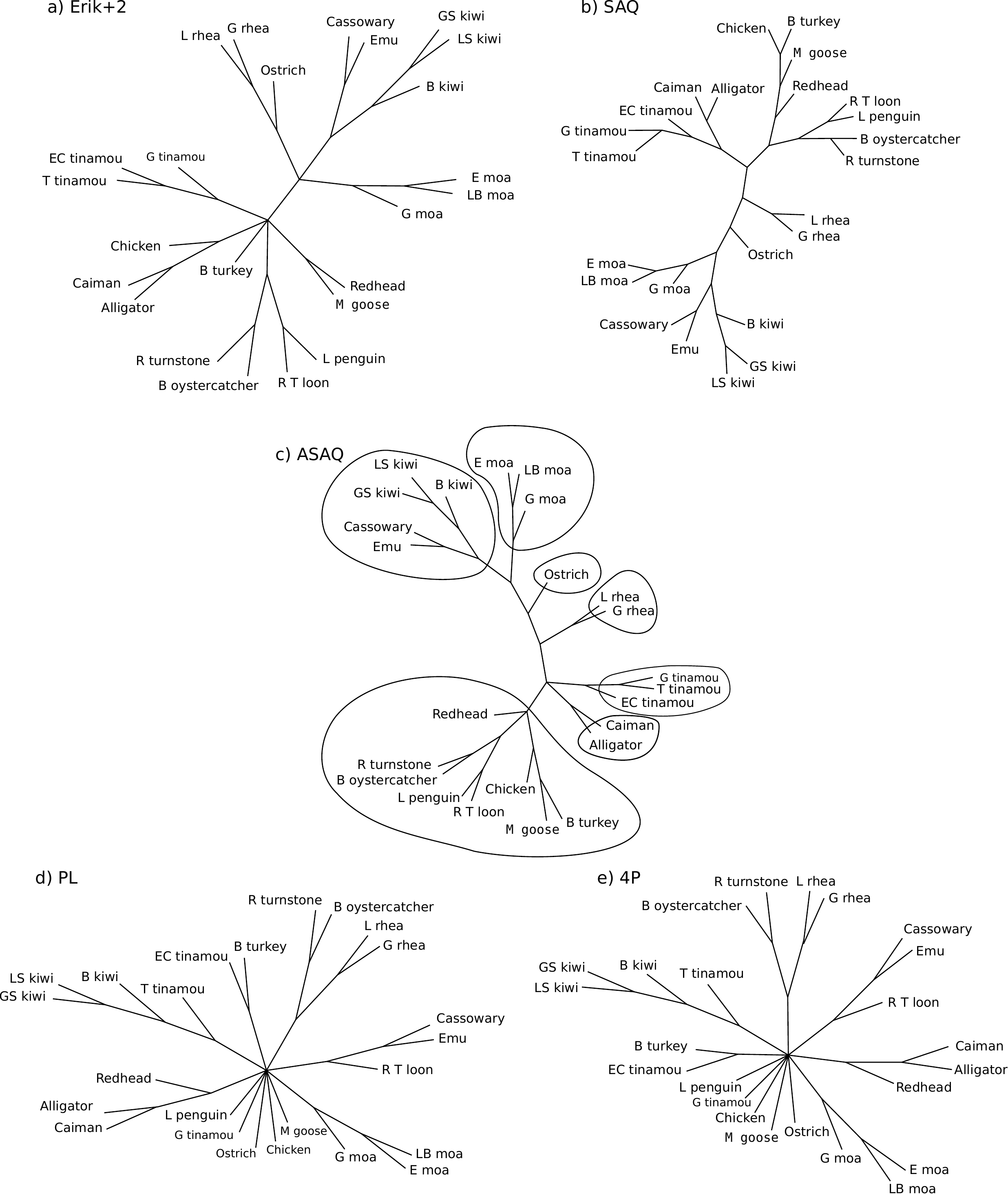}
	\end{center}

	\caption{\label{fig:5_consensustrees}  \footnotesize
		MRCTs obtained for the real data on ratites and tinamous (see Section \ref{sec_realdata}) with WO and different weights and applied to 2125 initial quartets and $m=1$: a) \eri, b) \saq, c) \asaq, d) \pl and d) \fp. The trees obtained with these weights from 5313 initial quartets or with both $m=1$ and $m=2$ are the same as shown. All edges are shown with the same length and they do not represent evolutionary distance.
		The tree c) (\asaq) corresponds to the phylogeny D: (outgroup, neognathus, tinamous, (rheas, (ostrich, (moas, CEK)))).
	}
\end{figure}

\newpage
\section{Appendix C. Performance of \asaq, \pl and Q-methods}



\setcounter{figure}{0}
\setcounter{table}{0}

\subsection{Performance of \asaq and \pl on the treespace under the GTR model}

\begin{figure}[]
\centering
\addtolength{\leftskip} {-2cm}
\addtolength{\rightskip}{-2cm}
\begin{center}
	\hspace{0.4cm} Performance on GTR data for the treespace \\
	\vspace{0.1cm}
	\hspace{0.3cm} \asaq \hspace{6.1cm} \pl \\
	\vspace{-0.5cm}
	\includegraphics[scale=0.4]{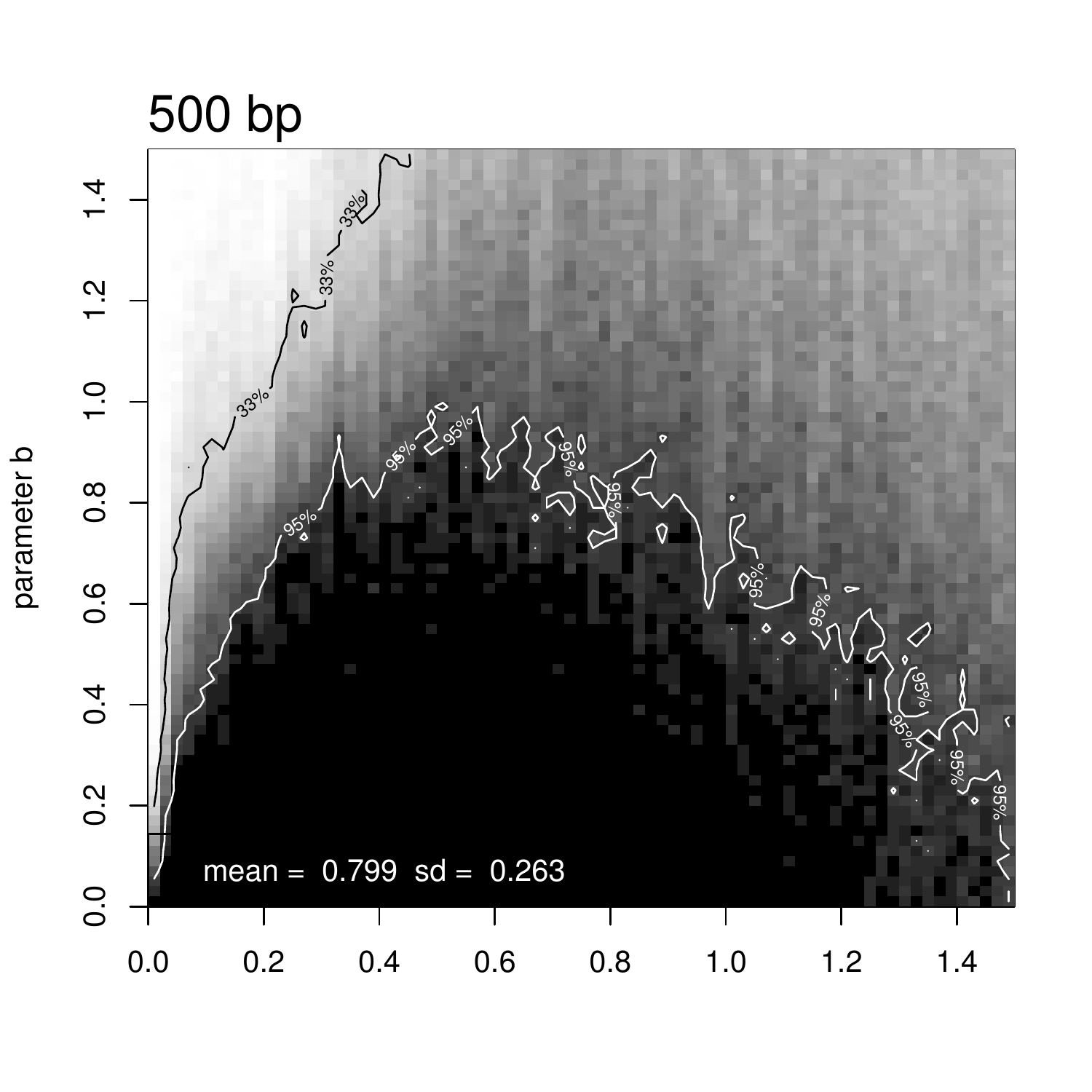}
	\includegraphics[scale=0.4]{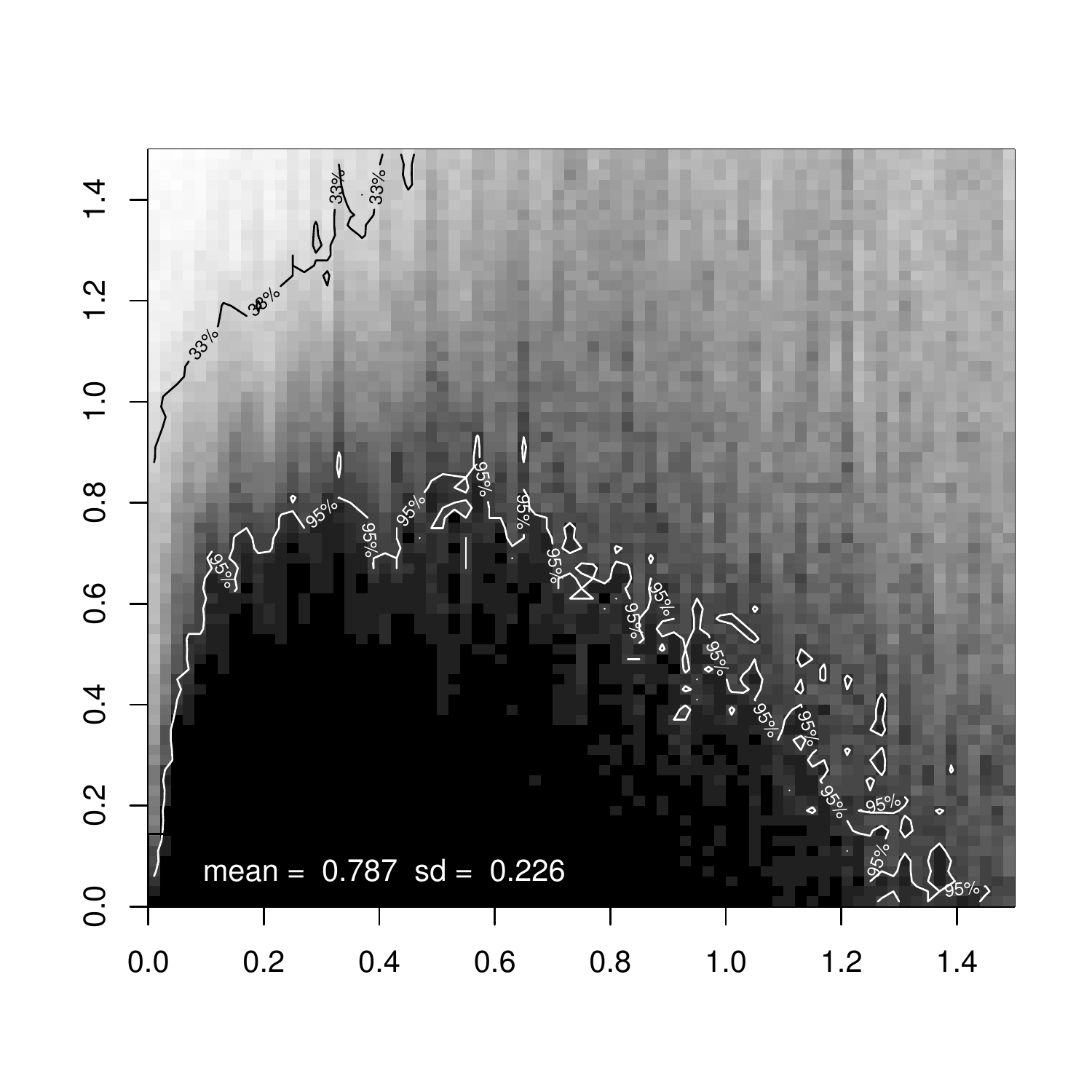}\\
	\vspace{-0.6cm}
	\includegraphics[scale=0.4]{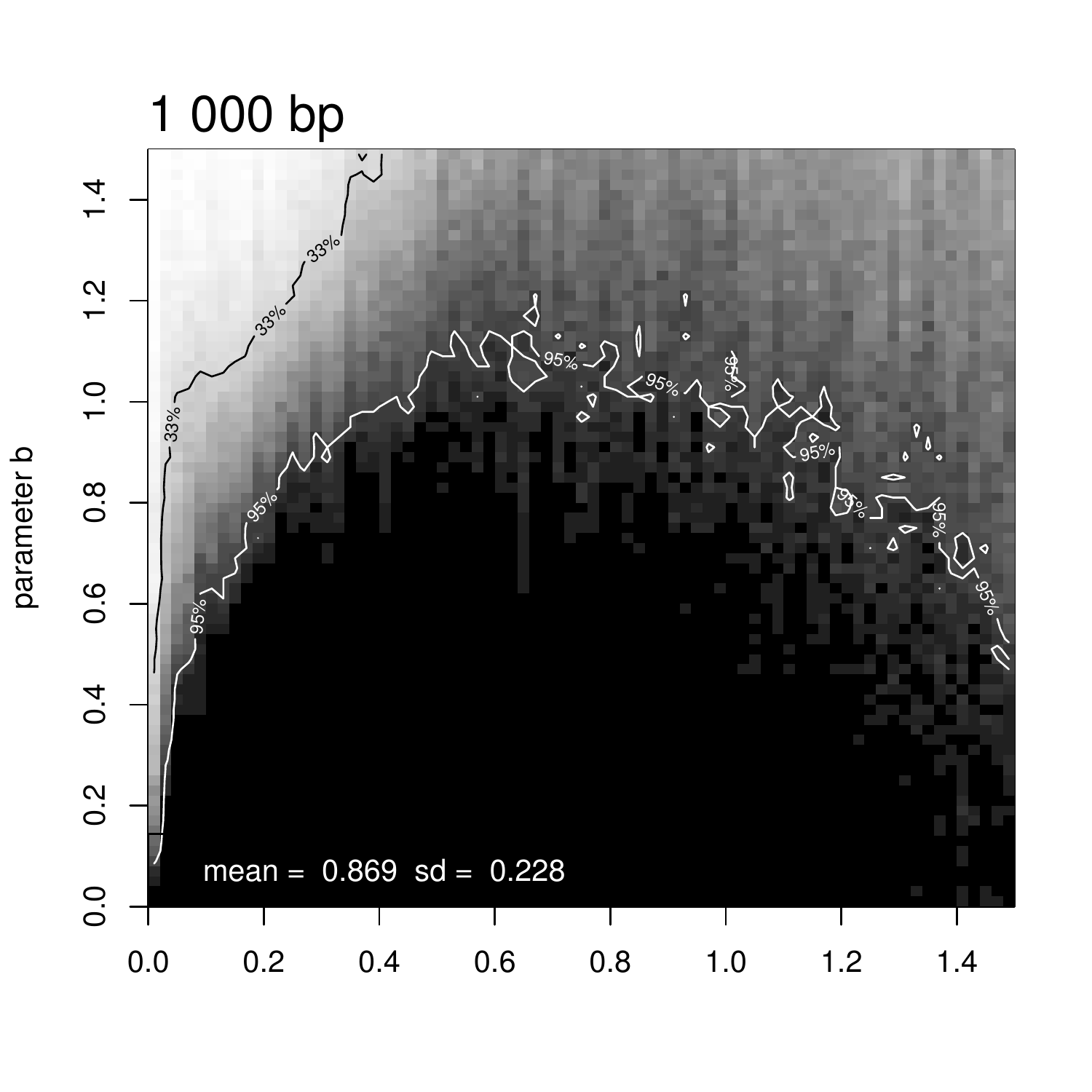}
	\includegraphics[scale=0.4]{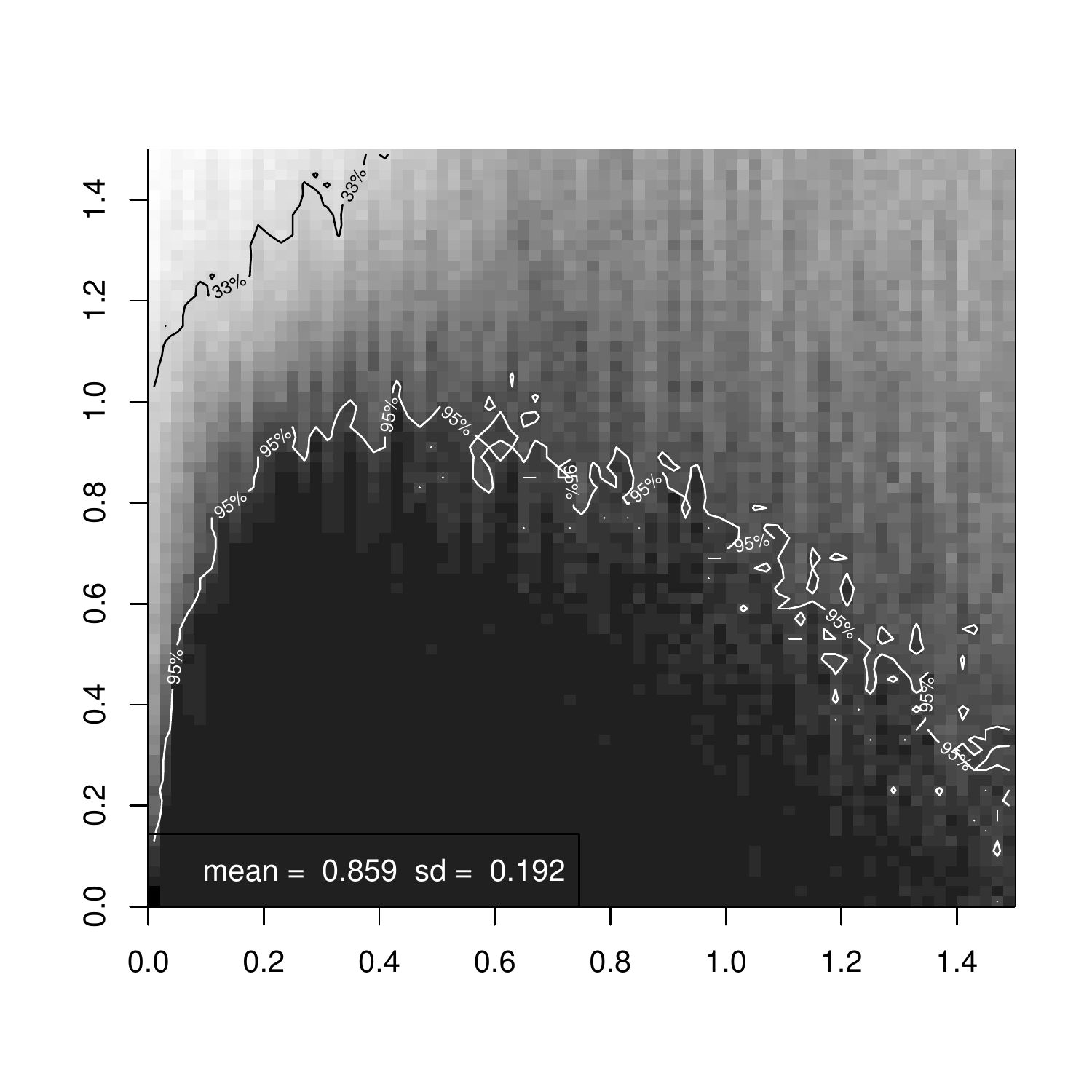}\\
	\vspace{-0.6cm}
	\includegraphics[scale=0.4]{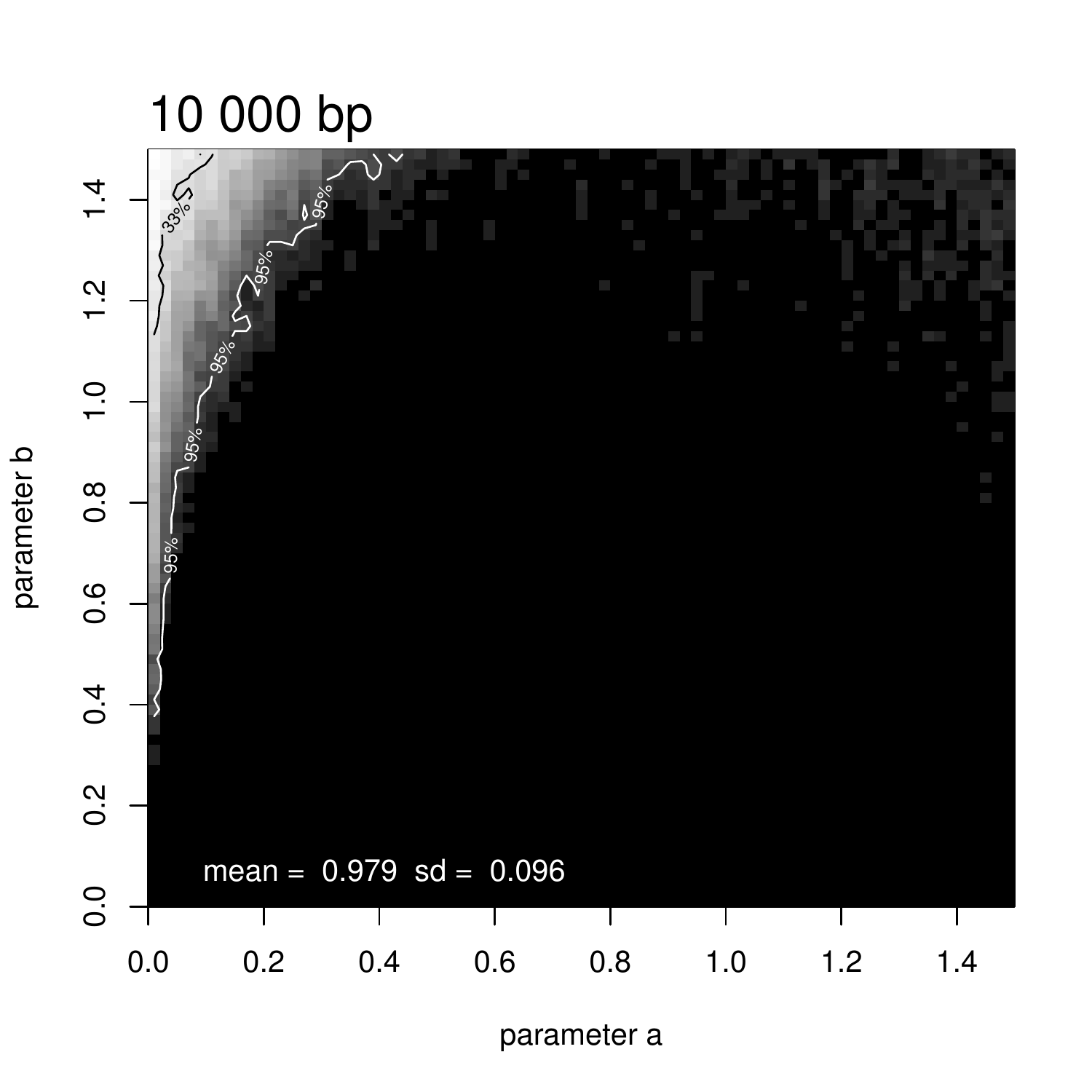}
	\includegraphics[scale=0.4]{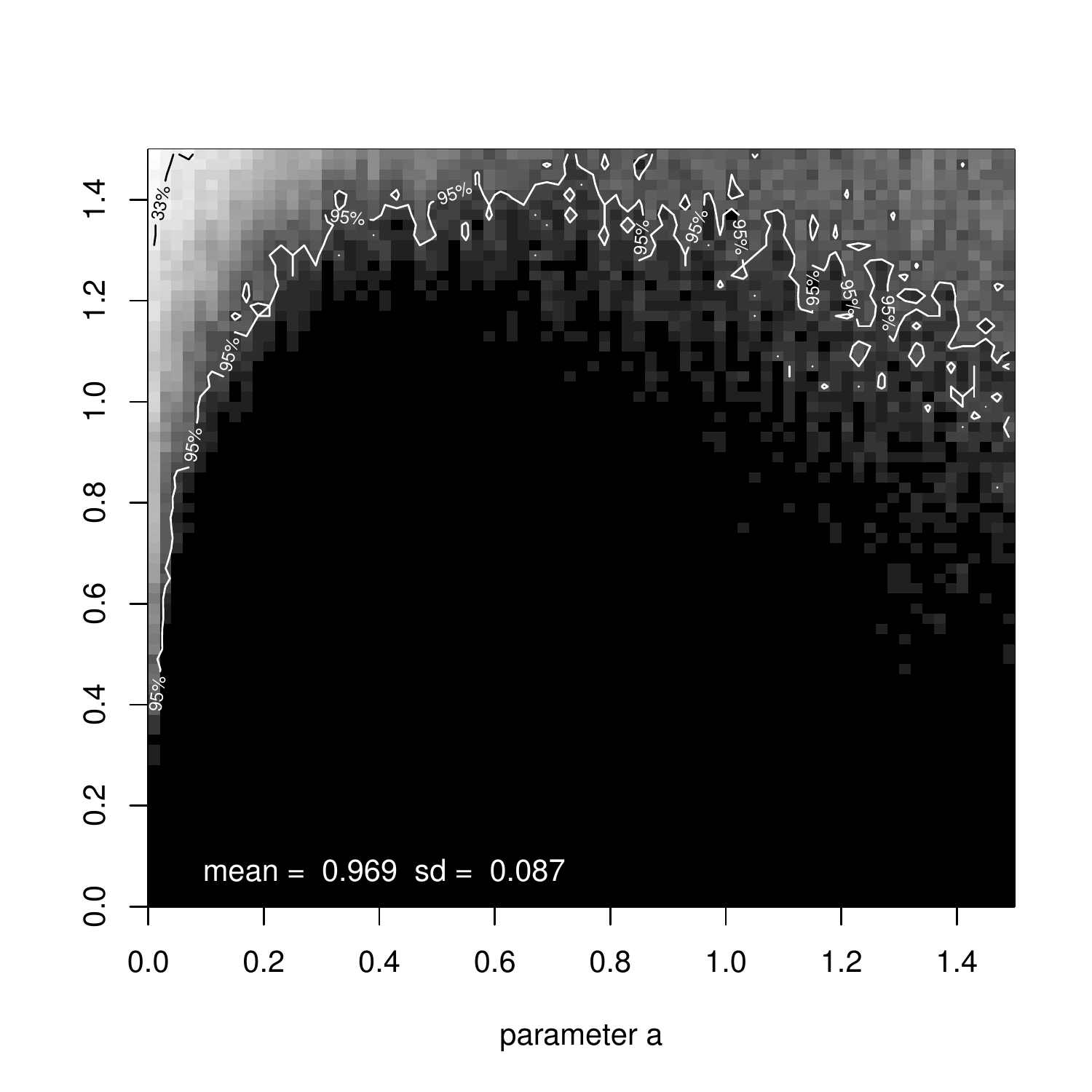}\\
	
\end{center}
\caption{\label{treespaces_GTR} \footnotesize
	Performance of \asaq (left) and \pl (right) in the tree space of Figure \ref{tree} b) on alignments of length  500 bp (top), 1 000 bp (middle) and 10 000 bp (bottom) generated under the GTR model. Black is used to represent 100\% of successful quartet reconstruction, white to represent 0\%, and different tones of gray the intermediate frequencies. The 95\% contour line
	is drawn in white, whereas the 33\% contour line is drawn in black. 
}
\end{figure}

Figure \ref{treespaces_GTR} represents the results of \asaq and \pl in recovering the correct quartet on data generated under the GTR model on the tree space described in section \ref{sec:simdata_quartet}.



\subsection{Performance of \asaq on quartets with random branch lengths under the GTR model}

\begin{figure}[H]
\centering
Performance on GTR data for random branches\\
\includegraphics[scale=0.25]{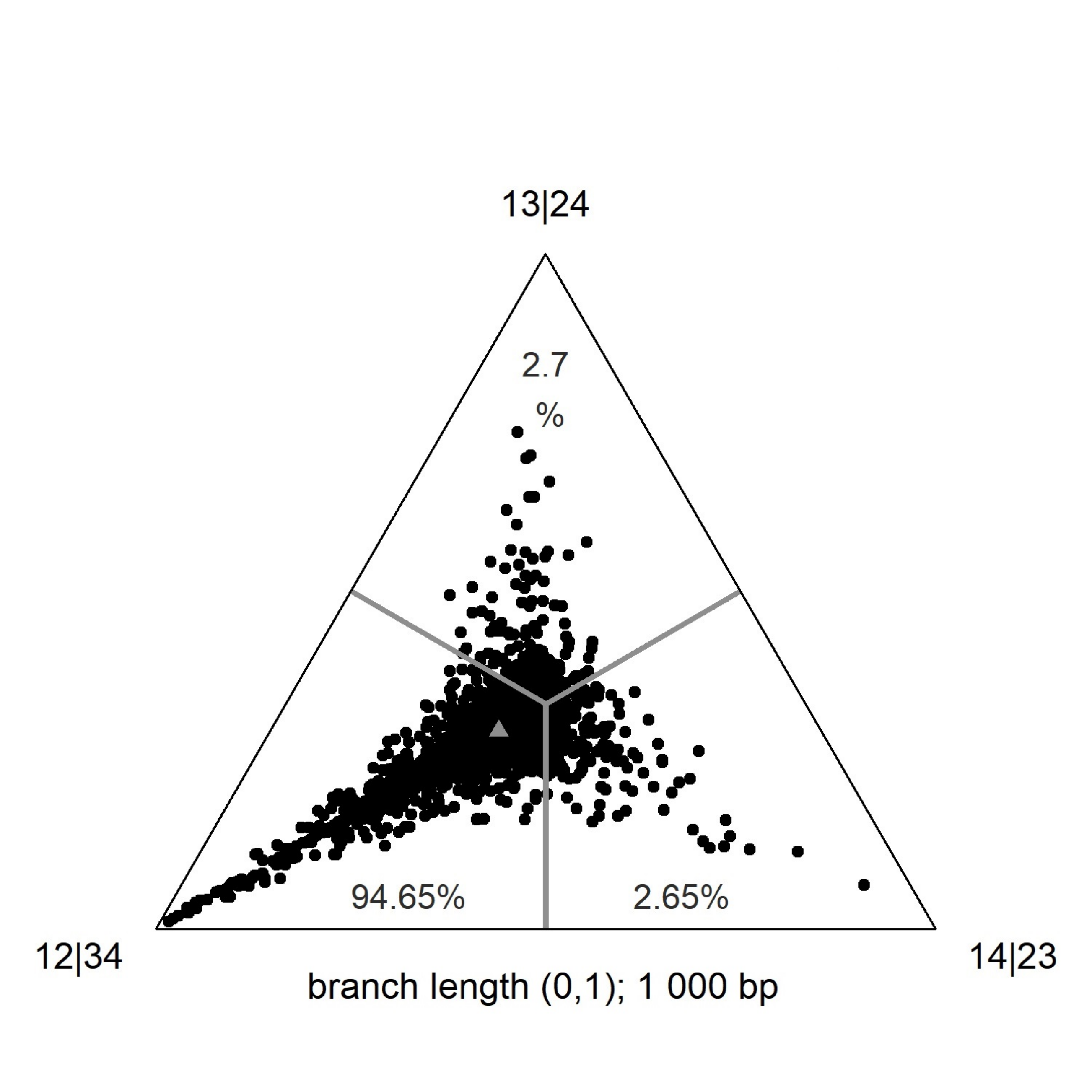}
\includegraphics[scale=0.25]{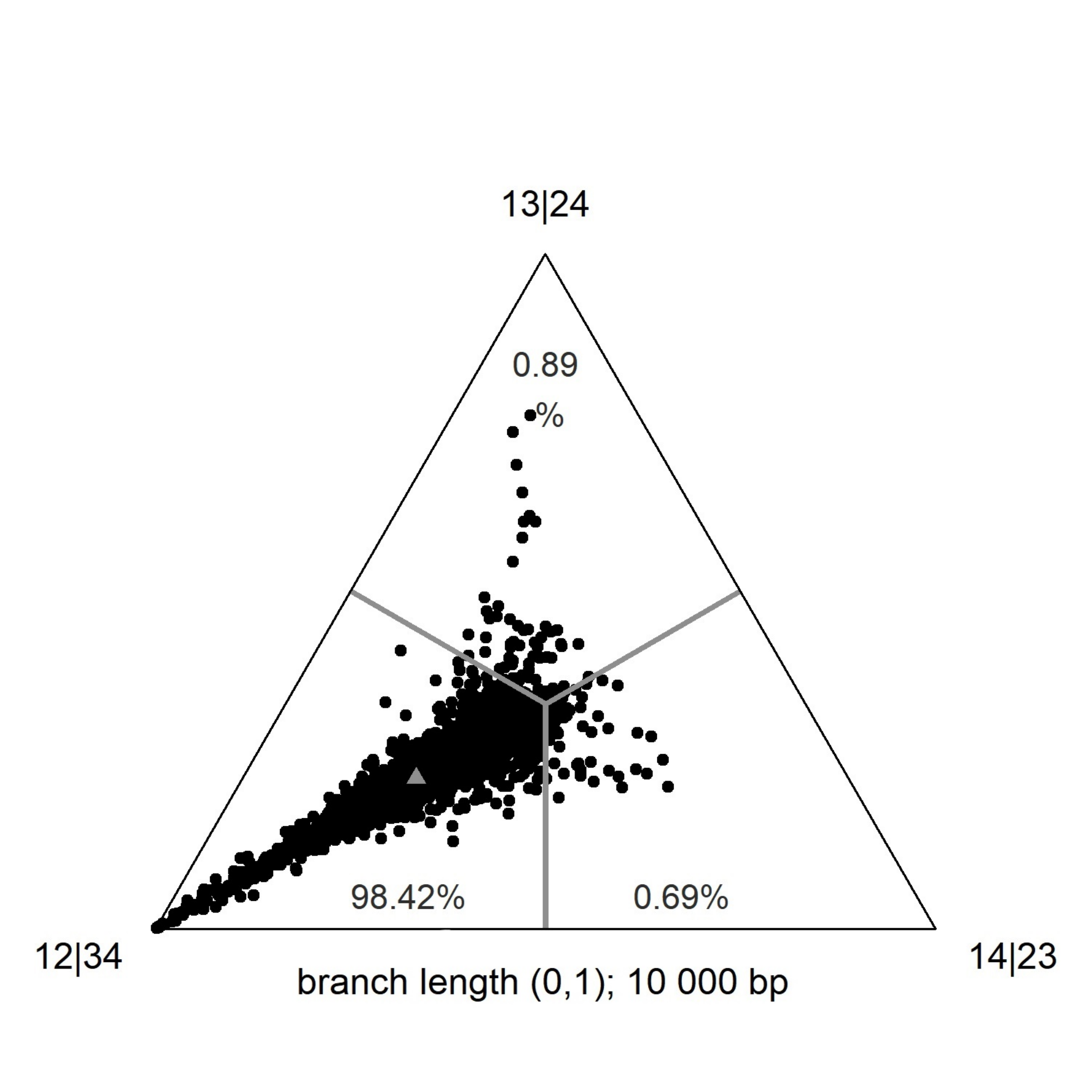} \\
\includegraphics[scale=0.25]{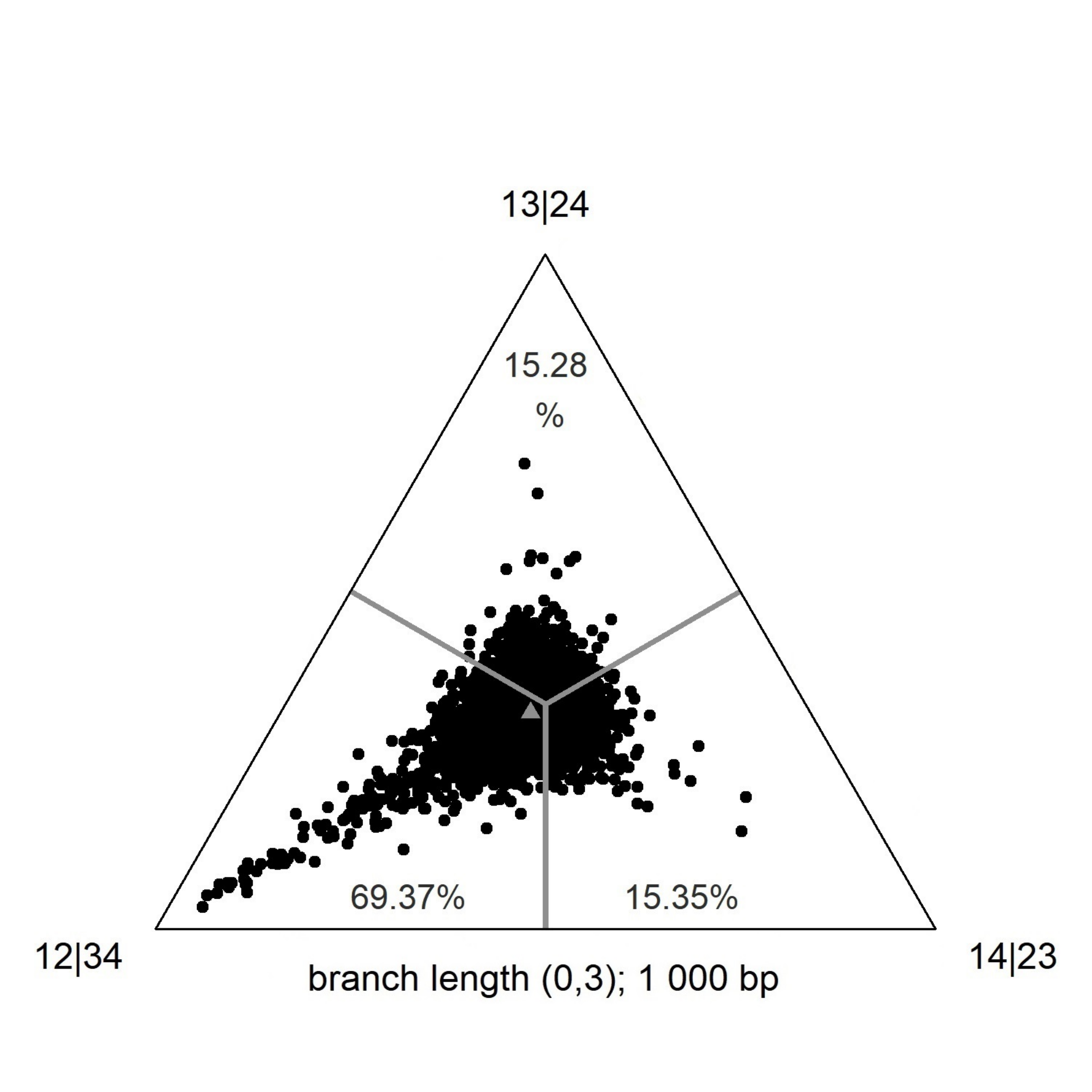}
\includegraphics[scale=0.25]{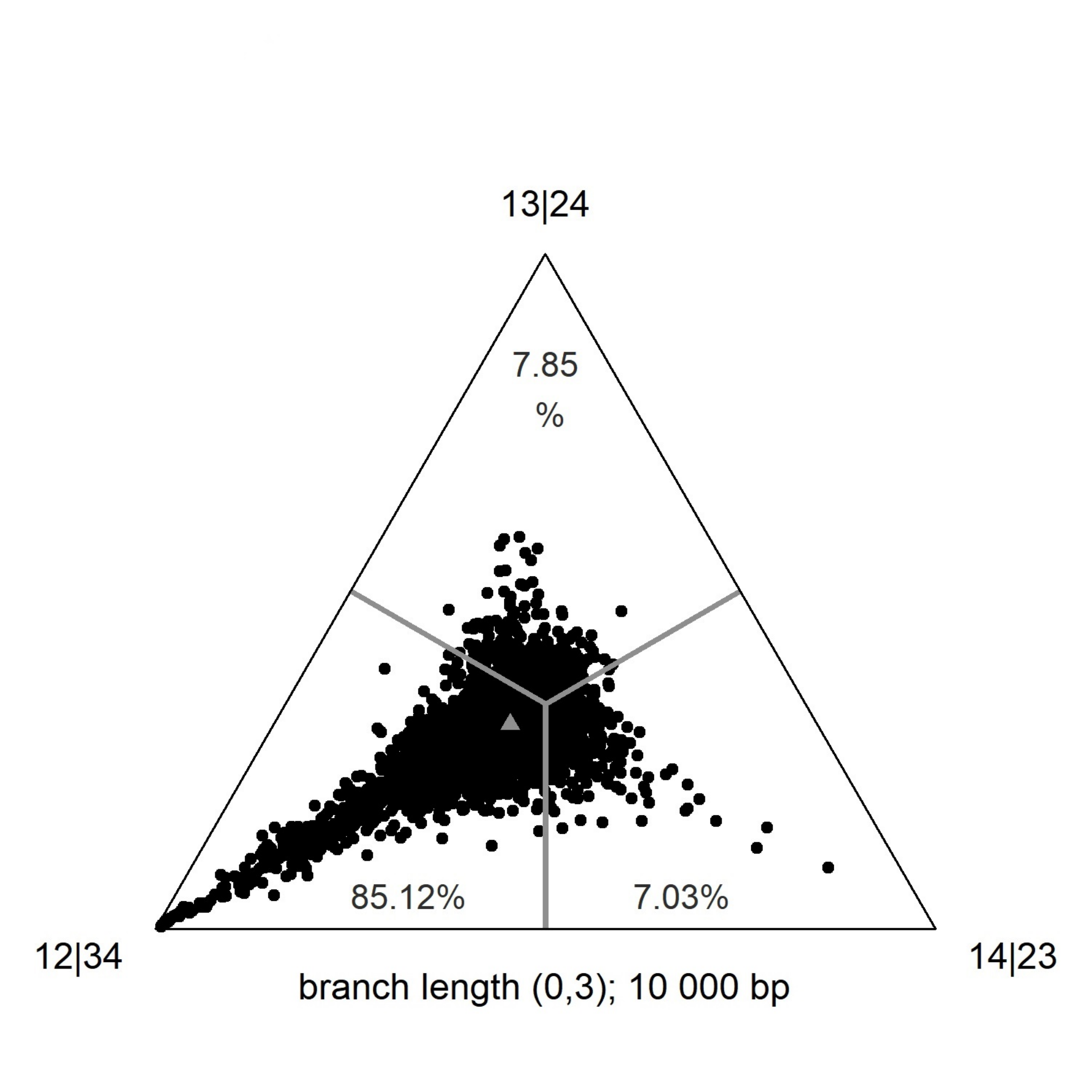}

\caption{\label{ternary_GTR} \footnotesize
	Ternary plots corresponding to the weights of \asaq applied to 10000 alignments generated under the GTR model on the  $12|34$ tree. On each triangle the bottom-left vertex represents the underlying tree $12|34$, the bottom-right vertex is the tree $13|24$ and the top vertex is $14|23$. 
	The small grey triangle point depicted represents the average point of all the dots in the figure. 
	Top: correspond to trees with random branch lengths uniformly distributed between $0$ and $1$; bottom: random branch lengths uniformly distributed between $0$ and $3$. Left : 1 000 bp; Right: 10 000 bp.}
\end{figure}

Figure \ref{ternary_GTR} shows the ternary plots corresponding to the \asaq method applied to GTR data on trees with branches of random length.

\subsection{Performance of Q-methods with different systems of weights}

\begin{figure}[H]
\centering
\includegraphics[scale=0.5]{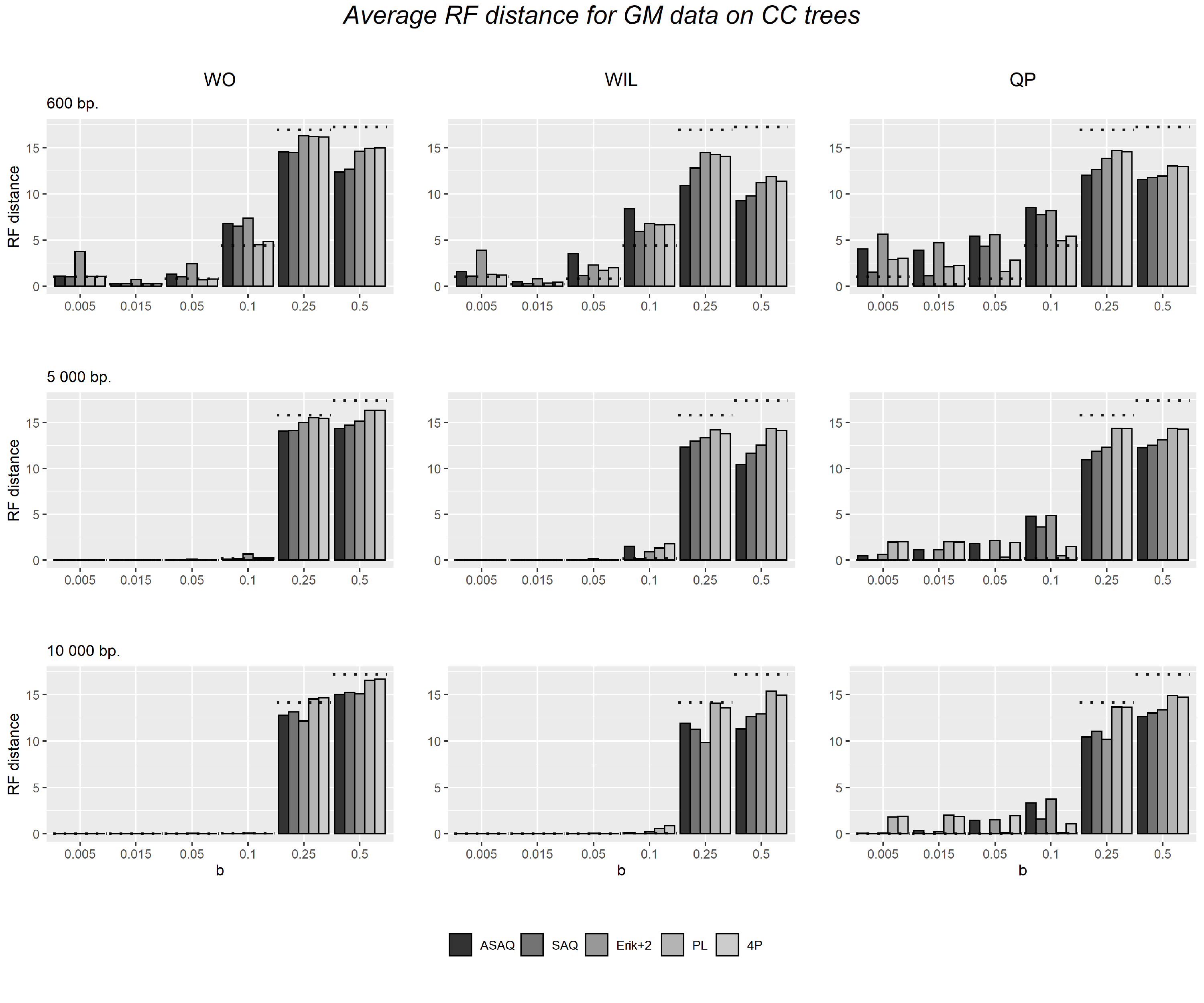}
\captionsetup{width=1\textwidth}
\caption{\label{fig:CCgennonh}  \footnotesize 
	Average Robinson-Foulds distance for GM data simulated on the tree \CC with alignment length 600 bp. (above), 5 000 bp. (middle) and 10 000 bp (below). 
	The Q-methods WO (left), WIL (center) and QP (right) are applied with different systems of weights, namely \asaq, \eri, \pl and \fp. 
	The horizontal dotted lines represent the average RF distance of the tree reconstructed using a \emph{global} \nj with paralinear distance.
	Concrete values of these results are detailed in Table \ref{tab_CCgennonh}.}	
\end{figure}

\begin{figure}
\centering
%
\includegraphics[scale=0.5]{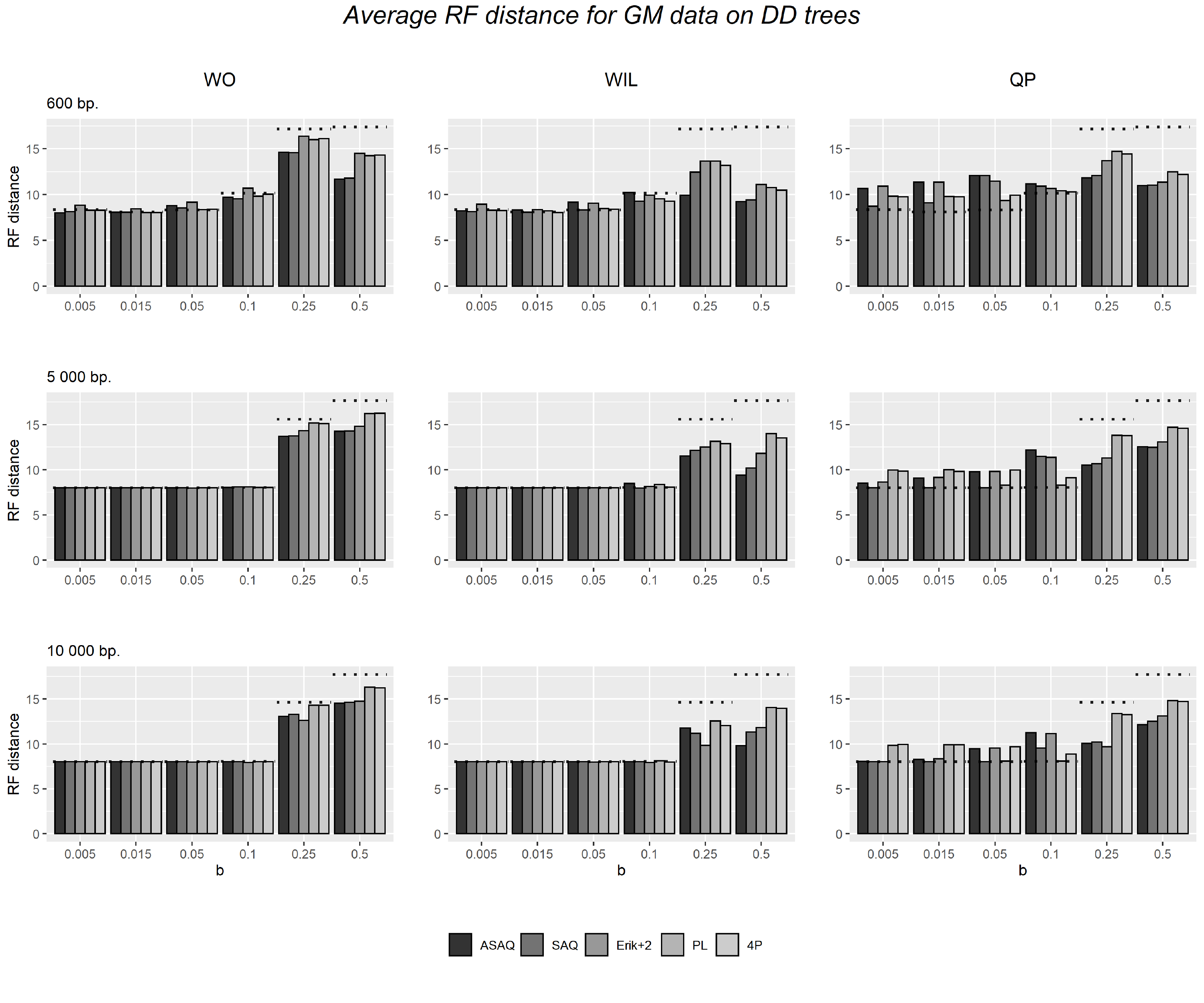}
\captionsetup{width=1\textwidth}
\caption{\label{fig:DDgennonh}  \footnotesize  
	Average Robinson-Foulds distance for GM data simulated on the tree \DD with alignment length 600 bp. (above), 5 000 bp. (middle) and 10 000 bp (below). 
	The Q-methods WO (left), WIL (center) and QP (right) are applied with different systems of weights, namely \asaq, \eri, \pl and \fp. 
	The horizontal dotted lines represent the average RF distance of the tree reconstructed using a \emph{global} \nj with paralinear distance. 
	Concrete values of these results are detailed in Table \ref{tab_DDgennonh}.
}
\end{figure}

Figures \ref{fig:CCgennonh} and \ref{fig:DDgennonh} represent the average Robinson-Foulds distance for GM data on $CC$ and $DD$ trees of Q-methods with \asaq, \eri, \pl and \fp weights. 
Similarly, Figure \ref{fig:DDgtr} shows the average RF distance of Q-methods applied to GTR data on $DD$ trees. 

\subsubsection{Results on the general Markov model}

\newpage

\subsubsection{Results on the GTR model}

\begin{figure}[H]
\centering
\includegraphics[scale=0.25]{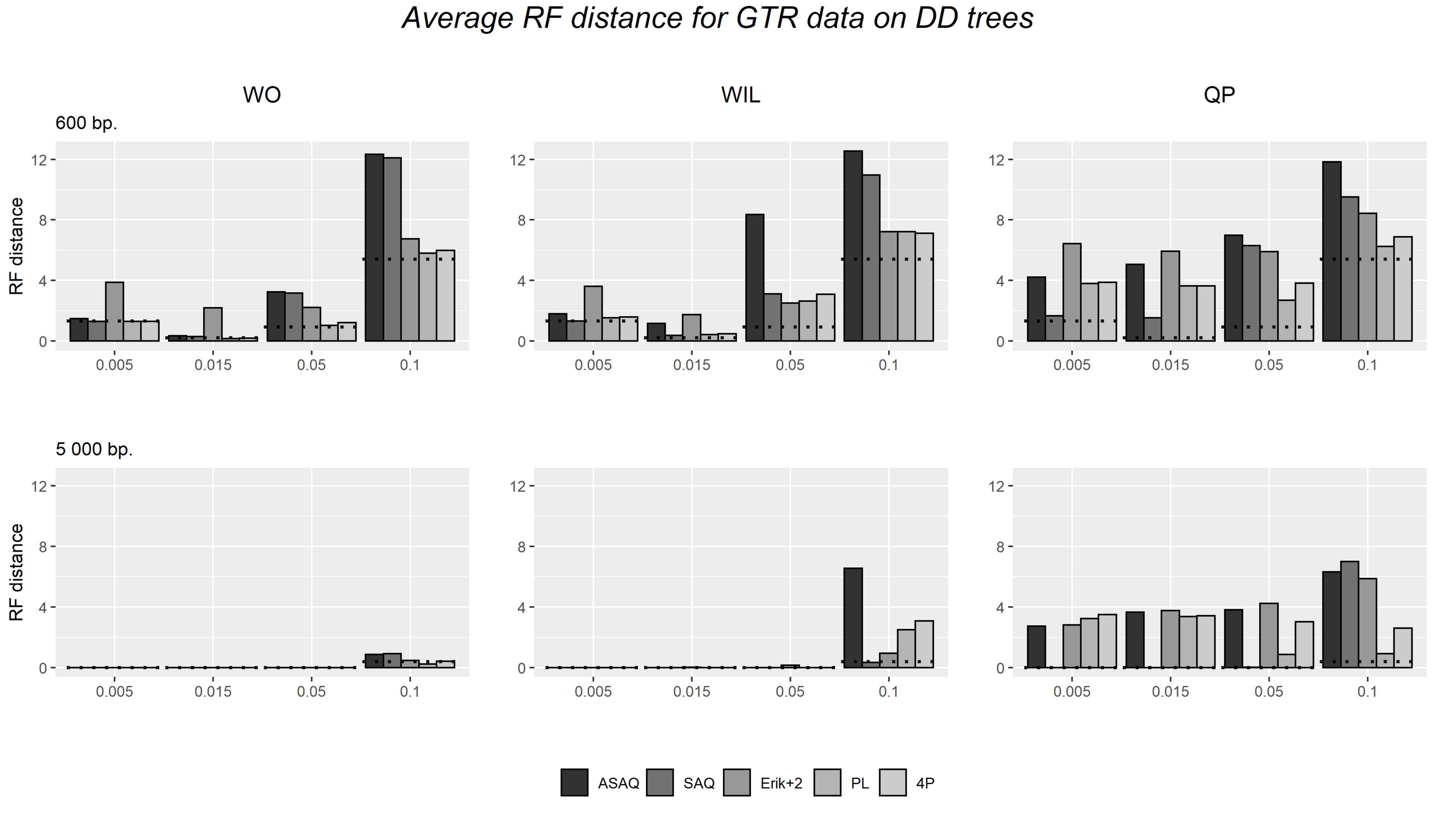}\\
\captionsetup{width=1\textwidth}
\caption{\label{fig:DDgtr}  \footnotesize 	Average Robinson-Foulds distance for GTR data simulated on the tree \DD with alignment length 600 bp. (above)
	and 5 000 bp. (below). 
	The Q-methods WO (left), WIL (center) and QP (right) are applied with different systems of weights, namely \asaq, \eri, \pl and \fp. 
	The The horizontal dotted lines represent the average RF distance of the tree reconstructed using a \emph{global} \nj with paralinear distance.
	Concrete values of these results are detailed in Table \ref{tab_DDgtr}.
}
\end{figure}

The results obtained by Quartet-Puzzling, Weight Optimization and Willson methods applied to the input weights of the different methods on GTR data from the DD trees are shown in Figure \ref{fig:DDgtr} (lenghts 600 bp. and 5 000 bpp.) and summarized in Table \ref{tab_CCgennonh} (also including the results for 10 000 bp.). 
It is remarkable that in these simulations, it is enough to consider alignments of length 5~000 bp. to obtain an almost perfect reconstruction of the original tree using the WO or Willson methods. It is also remarkable that for this particular topology the results described in section 3.2.1, where the general Markov model was assumed, were not this good, obtaining distance values around 8 (see Figure \ref{fig:DDgennonh}). In this case, assuming the GTR these reconstruction methods manage to correctly infer the splits of the tree topology.
Compared with these results, the performance of QP is poor, although it also improves the results obtained when applied to general Markov data.  
For short alignments (600 bp.) \pl and \fp weights obtained slightly better results overall than the other weights.

Note also the bad performance of WO when combined with \asaq weights when applied to alignments of 600 bp. and parameter $b=0.1$. For short alignments, \asaq tends to choose the weights of \saq, so this poor perfomance is probably caused by the bad results of \saq when applied to GTR data (see \cite{casfergar2020}). 

The performance of every Q-method applied to \ml weights for these data was quite poor and the resuling RF distance was never smaller than 8.95 (see Table \ref{tab_DDgtr}.)

\subsection{Tables of figures}
Here we present tables with the figures of the performance of the Q-methods QP, WIL and WO when applied to different weighing systems, namely \asaq, \eri, Maximum likelihood and Neighbour-joining, when applied to different data.
The data has been simulated according to a general Markov model and a general time-reversible model on the trees $CC$, $CD$ and $DD$. 

The values of the 3rd to 6th columns correspond to the average of the \textit{Robinson Foulds distance} to the original tree of the consensus
tree of 100 replicates for each of the 100 generated alignments. In case the reconstruction method failed to provide a weight configuration
for some of the $\binom{12}{4}$ quartets, a resulting tree cannot be constructed and the corresponding alignment has been neglected.
The number in parentheses represents the number of consensus trees that we have been able to reconstruct. The value of the 7th column is the average value of its correspondent row. It indicates the success of the performance of a reconstruction method combined with a quartet-based method, independently of the branch length parameter $b$. Finally, for each quartet-based method and for each value of $b$, the minimum average distance among the four computed by the weighted methods is highlighted in bold type. Thus, this data provides information about which quartet-based method does better with each weighted method.

\subsubsection{Q-methods on GM data}

\newpage

\pagenumbering{gobble}

\begin{landscape}	
	\vspace*{\fill}
	\begin{table}[H]
		\centering
		\footnotesize
		\setlength{\tabcolsep}{3pt}
		
		\begin{tabular}{ccc|cccccc|c|cccccc|c|cccccc|} 
			\cline{4-9} \cline{11-16} \cline{18-23} 
			&    &     & \multicolumn{6}{c|}{Lenght 600 bp}        &  & \multicolumn{6}{c|}{Lenght 5 000 bp}      &  & \multicolumn{6}{c|}{Lenght 10 000 bp}   \\ \hhline{~~~|======|~|======|~|======|} 
			& & & \textbf{b=0.005} & \textbf{0.015} & \textbf{0.05} & \textbf{0.1}  & \textbf{0.25} & \textbf{0.5}  &  & \textbf{0.005} & \textbf{0.015} & \textbf{0.05} & \textbf{0.1}  & \textbf{0.25} & \textbf{0.5}  &  & \textbf{0.005} & \textbf{0.015} & \textbf{0.05} & \textbf{0.1}  & \textbf{0.25} & \textbf{0.5}  \\ \hhline{|=|=|~|======|~|======|~|======|} 
			\multicolumn{1}{|c|}{\multirow{6}{*}{WO}}  & \multicolumn{1}{c|}{\asaq}  & & 1.1 & 0.26 & 1.29 & 6.79 & 14.52 & 12.36  & & 0 & 0 & 0 & 0.08 & 14.08 & 14.33  & & 0 & 0 & 0 & 0 & 12.79 & 15.01  \\ \cline{2-2} \cline{4-9} \cline{11-16} \cline{18-23}  
			\multicolumn{1}{|c|}{\multirow{6}{*}{}}  & \multicolumn{1}{c|}{\saq}  & & 1 & 0.28 & 1.02 & 6.5 & 14.44 & 12.67  & & 0 & 0 & 0 & 0.14 & 14.11 & 14.68  & & 0 & 0 & 0 & 0 & 13.14 & 15.23  \\ \cline{2-2} \cline{4-9} \cline{11-16} \cline{18-23}  
			\multicolumn{1}{|c|}{\multirow{6}{*}{}}  & \multicolumn{1}{c|}{\eri}  & & 3.75 & 0.74 & 2.42 & 7.37 & 16.29 & 14.61  & & 0 & 0 & 0.1 & 0.65 & 14.98 & 15.12  & & 0 & 0 & 0.04 & 0.1 & 12.18 & 15.08  \\ \cline{2-2} \cline{4-9} \cline{11-16} \cline{18-23}  
			\multicolumn{1}{|c|}{\multirow{6}{*}{}}  & \multicolumn{1}{c|}{\pl}  & & 1.06 & 0.26 & 0.69 & 4.5 & 16.19 & 14.92  & & 0 & 0 & 0 & 0.19 & 15.52 & 16.32  & & 0 & 0 & 0 & 0 & 14.54 & 16.57  \\ \cline{2-2} \cline{4-9} \cline{11-16} \cline{18-23}  
			\multicolumn{1}{|c|}{\multirow{6}{*}{}}  & \multicolumn{1}{c|}{\fp}  & & 1.06 & 0.26 & 0.76 & 4.84 & 16.16 & 14.96  & & 0 & 0 & 0 & 0.19 & 15.45 & 16.35  & & 0 & 0 & 0 & 0.02 & 14.65 & 16.68  \\ \cline{2-2} \cline{4-9} \cline{11-16} \cline{18-23}  
			\multicolumn{1}{|c|}{\multirow{6}{*}{}}  & \multicolumn{1}{c|}{\ml}  & & 12.71 & 12.25 & 12.75 & 11 \scriptsize{(3)} & - & -  & & 11.03 & 11.18 \scriptsize{(93)} & 11.83 \scriptsize{(93)} & - \scriptsize{(0)} & - & -  & & 10.97 & 11.01 \scriptsize{(96)} & 11.69 \scriptsize{(91)} & 13 \scriptsize{(1)} & - & -  \\ \hhline{|=|=|~|======|~|======|~|======|} 
			\multicolumn{1}{|c|}{\multirow{6}{*}{WIL}}  & \multicolumn{1}{c|}{\asaq}  & & 1.61 & 0.47 & 3.5 & 8.37 & 10.92 & 9.23  & & 0 & 0 & 0 & 1.47 & 12.31 & 10.41  & & 0 & 0 & 0 & 0.14 & 11.91 & 11.29  \\ \cline{2-2} \cline{4-9} \cline{11-16} \cline{18-23}  
			\multicolumn{1}{|c|}{\multirow{6}{*}{}}  & \multicolumn{1}{c|}{\saq}  & & 1.1 & 0.3 & 1.14 & 5.93 & 12.79 & 9.77  & & 0 & 0 & 0 & 0.12 & 12.97 & 11.63  & & 0 & 0 & 0 & 0 & 11.26 & 12.62  \\ \cline{2-2} \cline{4-9} \cline{11-16} \cline{18-23}  
			\multicolumn{1}{|c|}{\multirow{6}{*}{}}  & \multicolumn{1}{c|}{\eri}  & & 3.87 & 0.78 & 2.28 & 6.76 & 14.45 & 11.18  & & 0 & 0 & 0.15 & 0.89 & 13.35 & 12.54  & & 0 & 0 & 0.06 & 0.21 & 9.86 & 12.92  \\ \cline{2-2} \cline{4-9} \cline{11-16} \cline{18-23}  
			\multicolumn{1}{|c|}{\multirow{6}{*}{}}  & \multicolumn{1}{c|}{\pl}  & & 1.27 & 0.34 & 1.7 & 6.63 & 14.23 & 11.89  & & 0 & 0 & 0 & 1.3 & 14.18 & 14.31  & & 0 & 0 & 0 & 0.54 & 14.06 & 15.36  \\ \cline{2-2} \cline{4-9} \cline{11-16} \cline{18-23}  
			\multicolumn{1}{|c|}{\multirow{6}{*}{}}  & \multicolumn{1}{c|}{\fp}  & & 1.19 & 0.43 & 2 & 6.65 & 14.05 & 11.38  & & 0 & 0 & 0 & 1.78 & 13.78 & 14.09  & & 0 & 0 & 0 & 0.88 & 13.59 & 14.96  \\ \cline{2-2} \cline{4-9} \cline{11-16} \cline{18-23}  
			\multicolumn{1}{|c|}{\multirow{6}{*}{}}  & \multicolumn{1}{c|}{\ml}  & & 9.83 & 9.79 & 9.84 & 9 \scriptsize{(3)} & - & -  & & 9.97 & 9.28 \scriptsize{(93)} & 9.27 \scriptsize{(93)} & - \scriptsize{(0)} & - & -  & & 9.33 & 9 \scriptsize{(96)} & 8.98 \scriptsize{(91)} & 9 \scriptsize{(1)} & - & -  \\ \hhline{|=|=|~|======|~|======|~|======|} 
			\multicolumn{1}{|c|}{\multirow{6}{*}{QP}}  & \multicolumn{1}{c|}{\asaq}  & & 4.03 & 3.89 & 5.41 & 8.52 & 12.03 & 11.56  & & 0.47 & 1.11 & 1.81 & 4.76 & 10.94 & 12.24  & & 0.05 & 0.35 & 1.46 & 3.34 & 10.44 & 12.63  \\ \cline{2-2} \cline{4-9} \cline{11-16} \cline{18-23}  
			\multicolumn{1}{|c|}{\multirow{6}{*}{}}  & \multicolumn{1}{c|}{\saq}  & & 1.53 & 1.11 & 4.31 & 7.74 & 12.63 & 11.77  & & 0 & 0 & 0 & 3.59 & 11.85 & 12.49  & & 0 & 0 & 0 & 1.59 & 11.04 & 13.05  \\ \cline{2-2} \cline{4-9} \cline{11-16} \cline{18-23}  
			\multicolumn{1}{|c|}{\multirow{6}{*}{}}  & \multicolumn{1}{c|}{\eri}  & & 5.6 & 4.72 & 5.58 & 8.2 & 13.82 & 11.91  & & 0.62 & 1.13 & 2.12 & 4.86 & 12.27 & 13.07  & & 0.08 & 0.24 & 1.51 & 3.72 & 10.19 & 13.36  \\ \cline{2-2} \cline{4-9} \cline{11-16} \cline{18-23}  
			\multicolumn{1}{|c|}{\multirow{6}{*}{}}  & \multicolumn{1}{c|}{\pl}  & & 2.88 & 2.1 & 1.6 & 4.92 & 14.69 & 13.02  & & 1.95 & 2.01 & 0.32 & 0.47 & 14.38 & 14.38  & & 1.83 & 1.98 & 0.13 & 0.13 & 13.69 & 14.92  \\ \cline{2-2} \cline{4-9} \cline{11-16} \cline{18-23}  
			\multicolumn{1}{|c|}{\multirow{6}{*}{}}  & \multicolumn{1}{c|}{\fp}  & & 3 & 2.24 & 2.82 & 5.4 & 14.55 & 12.95  & & 1.99 & 1.94 & 1.9 & 1.44 & 14.31 & 14.26  & & 1.89 & 1.87 & 1.96 & 1.05 & 13.65 & 14.73  \\ \cline{2-2} \cline{4-9} \cline{11-16} \cline{18-23}  
			\multicolumn{1}{|c|}{\multirow{6}{*}{}}  & \multicolumn{1}{c|}{\ml}  & & 9.55 & 9.58 & 9.67 & 10.67 \scriptsize{(3)} & - & -  & & 9 & 9.08 \scriptsize{(93)} & 9.65 \scriptsize{(93)} & - \scriptsize{(0)} & - & -  & & 9 & 9.11 \scriptsize{(96)} & 9.6 \scriptsize{(91)} & 9 \scriptsize{(1)} & - & -  \\ \hhline{|=|=|~|======|~|======|~|======|} 
			\multicolumn{2}{|c|}{Global \nj} & & 1.02 & 0.2 & 0.8 & 4.38 & 16.92 & 17.26  & & 0 & 0 & 0 & 0.14 & 15.8 & 17.38  & & 0 & 0 & 0 & 0.04 & 14.14 & 17.18  \\ \cline{1-9}  \cline{4-9} \cline{11-16} \cline{18-23}  
		\end{tabular}
		\vspace*{2mm}
		\captionsetup{width=1.4\textwidth}
		
		\caption{ 
			\label{tab_CCgennonh}  \small Average Robinson-Foulds distance for tree \textit{CC}. 
			The number in parentheses represents the number of consensus trees that we have been able to reconstruct (\mll(homGMc) may not converge for some quartets and in this case we cannot reconstruct the tree). If it is missing. then we have been able to reconstruct the $100$ simulated trees. }
	\end{table}
	\vspace*{\fill}
\end{landscape}

\begin{landscape}
	\vspace*{\fill}
	\begin{table}[H]
		\centering
		\footnotesize
		\setlength{\tabcolsep}{3pt}
		\begin{tabular}{ccc|cccccc|c|cccccc|c|cccccc|} 
			\cline{4-9} \cline{11-16} \cline{18-23} 
			&    &     & \multicolumn{6}{c|}{Lenght 600 bp}        &  & \multicolumn{6}{c|}{Lenght 5 000 bp}      &  & \multicolumn{6}{c|}{Lenght 10 000 bp}   \\ \hhline{~~~|======|~|======|~|======|} 
			& & & \textbf{b=0.005} & \textbf{0.015} & \textbf{0.05} & \textbf{0.1}  & \textbf{0.25} & \textbf{0.5}  &  & \textbf{0.005} & \textbf{0.015} & \textbf{0.05} & \textbf{0.1}  & \textbf{0.25} & \textbf{0.5}  &  & \textbf{0.005} & \textbf{0.015} & \textbf{0.05} & \textbf{0.1}  & \textbf{0.25} & \textbf{0.5}  \\ \hhline{|=|=|~|======|~|======|~|======|} 
			\multicolumn{1}{|c|}{\multirow{6}{*}{WO}}  & \multicolumn{1}{c|}{\asaq}  & & 4.4 & 4.22 & 4.84 & 8.46 & 14.47 & 11.83  & & 4 & 4 & 4 & 4.06 & 13.43 & 14.26  & & 4 & 4 & 4 & 4 & 12.69 & 14.9  \\ \cline{2-2} \cline{4-9} \cline{11-16} \cline{18-23}  
			\multicolumn{1}{|c|}{\multirow{6}{*}{}}  & \multicolumn{1}{c|}{\saq}  & & 4.5 & 4.18 & 4.58 & 8.37 & 14.54 & 11.95  & & 4 & 4 & 4 & 4.14 & 13.34 & 14.73  & & 4 & 4 & 4 & 4 & 12.68 & 15.2  \\ \cline{2-2} \cline{4-9} \cline{11-16} \cline{18-23}  
			\multicolumn{1}{|c|}{\multirow{6}{*}{}}  & \multicolumn{1}{c|}{\eri}  & & 6.06 & 4.58 & 5.58 & 9.73 & 16.29 & 14.62  & & 4 & 4 & 4.06 & 4.41 & 14.25 & 14.83  & & 4 & 4 & 4 & 4 & 12.45 & 15.49  \\ \cline{2-2} \cline{4-9} \cline{11-16} \cline{18-23}  
			\multicolumn{1}{|c|}{\multirow{6}{*}{}}  & \multicolumn{1}{c|}{\pl}  & & 4.44 & 4.22 & 4.3 & 7.46 & 16.06 & 13.98  & & 4 & 4 & 4 & 4.09 & 15.07 & 16.32  & & 4 & 4 & 4 & 4 & 14.72 & 16.51  \\ \cline{2-2} \cline{4-9} \cline{11-16} \cline{18-23}  
			\multicolumn{1}{|c|}{\multirow{6}{*}{}}  & \multicolumn{1}{c|}{\fp}  & & 4.4 & 4.21 & 4.33 & 7.71 & 16.21 & 14.16  & & 4 & 4 & 4 & 4.1 & 15.06 & 16.27  & & 4 & 4 & 4 & 4 & 14.71 & 16.4  \\ \cline{2-2} \cline{4-9} \cline{11-16} \cline{18-23}  
			\multicolumn{1}{|c|}{\multirow{6}{*}{}}  & \multicolumn{1}{c|}{\ml}  & & 12.28 \scriptsize{(78)} & 12.04 \scriptsize{(98)} & 12.32 \scriptsize{(95)} & 10 \scriptsize{(1)} & - & -  & & 11.04 & 11.18 \scriptsize{(96)} & 11.76 \scriptsize{(96)} & 10 \scriptsize{(1)} & - & -  & & 10.96 & 10.96 \scriptsize{(96)} & 11.55 \scriptsize{(98)} & 9 \scriptsize{(1)} & - & -  \\ \hhline{|=|=|~|======|~|======|~|======|} 
			\multicolumn{1}{|c|}{\multirow{6}{*}{WIL}}  & \multicolumn{1}{c|}{\asaq}  & & 4.73 & 4.34 & 6.12 & 9.59 & 10.2 & 9.2  & & 4 & 4 & 4 & 5.11 & 12.1 & 10  & & 4 & 4 & 4 & 4.09 & 11.65 & 10.77  \\ \cline{2-2} \cline{4-9} \cline{11-16} \cline{18-23}  
			\multicolumn{1}{|c|}{\multirow{6}{*}{}}  & \multicolumn{1}{c|}{\saq}  & & 4.46 & 4.13 & 4.39 & 7.99 & 12.74 & 9.54  & & 4 & 4 & 4 & 3.98 & 12.21 & 11.22  & & 4 & 4 & 4 & 4 & 11.05 & 12.15  \\ \cline{2-2} \cline{4-9} \cline{11-16} \cline{18-23}  
			\multicolumn{1}{|c|}{\multirow{6}{*}{}}  & \multicolumn{1}{c|}{\eri}  & & 6.1 & 4.59 & 5.53 & 8.38 & 13.89 & 10.98  & & 4 & 4 & 4.04 & 4.63 & 12.82 & 11.94  & & 4 & 4 & 4.02 & 4.05 & 10.15 & 12.42  \\ \cline{2-2} \cline{4-9} \cline{11-16} \cline{18-23}  
			\multicolumn{1}{|c|}{\multirow{6}{*}{}}  & \multicolumn{1}{c|}{\pl}  & & 4.69 & 4.33 & 4.69 & 8.08 & 14.17 & 11.07  & & 4 & 4 & 4 & 4.68 & 13.64 & 14.25  & & 4 & 4 & 4 & 4.36 & 13.32 & 14.85  \\ \cline{2-2} \cline{4-9} \cline{11-16} \cline{18-23}  
			\multicolumn{1}{|c|}{\multirow{6}{*}{}}  & \multicolumn{1}{c|}{\fp}  & & 4.64 & 4.36 & 4.97 & 8.17 & 13.84 & 10.79  & & 4 & 4 & 4 & 4.85 & 13.57 & 14.04  & & 4 & 4 & 4 & 4.42 & 13.05 & 14.36  \\ \cline{2-2} \cline{4-9} \cline{11-16} \cline{18-23}  
			\multicolumn{1}{|c|}{\multirow{6}{*}{}}  & \multicolumn{1}{c|}{\ml}  & & 9.87 \scriptsize{(78)} & 9.9 \scriptsize{(98)} & 9.61 \scriptsize{(95)} & 9 \scriptsize{(1)} & - & -  & & 10 & 9.34 \scriptsize{(96)} & 9.34 \scriptsize{(96)} & 9 \scriptsize{(1)} & - & -  & & 9.33 & 8.99 \scriptsize{(96)} & 8.99 \scriptsize{(98)} & 9 \scriptsize{(1)} & - & -  \\ \hhline{|=|=|~|======|~|======|~|======|} 
			\multicolumn{1}{|c|}{\multirow{6}{*}{QP}}  & \multicolumn{1}{c|}{\asaq}  & & 7.02 & 7.83 & 8.57 & 10.3 & 11.9 & 11.23  & & 4.51 & 5.12 & 5.83 & 8.67 & 10.67 & 12.22  & & 4.05 & 4.36 & 5.5 & 7.31 & 9.96 & 12.81  \\ \cline{2-2} \cline{4-9} \cline{11-16} \cline{18-23}  
			\multicolumn{1}{|c|}{\multirow{6}{*}{}}  & \multicolumn{1}{c|}{\saq}  & & 4.9 & 5.12 & 8.11 & 9.66 & 12.3 & 11.45  & & 4 & 4 & 4 & 7.61 & 11.34 & 12.44  & & 4 & 4 & 4 & 5.49 & 10.52 & 12.9  \\ \cline{2-2} \cline{4-9} \cline{11-16} \cline{18-23}  
			\multicolumn{1}{|c|}{\multirow{6}{*}{}}  & \multicolumn{1}{c|}{\eri}  & & 8.17 & 8.06 & 8.77 & 9.65 & 13.83 & 11.64  & & 4.66 & 5.07 & 5.94 & 8.1 & 11.97 & 13.1  & & 4.04 & 4.32 & 5.56 & 7.27 & 9.79 & 13.54  \\ \cline{2-2} \cline{4-9} \cline{11-16} \cline{18-23}  
			\multicolumn{1}{|c|}{\multirow{6}{*}{}}  & \multicolumn{1}{c|}{\pl}  & & 6.12 & 6.06 & 5.4 & 8.05 & 14.24 & 12.43  & & 5.9 & 5.89 & 4.39 & 4.46 & 14.19 & 14.56  & & 5.95 & 5.9 & 4.16 & 4.09 & 13.5 & 14.9  \\ \cline{2-2} \cline{4-9} \cline{11-16} \cline{18-23}  
			\multicolumn{1}{|c|}{\multirow{6}{*}{}}  & \multicolumn{1}{c|}{\fp}  & & 6.15 & 6.12 & 6.17 & 8.25 & 14.32 & 12.23  & & 5.93 & 5.97 & 5.9 & 5.36 & 13.94 & 14.55  & & 5.88 & 5.99 & 5.89 & 4.97 & 13.52 & 14.66  \\ \cline{2-2} \cline{4-9} \cline{11-16} \cline{18-23}  
			\multicolumn{1}{|c|}{\multirow{6}{*}{}}  & \multicolumn{1}{c|}{\ml}  & & 9.53 \scriptsize{(78)} & 9.54 \scriptsize{(98)} & 9.67 \scriptsize{(95)} & 10 \scriptsize{(1)} & - & -  & & 9 & 9.08 \scriptsize{(96)} & 9.58 \scriptsize{(96)} & 10 \scriptsize{(1)} & - & -  & & 9 & 9.09 \scriptsize{(96)} & 9.6 \scriptsize{(98)} & 10 \scriptsize{(1)} & - & -  \\ \hhline{|=|=|~|======|~|======|~|======|} 
			\multicolumn{2}{|c|}{Global \nj} & & 4.42 & 4.2 & 4.34 & 7.7 & 17.28 & 17.41  & & 4 & 4 & 4 & 4.12 & 15.84 & 17.5  & & 4 & 4 & 4 & 4.04 & 14.8 & 17.38  \\ \cline{1-9}  \cline{4-9} \cline{11-16} \cline{18-23}  
		\end{tabular}
		
		\vspace*{2mm}
		\captionsetup{width=1.4\textwidth}
		
		\caption{ \label{tab_CDgennonh}
			\small Average Robinson-Foulds distance for tree \textit{CD}. 
			The number in parentheses represents the number of consensus trees that we have been able to reconstruct (\mll(homGMc) may not converge for some quartets and in this case we cannot reconstruct the tree). If it is missing. then we have been able to reconstruct the $100$ simulated trees. }
	\end{table}
	\vspace*{\fill}
\end{landscape}

\begin{landscape}
	\vspace*{\fill}
	\begin{table}[H]
		\centering
		\footnotesize
		\setlength{\tabcolsep}{2.8pt}
		\begin{tabular}{ccc|cccccc|c|cccccc|c|cccccc|} 
			\cline{4-9} \cline{11-16} \cline{18-23} 
			&    &     & \multicolumn{6}{c|}{Lenght 600 bp}        &  & \multicolumn{6}{c|}{Lenght 5 000 bp}      &  & \multicolumn{6}{c|}{Lenght 10 000 bp}   \\ \hhline{~~~|======|~|======|~|======|} 
			& & & \textbf{b=0.005} & \textbf{0.015} & \textbf{0.05} & \textbf{0.1}  & \textbf{0.25} & \textbf{0.5}  &  & \textbf{0.005} & \textbf{0.015} & \textbf{0.05} & \textbf{0.1}  & \textbf{0.25} & \textbf{0.5}  &  & \textbf{0.005} & \textbf{0.015} & \textbf{0.05} & \textbf{0.1}  & \textbf{0.25} & \textbf{0.5}  \\ \hhline{|=|=|~|======|~|======|~|======|} 
			\multicolumn{1}{|c|}{\multirow{6}{*}{WO}}  & \multicolumn{1}{c|}{\asaq}  & & 7.99 & 8.06 & 8.81 & 9.72 & 14.62 & 11.67  & & 8 & 8 & 8 & 8.04 & 13.69 & 14.27  & & 8 & 8 & 8 & 8 & 13.08 & 14.51  \\ \cline{2-2} \cline{4-9} \cline{11-16} \cline{18-23}  
			\multicolumn{1}{|c|}{\multirow{6}{*}{}}  & \multicolumn{1}{c|}{\saq}  & & 8.13 & 8.07 & 8.56 & 9.52 & 14.57 & 11.81  & & 8 & 8 & 8 & 8.1 & 13.72 & 14.28  & & 8 & 8 & 8 & 8 & 13.28 & 14.65  \\ \cline{2-2} \cline{4-9} \cline{11-16} \cline{18-23}  
			\multicolumn{1}{|c|}{\multirow{6}{*}{}}  & \multicolumn{1}{c|}{\eri}  & & 8.85 & 8.43 & 9.17 & 10.7 & 16.37 & 14.49  & & 8 & 8 & 7.96 & 8.12 & 14.34 & 14.81  & & 8 & 8 & 7.98 & 7.94 & 12.64 & 14.73  \\ \cline{2-2} \cline{4-9} \cline{11-16} \cline{18-23}  
			\multicolumn{1}{|c|}{\multirow{6}{*}{}}  & \multicolumn{1}{c|}{\pl}  & & 8.3 & 8.04 & 8.35 & 9.83 & 16.01 & 14.26  & & 8 & 8 & 8 & 8.02 & 15.19 & 16.22  & & 8 & 8 & 8 & 8 & 14.3 & 16.31  \\ \cline{2-2} \cline{4-9} \cline{11-16} \cline{18-23}  
			\multicolumn{1}{|c|}{\multirow{6}{*}{}}  & \multicolumn{1}{c|}{\fp}  & & 8.3 & 8.02 & 8.4 & 10.05 & 16.1 & 14.32  & & 8 & 8 & 8 & 8.03 & 15.12 & 16.25  & & 8 & 8 & 8 & 8.01 & 14.31 & 16.25  \\ \cline{2-2} \cline{4-9} \cline{11-16} \cline{18-23}  
			\multicolumn{1}{|c|}{\multirow{6}{*}{}}  & \multicolumn{1}{c|}{\ml}  & & 12.26 \scriptsize{(74)} & 11.92 \scriptsize{(93)} & 11.89 \scriptsize{(89)} & 11 \scriptsize{(8)} & - & -  & & 11.04 & 11.18 \scriptsize{(85)} & 11.79 \scriptsize{(87)} & - \scriptsize{(0)} & - & -  & & 10.98 & 10.91 \scriptsize{(90)} & 11.52 \scriptsize{(90)} & - \scriptsize{(0)} & - & -  \\ \hhline{|=|=|~|======|~|======|~|======|} 
			\multicolumn{1}{|c|}{\multirow{6}{*}{WIL}}  & \multicolumn{1}{c|}{\asaq}  & & 8.21 & 8.34 & 9.16 & 10.24 & 9.93 & 9.24  & & 8 & 8 & 8 & 8.49 & 11.53 & 9.4  & & 8 & 8 & 8 & 8.02 & 11.75 & 9.81  \\ \cline{2-2} \cline{4-9} \cline{11-16} \cline{18-23}  
			\multicolumn{1}{|c|}{\multirow{6}{*}{}}  & \multicolumn{1}{c|}{\saq}  & & 8.16 & 8.08 & 8.32 & 9.26 & 12.44 & 9.41  & & 8 & 8 & 8 & 7.94 & 12.14 & 10.19  & & 8 & 8 & 8 & 8 & 11.16 & 11.34  \\ \cline{2-2} \cline{4-9} \cline{11-16} \cline{18-23}  
			\multicolumn{1}{|c|}{\multirow{6}{*}{}}  & \multicolumn{1}{c|}{\eri}  & & 8.93 & 8.36 & 9.05 & 9.92 & 13.65 & 11.1  & & 8 & 8 & 7.98 & 8.13 & 12.51 & 11.82  & & 8 & 8 & 7.98 & 7.95 & 9.83 & 11.81  \\ \cline{2-2} \cline{4-9} \cline{11-16} \cline{18-23}  
			\multicolumn{1}{|c|}{\multirow{6}{*}{}}  & \multicolumn{1}{c|}{\pl}  & & 8.29 & 8.21 & 8.48 & 9.54 & 13.65 & 10.79  & & 8 & 8 & 8 & 8.37 & 13.16 & 13.99  & & 8 & 8 & 8 & 8.14 & 12.56 & 14.04  \\ \cline{2-2} \cline{4-9} \cline{11-16} \cline{18-23}  
			\multicolumn{1}{|c|}{\multirow{6}{*}{}}  & \multicolumn{1}{c|}{\fp}  & & 8.24 & 8.05 & 8.4 & 9.26 & 13.17 & 10.49  & & 8 & 8 & 8 & 8.08 & 12.88 & 13.52  & & 8 & 8 & 8 & 7.98 & 12.02 & 13.97  \\ \cline{2-2} \cline{4-9} \cline{11-16} \cline{18-23}  
			\multicolumn{1}{|c|}{\multirow{6}{*}{}}  & \multicolumn{1}{c|}{\ml}  & & 9.88 \scriptsize{(74)} & 9.72 \scriptsize{(93)} & 9.6 \scriptsize{(89)} & 9.38 \scriptsize{(8)} & - & -  & & 10.04 & 9.31 \scriptsize{(85)} & 9.36 \scriptsize{(87)} & - \scriptsize{(0)} & - & -  & & 9.34 & 8.97 \scriptsize{(90)} & 8.99 \scriptsize{(90)} & - \scriptsize{(0)} & - & -  \\ \hhline{|=|=|~|======|~|======|~|======|} 
			\multicolumn{1}{|c|}{\multirow{6}{*}{QP}}  & \multicolumn{1}{c|}{\asaq}  & & 10.68 & 11.34 & 12.09 & 11.16 & 11.83 & 11.01  & & 8.53 & 9.07 & 9.77 & 12.17 & 10.51 & 12.54  & & 8.06 & 8.29 & 9.45 & 11.24 & 10.06 & 12.13  \\ \cline{2-2} \cline{4-9} \cline{11-16} \cline{18-23}  
			\multicolumn{1}{|c|}{\multirow{6}{*}{}}  & \multicolumn{1}{c|}{\saq}  & & 8.72 & 9.09 & 12.09 & 10.91 & 12.08 & 11.03  & & 8 & 8 & 8 & 11.46 & 10.68 & 12.47  & & 8 & 8 & 8 & 9.54 & 10.21 & 12.51  \\ \cline{2-2} \cline{4-9} \cline{11-16} \cline{18-23}  
			\multicolumn{1}{|c|}{\multirow{6}{*}{}}  & \multicolumn{1}{c|}{\eri}  & & 10.92 & 11.36 & 11.48 & 10.65 & 13.7 & 11.35  & & 8.63 & 9.13 & 9.81 & 11.37 & 11.29 & 13.07  & & 8.01 & 8.33 & 9.54 & 11.12 & 9.67 & 13.1  \\ \cline{2-2} \cline{4-9} \cline{11-16} \cline{18-23}  
			\multicolumn{1}{|c|}{\multirow{6}{*}{}}  & \multicolumn{1}{c|}{\pl}  & & 9.82 & 9.77 & 9.36 & 10.39 & 14.72 & 12.49  & & 9.95 & 9.99 & 8.28 & 8.3 & 13.83 & 14.69  & & 9.83 & 9.91 & 8.1 & 8.1 & 13.38 & 14.81  \\ \cline{2-2} \cline{4-9} \cline{11-16} \cline{18-23}  
			\multicolumn{1}{|c|}{\multirow{6}{*}{}}  & \multicolumn{1}{c|}{\fp}  & & 9.75 & 9.74 & 9.94 & 10.29 & 14.44 & 12.2  & & 9.85 & 9.82 & 9.97 & 9.12 & 13.77 & 14.58  & & 9.94 & 9.9 & 9.69 & 8.86 & 13.26 & 14.69  \\ \cline{2-2} \cline{4-9} \cline{11-16} \cline{18-23}  
			\multicolumn{1}{|c|}{\multirow{6}{*}{}}  & \multicolumn{1}{c|}{\ml}  & & 9.47 \scriptsize{(74)} & 9.6 \scriptsize{(93)} & 9.73 \scriptsize{(89)} & 9.62 \scriptsize{(8)} & - & -  & & 9 & 9.11 \scriptsize{(85)} & 9.59 \scriptsize{(87)} & - \scriptsize{(0)} & - & -  & & 9 & 9.08 \scriptsize{(90)} & 9.64 \scriptsize{(90)} & - \scriptsize{(0)} & - & -  \\ \hhline{|=|=|~|======|~|======|~|======|} 
			\multicolumn{2}{|c|}{Global \nj} & & 8.38 & 8.1 & 8.32 & 10.14 & 17.17 & 17.4  & & 8 & 8 & 8 & 8.04 & 15.6 & 17.66  & & 8 & 8 & 8 & 8.04 & 14.62 & 17.72  \\ \cline{1-9}  \cline{4-9} \cline{11-16} \cline{18-23}  
		\end{tabular}
		
		\vspace*{2mm}
		\captionsetup{width=1.4\textwidth}
		
		\caption{ \label{tab_DDgennonh}
			\small Average Robinson-Foulds distance for tree \textit{DD}. 
			The number in parentheses represents the number of consensus trees that we have been able to reconstruct (\mll(homGMc) may not converge for some quartets and in this case we cannot reconstruct the tree). If it is missing. then we have been able to reconstruct the $100$ simulated trees. }
	\end{table}
	\vspace*{\fill}
\end{landscape}

%
\begin{landscape}
	\vspace*{\fill}
	\subsubsection{Q-methods on GTR data}
	\begin{table}[H]
		\centering
		\footnotesize
		\setlength{\tabcolsep}{2.8pt}
		\begin{tabular}{ccc|cccc|c|cccc|c|cccc|} 
			\cline{4-7} \cline{9-12} \cline{14-17} 
			&    &     & \multicolumn{4}{c|}{Lenght 600 bp}        &  & \multicolumn{4}{c|}{Lenght 5 000 bp}      &  & \multicolumn{4}{c|}{Lenght 10 000 bp}   \\ \hhline{~~~|====|~|====|~|====|} 
			& & & \textbf{b=0.005} & \textbf{0.015} & \textbf{0.05} & \textbf{0.1}  &  & \textbf{0.005} & \textbf{0.015} & \textbf{0.05} & \textbf{0.1}   &  & \textbf{0.005} & \textbf{0.015} & \textbf{0.05} & \textbf{0.1}  \\ \hhline{|=|=|~|====|~|====|~|====|} 
			\multicolumn{1}{|c|}{\multirow{6}{*}{WO}}  & \multicolumn{1}{c|}{\asaq}  & & 1.48 & 0.35 & 3.24 & 12.33  & & 0 & 0 & 0 & 0.87  & & 0 & 0 & 0 & 0  \\ \cline{2-2} \cline{4-7} \cline{9-12} \cline{14-17}  
			\multicolumn{1}{|c|}{\multirow{6}{*}{}}  & \multicolumn{1}{c|}{\saq}  & & 1.28 & 0.28 & 3.16 & 12.09  & & 0 & 0 & 0 & 0.92  & & 0 & 0 & 0 & 0  \\ \cline{2-2} \cline{4-7} \cline{9-12} \cline{14-17}  
			\multicolumn{1}{|c|}{\multirow{6}{*}{}}  & \multicolumn{1}{c|}{\eri}  & & 3.87 & 2.19 & 2.21 & 6.74  & & 0 & 0 & 0 & 0.46  & & 0 & 0 & 0 & 0.02  \\ \cline{2-2} \cline{4-7} \cline{9-12} \cline{14-17}  
			\multicolumn{1}{|c|}{\multirow{6}{*}{}}  & \multicolumn{1}{c|}{\pl}  & & 1.29 & 0.16 & 1.03 & 5.8  & & 0 & 0 & 0 & 0.22  & & 0 & 0 & 0 & 0.04  \\ \cline{2-2} \cline{4-7} \cline{9-12} \cline{14-17}  
			\multicolumn{1}{|c|}{\multirow{6}{*}{}}  & \multicolumn{1}{c|}{\fp}  & & 1.3 & 0.19 & 1.2 & 5.99  & & 0 & 0 & 0 & 0.4  & & 0 & 0 & 0 & 0.12  \\ \cline{2-2} \cline{4-7} \cline{9-12} \cline{14-17}  
			\multicolumn{1}{|c|}{\multirow{6}{*}{}}  & \multicolumn{1}{c|}{\ml}  & & 12.71 \scriptsize{(77)} & 12.45 \scriptsize{(99)} & 11.98 \scriptsize{(98)} & - \scriptsize{(0)}  & & 11.88 \scriptsize{(64)} & 11.81 \scriptsize{(86)} & 12.04 \scriptsize{(97)} & 11.88 \scriptsize{(33)}  & & 11.08 \scriptsize{(71)} & 11.51 \scriptsize{(85)} & 11.69 \scriptsize{(96)} & 11.96 \scriptsize{(47)}  \\ \hhline{|=|=|~|====|~|====|~|====|} 
			\multicolumn{1}{|c|}{\multirow{6}{*}{WIL}}  & \multicolumn{1}{c|}{\asaq}  & & 1.8 & 1.15 & 8.36 & 12.55  & & 0 & 0 & 0 & 6.56  & & 0 & 0 & 0 & 1.94  \\ \cline{2-2} \cline{4-7} \cline{9-12} \cline{14-17}  
			\multicolumn{1}{|c|}{\multirow{6}{*}{}}  & \multicolumn{1}{c|}{\saq}  & & 1.31 & 0.37 & 3.11 & 10.96  & & 0 & 0 & 0 & 0.33  & & 0 & 0 & 0 & 0  \\ \cline{2-2} \cline{4-7} \cline{9-12} \cline{14-17}  
			\multicolumn{1}{|c|}{\multirow{6}{*}{}}  & \multicolumn{1}{c|}{\eri}  & & 3.61 & 1.74 & 2.5 & 7.23  & & 0 & 0.02 & 0.16 & 0.93  & & 0 & 0 & 0 & 0.21  \\ \cline{2-2} \cline{4-7} \cline{9-12} \cline{14-17}  
			\multicolumn{1}{|c|}{\multirow{6}{*}{}}  & \multicolumn{1}{c|}{\pl}  & & 1.52 & 0.42 & 2.63 & 7.23  & & 0 & 0 & 0 & 2.49  & & 0 & 0 & 0 & 1.47  \\ \cline{2-2} \cline{4-7} \cline{9-12} \cline{14-17}  
			\multicolumn{1}{|c|}{\multirow{6}{*}{}}  & \multicolumn{1}{c|}{\fp}  & & 1.57 & 0.48 & 3.08 & 7.12  & & 0 & 0 & 0 & 3.07  & & 0 & 0 & 0 & 1.77  \\ \cline{2-2} \cline{4-7} \cline{9-12} \cline{14-17}  
			\multicolumn{1}{|c|}{\multirow{6}{*}{}}  & \multicolumn{1}{c|}{\ml}  & & 9.71 \scriptsize{(77)} & 9.57 \scriptsize{(99)} & 9.43 \scriptsize{(98)} & - \scriptsize{(0)}  & & 9.6 \scriptsize{(64)} & 9.29 \scriptsize{(86)} & 9.38 \scriptsize{(97)} & 9.3 \scriptsize{(33)}  & & 9.11 \scriptsize{(71)} & 8.95 \scriptsize{(85)} & 9 \scriptsize{(96)} & 9.15 \scriptsize{(47)}  \\ \hhline{|=|=|~|====|~|====|~|====|} 
			\multicolumn{1}{|c|}{\multirow{6}{*}{QP}}  & \multicolumn{1}{c|}{\asaq}  & & 4.21 & 5.05 & 6.98 & 11.83  & & 2.73 & 3.65 & 3.8 & 6.31  & & 2.74 & 3.74 & 4.03 & 4.47  \\ \cline{2-2} \cline{4-7} \cline{9-12} \cline{14-17}  
			\multicolumn{1}{|c|}{\multirow{6}{*}{}}  & \multicolumn{1}{c|}{\saq}  & & 1.67 & 1.53 & 6.3 & 9.52  & & 0 & 0 & 0.02 & 6.99  & & 0 & 0 & 0 & 4.18  \\ \cline{2-2} \cline{4-7} \cline{9-12} \cline{14-17}  
			\multicolumn{1}{|c|}{\multirow{6}{*}{}}  & \multicolumn{1}{c|}{\eri}  & & 6.43 & 5.93 & 5.89 & 8.42  & & 2.81 & 3.77 & 4.23 & 5.86  & & 2.73 & 3.63 & 4.15 & 4.96  \\ \cline{2-2} \cline{4-7} \cline{9-12} \cline{14-17}  
			\multicolumn{1}{|c|}{\multirow{6}{*}{}}  & \multicolumn{1}{c|}{\pl}  & & 3.8 & 3.64 & 2.69 & 6.24  & & 3.24 & 3.37 & 0.87 & 0.91  & & 3.34 & 3.37 & 0.46 & 0.34  \\ \cline{2-2} \cline{4-7} \cline{9-12} \cline{14-17}  
			\multicolumn{1}{|c|}{\multirow{6}{*}{}}  & \multicolumn{1}{c|}{\fp}  & & 3.88 & 3.64 & 3.82 & 6.88  & & 3.49 & 3.41 & 3.01 & 2.59  & & 3.38 & 3.44 & 3.14 & 1.79  \\ \cline{2-2} \cline{4-7} \cline{9-12} \cline{14-17}  
			\multicolumn{1}{|c|}{\multirow{6}{*}{}}  & \multicolumn{1}{c|}{\ml}  & & 10.1 \scriptsize{(77)} & 9.88 \scriptsize{(99)} & 9.65 \scriptsize{(98)} & - \scriptsize{(0)}  & & 9.19 \scriptsize{(64)} & 9.28 \scriptsize{(86)} & 9.28 \scriptsize{(97)} & 9.82 \scriptsize{(33)}  & & 9.1 \scriptsize{(71)} & 9.11 \scriptsize{(85)} & 9.32 \scriptsize{(96)} & 9.38 \scriptsize{(47)}  \\ \hhline{|=|=|~|====|~|====|~|====|} 
			\multicolumn{2}{|c|}{Global \nj} & & 1.32 & 0.2 & 0.92 & 5.4  & & 0 & 0 & 0 & 0.38  & & 0 & 0 & 0 & 0  \\ \cline{1-9}  \cline{4-7} \cline{9-12} \cline{14-17}  
		\end{tabular}

		\vspace*{2mm}
		\captionsetup{width=1.2\textwidth}
		\caption{\label{tab_DDgtr} \small Average Robinson-Foulds distance for tree \textit{DD} on GTR data. 
			The number in parentheses represents the number of consensus trees that we have been able to reconstruct (\mll(homGMc) may not converge for some quartets and in this case we cannot reconstruct the tree). If it is missing. then we have been able to reconstruct the $100$ simulated trees. }
	\end{table}
	\vspace*{\fill}
\end{landscape}

\subsubsection{Q-methods on mixture data}

\begin{table}[H]
	\centering
	\footnotesize
	\setlength{\tabcolsep}{2.8pt}
	\begin{tabular}{ccc|cccc|c|cccc|c|cccc|} 
		\cline{4-7} \cline{9-12} \cline{14-17} 
		&    &     & \multicolumn{4}{c|}{Lenght 600 bp}        &  & \multicolumn{4}{c|}{Lenght 5 000 bp}      &  & \multicolumn{4}{c|}{Lenght 10 000 bp}   \\ \hhline{~~~|====|~|====|~|====|} 
		& & & \textbf{b=0.005} & \textbf{0.015} & \textbf{0.05} & \textbf{0.1}  &  & \textbf{0.005} & \textbf{0.015} & \textbf{0.05} & \textbf{0.1}   &  & \textbf{0.005} & \textbf{0.015} & \textbf{0.05} & \textbf{0.1}  \\ \hhline{|=|=|~|====|~|====|~|====|} 
		\multicolumn{1}{|c|}{\multirow{6}{*}{WO}}  & \multicolumn{1}{c|}{\asaq}  & & 1.74 & 0.46 & 1.73 & 6.68  & & 0 & 0 & 0 & 0.95  & & 0 & 0 & 0.02 & 1.31  \\ \cline{2-2} \cline{4-7} \cline{9-12} \cline{14-17}  
		\multicolumn{1}{|c|}{\multirow{6}{*}{}}  & \multicolumn{1}{c|}{\asaq (m=2)}  & & 1.63 & 0.48 & 1.24 & 5.8  & & 0 & 0 & 0 & 0.96  & & 0 & 0 & 0.02 & 0.98  \\ \cline{2-2} \cline{4-7} \cline{9-12} \cline{14-17}  
		\multicolumn{1}{|c|}{\multirow{6}{*}{}}  & \multicolumn{1}{c|}{\eri (m=2)}  & & 3.49 & 1.36 & 2.1 & 6.94  & & 0.06 & 0.02 & 0.04 & 0.73  & & 0.02 & 0 & 0.04 & 0.58  \\ \cline{2-2} \cline{4-7} \cline{9-12} \cline{14-17}  
		\multicolumn{1}{|c|}{\multirow{6}{*}{}}  & \multicolumn{1}{c|}{\pl}  & & 1.72 & 0.16 & 1.22 & 5.13  & & 0 & 0 & 0.04 & 1.64  & & 0 & 0 & 0.06 & 1.65  \\ \cline{2-2} \cline{4-7} \cline{9-12} \cline{14-17}  
		\multicolumn{1}{|c|}{\multirow{6}{*}{}}  & \multicolumn{1}{c|}{\ml}  & & 12.84 & 12.1 & 11.84 & 12 \scriptsize{(1)}  & & 11.32 & 11.21 & 11.29 & 10 \scriptsize{(1)}  & & 11 & 11.07 & 11.27 & 10.75 \scriptsize{(4)}  \\ \hhline{|=|=|~|====|~|====|~|====|} 
		\multicolumn{1}{|c|}{\multirow{6}{*}{WIL}}  & \multicolumn{1}{c|}{\asaq}  & & 1.96 & 0.96 & 3.14 & 7.46  & & 0 & 0 & 0.3 & 2.3  & & 0 & 0 & 0.16 & 1.92  \\ \cline{2-2} \cline{4-7} \cline{9-12} \cline{14-17}  
		\multicolumn{1}{|c|}{\multirow{6}{*}{}}  & \multicolumn{1}{c|}{\asaq (m=2)}  & & 3.04 & 0.88 & 2.26 & 6.88  & & 0 & 0 & 0.35 & 2.3  & & 0 & 0 & 0.19 & 2.49  \\ \cline{2-2} \cline{4-7} \cline{9-12} \cline{14-17}  
		\multicolumn{1}{|c|}{\multirow{6}{*}{}}  & \multicolumn{1}{c|}{\eri (m=2)}  & & 4 & 1.57 & 2.71 & 6.98  & & 0.04 & 0.08 & 0.08 & 1.3  & & 0.02 & 0 & 0.06 & 0.83  \\ \cline{2-2} \cline{4-7} \cline{9-12} \cline{14-17}  
		\multicolumn{1}{|c|}{\multirow{6}{*}{}}  & \multicolumn{1}{c|}{\pl}  & & 1.83 & 0.41 & 1.8 & 6.67  & & 0.02 & 0 & 0.29 & 2.81  & & 0 & 0 & 0.31 & 2.81  \\ \cline{2-2} \cline{4-7} \cline{9-12} \cline{14-17}  
		\multicolumn{1}{|c|}{\multirow{6}{*}{}}  & \multicolumn{1}{c|}{\ml}  & & 10.07 & 9.57 & 9.49 & 9 \scriptsize{(1)}  & & 9.95 & 9.53 & 9.32 & 10 \scriptsize{(1)}  & & 9.24 & 8.98 & 9.03 & 9 \scriptsize{(4)}  \\ \hhline{|=|=|~|====|~|====|~|====|} 
		\multicolumn{1}{|c|}{\multirow{6}{*}{QP}}  & \multicolumn{1}{c|}{\asaq}  & & 3.91 & 4.01 & 5.03 & 7.95  & & 1.55 & 2.08 & 2.93 & 5.11  & & 0.88 & 1.43 & 2.54 & 4.47  \\ \cline{2-2} \cline{4-7} \cline{9-12} \cline{14-17}  
		\multicolumn{1}{|c|}{\multirow{6}{*}{}}  & \multicolumn{1}{c|}{\asaq (m=2)}  & & 1.93 & 3.7 & 4.59 & 7.48  & & 0.88 & 2.24 & 2.59 & 5.02  & & 0.96 & 1.76 & 2.47 & 4.7  \\ \cline{2-2} \cline{4-7} \cline{9-12} \cline{14-17}  
		\multicolumn{1}{|c|}{\multirow{6}{*}{}}  & \multicolumn{1}{c|}{\eri (m=2)}  & & 3.54 & 3.23 & 4.19 & 7.36  & & 0.14 & 0.63 & 2.33 & 4.33  & & 0.04 & 0.08 & 1.52 & 3.69  \\ \cline{2-2} \cline{4-7} \cline{9-12} \cline{14-17}  
		\multicolumn{1}{|c|}{\multirow{6}{*}{}}  & \multicolumn{1}{c|}{\pl}  & & 3.52 & 3.02 & 3.37 & 6.12  & & 2.81 & 2.61 & 2.57 & 3.11  & & 2.77 & 2.6 & 2.58 & 2.95  \\ \cline{2-2} \cline{4-7} \cline{9-12} \cline{14-17}  
		\multicolumn{1}{|c|}{\multirow{6}{*}{}}  & \multicolumn{1}{c|}{\ml}  & & 9.66 & 10.24 & 9.65 & 9 \scriptsize{(1)}  & & 9.11 & 9.29 & 9.87 & 9 \scriptsize{(1)}  & & 9.28 & 9.15 & 9.8 & 10.25 \scriptsize{(4)}  \\ \hhline{|=|=|~|====|~|====|~|====|} 
		\multicolumn{2}{|c|}{Global \nj} & & 1.84 & 0.16 & 0.96 & 4.98  & & 0 & 0 & 0.06 & 1.68  & & 0 & 0 & 0.04 & 1.64  \\ \cline{1-9}  \cline{4-7} \cline{9-12} \cline{14-17}  
	\end{tabular}
	\vspace*{2mm}
	\captionsetup{width=1.1\textwidth}
	\caption{ \small\label{tab_mixt_p25}  Average Robinson-Foulds distance on mixture data for $p=0.25$. The number in parentheses represents the number of consensus trees that we have been able to reconstruct (\mll(homGMc) may not converge for some 4-uples and in this case we cannot reconstruct the tree). %
		Figure \ref{fig:mixt_p} represents the bar plots of these values. 
	}
\end{table}

\begin{table}[H]
	\centering
	\footnotesize
	\setlength{\tabcolsep}{2.8pt}
	\begin{tabular}{ccc|cccc|c|cccc|c|cccc|} 
		\cline{4-7} \cline{9-12} \cline{14-17} 
		&    &     & \multicolumn{4}{c|}{Lenght 600 bp}        &  & \multicolumn{4}{c|}{Lenght 5 000 bp}      &  & \multicolumn{4}{c|}{Lenght 10 000 bp}   \\ \hhline{~~~|====|~|====|~|====|} 
		& & & \textbf{b=0.005} & \textbf{0.015} & \textbf{0.05} & \textbf{0.1}  &  & \textbf{0.005} & \textbf{0.015} & \textbf{0.05} & \textbf{0.1}   &  & \textbf{0.005} & \textbf{0.015} & \textbf{0.05} & \textbf{0.1}  \\ \hhline{|=|=|~|====|~|====|~|====|} 
		\multicolumn{1}{|c|}{\multirow{6}{*}{WO}}  & \multicolumn{1}{c|}{\asaq}  & & 2.95 & 0.86 & 1.83 & 5.58  & & 0 & 0 & 0.16 & 1.81  & & 0 & 0 & 0.2 & 2.28  \\ \cline{2-2} \cline{4-7} \cline{9-12} \cline{14-17}  
		\multicolumn{1}{|c|}{\multirow{6}{*}{}}  & \multicolumn{1}{c|}{\asaq (m=2)}  & & 3 & 0.86 & 1.3 & 4.89  & & 0 & 0 & 0.1 & 1.24  & & 0 & 0 & 0.07 & 1.45  \\ \cline{2-2} \cline{4-7} \cline{9-12} \cline{14-17}  
		\multicolumn{1}{|c|}{\multirow{6}{*}{}}  & \multicolumn{1}{c|}{\eri (m=2)}  & & 5.08 & 2.56 & 2.57 & 6.29  & & 0.1 & 0 & 0.1 & 1.08  & & 0.02 & 0 & 0.02 & 0.42  \\ \cline{2-2} \cline{4-7} \cline{9-12} \cline{14-17}  
		\multicolumn{1}{|c|}{\multirow{6}{*}{}}  & \multicolumn{1}{c|}{\pl}  & & 2.96 & 0.82 & 1.5 & 6.15  & & 0 & 0 & 0.21 & 1.88  & & 0 & 0 & 0.16 & 2.01  \\ \cline{2-2} \cline{4-7} \cline{9-12} \cline{14-17}  
		\multicolumn{1}{|c|}{\multirow{6}{*}{}}  & \multicolumn{1}{c|}{\ml}  & & 12.18 & 12.13 & 12.57 & 11.67 \scriptsize{(3)}  & & 11.18 & 11.28 & 11.64 & 11 \scriptsize{(1)}  & & 11.1 & 11.15 & 11.37 & 11 \scriptsize{(1)}  \\ \hhline{|=|=|~|====|~|====|~|====|} 
		\multicolumn{1}{|c|}{\multirow{6}{*}{WIL}}  & \multicolumn{1}{c|}{\asaq}  & & 3.29 & 1.43 & 3.46 & 6.27  & & 0 & 0 & 0.8 & 3.27  & & 0 & 0 & 0.7 & 2.97  \\ \cline{2-2} \cline{4-7} \cline{9-12} \cline{14-17}  
		\multicolumn{1}{|c|}{\multirow{6}{*}{}}  & \multicolumn{1}{c|}{\asaq (m=2)}  & & 3.99 & 1.05 & 2.92 & 6.3  & & 0.04 & 0 & 0.67 & 3.5  & & 0 & 0 & 0.57 & 3.16  \\ \cline{2-2} \cline{4-7} \cline{9-12} \cline{14-17}  
		\multicolumn{1}{|c|}{\multirow{6}{*}{}}  & \multicolumn{1}{c|}{\eri (m=2)}  & & 5.3 & 2.55 & 3.17 & 6.58  & & 0.1 & 0.02 & 0.25 & 1.7  & & 0 & 0 & 0.02 & 0.85  \\ \cline{2-2} \cline{4-7} \cline{9-12} \cline{14-17}  
		\multicolumn{1}{|c|}{\multirow{6}{*}{}}  & \multicolumn{1}{c|}{\pl}  & & 3.28 & 0.99 & 2.69 & 7.31  & & 0 & 0 & 0.66 & 3.36  & & 0 & 0 & 0.41 & 3.16  \\ \cline{2-2} \cline{4-7} \cline{9-12} \cline{14-17}  
		\multicolumn{1}{|c|}{\multirow{6}{*}{}}  & \multicolumn{1}{c|}{\ml}  & & 9.85 & 9.67 & 9.74 & 9 \scriptsize{(3)}  & & 9.82 & 9.33 & 9.4 & 9 \scriptsize{(1)}  & & 9.19 & 9.11 & 9 & 9 \scriptsize{(1)}  \\ \hhline{|=|=|~|====|~|====|~|====|} 
		\multicolumn{1}{|c|}{\multirow{6}{*}{QP}}  & \multicolumn{1}{c|}{\asaq}  & & 4.56 & 4.46 & 5.15 & 7.57  & & 1.43 & 2.35 & 3.28 & 5.75  & & 0.9 & 1.53 & 3.24 & 5.85  \\ \cline{2-2} \cline{4-7} \cline{9-12} \cline{14-17}  
		\multicolumn{1}{|c|}{\multirow{6}{*}{}}  & \multicolumn{1}{c|}{\asaq (m=2)}  & & 3 & 4.04 & 4.85 & 7.06  & & 0.97 & 2.37 & 2.77 & 5.4  & & 0.76 & 1.77 & 2.62 & 4.99  \\ \cline{2-2} \cline{4-7} \cline{9-12} \cline{14-17}  
		\multicolumn{1}{|c|}{\multirow{6}{*}{}}  & \multicolumn{1}{c|}{\eri (m=2)}  & & 4.89 & 3.88 & 4.54 & 7.21  & & 0.13 & 0.92 & 2.52 & 4.76  & & 0.02 & 0.16 & 1.6 & 4.04  \\ \cline{2-2} \cline{4-7} \cline{9-12} \cline{14-17}  
		\multicolumn{1}{|c|}{\multirow{6}{*}{}}  & \multicolumn{1}{c|}{\pl}  & & 4.29 & 3.36 & 3.52 & 6.44  & & 2.75 & 2.77 & 2.56 & 3.57  & & 2.83 & 2.71 & 2.72 & 3.37  \\ \cline{2-2} \cline{4-7} \cline{9-12} \cline{14-17}  
		\multicolumn{1}{|c|}{\multirow{6}{*}{}}  & \multicolumn{1}{c|}{\ml}  & & 9.67 & 9.89 & 9.7 & 10 \scriptsize{(3)}  & & 9.25 & 9.38 & 10.07 & 10 \scriptsize{(1)}  & & 9.13 & 9.17 & 9.63 & 9 \scriptsize{(1)}  \\ \hhline{|=|=|~|====|~|====|~|====|} 
		\multicolumn{2}{|c|}{Global \nj} & & 2.98 & 0.82 & 1.18 & 6.18  & & 0 & 0 & 0.18 & 1.84  & & 0 & 0 & 0.14 & 2.14  \\ \cline{1-9}  \cline{4-7} \cline{9-12} \cline{14-17}  
	\end{tabular}
	\vspace*{2mm}
	\captionsetup{width=1.1\textwidth}
	\caption{ \small\label{tab_mixt_p50} Average Robinson-Foulds distance on mixture data for $p=0.50$. The number in parentheses represents the number of consensus trees that we have been able to reconstruct (\mll(homGMc) may not converge for some 4-uples and in this case we cannot reconstruct the tree). %
		Figure \ref{fig:mixt_p} represents the bar plots of these values. }
\end{table}

\begin{table}[H]
	\centering
	\footnotesize
	\setlength{\tabcolsep}{2.8pt}
	\begin{tabular}{ccc|cccc|c|cccc|c|cccc|} 
		\cline{4-7} \cline{9-12} \cline{14-17} 
		&    &     & \multicolumn{4}{c|}{Lenght 600 bp}        &  & \multicolumn{4}{c|}{Lenght 5 000 bp}      &  & \multicolumn{4}{c|}{Lenght 10 000 bp}   \\ \hhline{~~~|====|~|====|~|====|} 
		& & & \textbf{b=0.005} & \textbf{0.015} & \textbf{0.05} & \textbf{0.1}  &  & \textbf{0.005} & \textbf{0.015} & \textbf{0.05} & \textbf{0.1}   &  & \textbf{0.005} & \textbf{0.015} & \textbf{0.05} & \textbf{0.1}  \\ \hhline{|=|=|~|====|~|====|~|====|} 
		\multicolumn{1}{|c|}{\multirow{6}{*}{WO}}  & \multicolumn{1}{c|}{\asaq}  & & 1.82 & 0.87 & 1.57 & 6.27  & & 0 & 0 & 0 & 1.44  & & 0 & 0 & 0.12 & 1.54  \\ \cline{2-2} \cline{4-7} \cline{9-12} \cline{14-17}  
		\multicolumn{1}{|c|}{\multirow{6}{*}{}}  & \multicolumn{1}{c|}{\asaq (m=2)}  & & 1.76 & 0.44 & 0.95 & 5.55  & & 0 & 0 & 0.04 & 1.01  & & 0 & 0 & 0.08 & 0.81  \\ \cline{2-2} \cline{4-7} \cline{9-12} \cline{14-17}  
		\multicolumn{1}{|c|}{\multirow{6}{*}{}}  & \multicolumn{1}{c|}{\eri (m=2)}  & & 4.62 & 1.7 & 2.06 & 6.87  & & 0.06 & 0 & 0.04 & 1.08  & & 0 & 0.02 & 0 & 0.28  \\ \cline{2-2} \cline{4-7} \cline{9-12} \cline{14-17}  
		\multicolumn{1}{|c|}{\multirow{6}{*}{}}  & \multicolumn{1}{c|}{\pl}  & & 1.61 & 0.56 & 1.09 & 5.22  & & 0 & 0 & 0.14 & 1.36  & & 0 & 0 & 0.02 & 1.5  \\ \cline{2-2} \cline{4-7} \cline{9-12} \cline{14-17}  
		\multicolumn{1}{|c|}{\multirow{6}{*}{}}  & \multicolumn{1}{c|}{\ml}  & & 12.81 & 12.2 & 12.15 & - \scriptsize{(0)} & & 12.11 & 11.43 & 11.68 & - \scriptsize{(0)} & & 11.25 & 11.28 & 11.41 & - \scriptsize{(0)} \\ \hhline{|=|=|~|====|~|====|~|====|} 
		\multicolumn{1}{|c|}{\multirow{6}{*}{WIL}}  & \multicolumn{1}{c|}{\asaq}  & & 2.07 & 0.99 & 3.63 & 7.9  & & 0 & 0 & 0.16 & 3.12  & & 0 & 0 & 0.32 & 2.87  \\ \cline{2-2} \cline{4-7} \cline{9-12} \cline{14-17}  
		\multicolumn{1}{|c|}{\multirow{6}{*}{}}  & \multicolumn{1}{c|}{\asaq (m=2)}  & & 2.87 & 0.77 & 2.56 & 7.15  & & 0 & 0 & 0.3 & 3.21  & & 0 & 0 & 0.22 & 2.93  \\ \cline{2-2} \cline{4-7} \cline{9-12} \cline{14-17}  
		\multicolumn{1}{|c|}{\multirow{6}{*}{}}  & \multicolumn{1}{c|}{\eri (m=2)}  & & 5.12 & 1.74 & 2.41 & 6.75  & & 0.02 & 0.01 & 0.07 & 1.45  & & 0 & 0.02 & 0.01 & 0.33  \\ \cline{2-2} \cline{4-7} \cline{9-12} \cline{14-17}  
		\multicolumn{1}{|c|}{\multirow{6}{*}{}}  & \multicolumn{1}{c|}{\pl}  & & 1.82 & 1 & 2.2 & 6.56  & & 0 & 0 & 0.6 & 2.97  & & 0 & 0 & 0.33 & 3.07  \\ \cline{2-2} \cline{4-7} \cline{9-12} \cline{14-17}  
		\multicolumn{1}{|c|}{\multirow{6}{*}{}}  & \multicolumn{1}{c|}{\ml}  & & 9.81 & 9.89 & 9.36 & - \scriptsize{(0)} & & 9.94 & 9.53 & 9.46 & - \scriptsize{(0)} & & 9.09 & 8.86 & 9.03 & - \scriptsize{(0)} \\ \hhline{|=|=|~|====|~|====|~|====|} 
		\multicolumn{1}{|c|}{\multirow{6}{*}{QP}}  & \multicolumn{1}{c|}{\asaq}  & & 4.15 & 4.35 & 4.97 & 8.13  & & 1.39 & 1.99 & 2.91 & 5.47  & & 0.73 & 1.49 & 2.73 & 5.63  \\ \cline{2-2} \cline{4-7} \cline{9-12} \cline{14-17}  
		\multicolumn{1}{|c|}{\multirow{6}{*}{}}  & \multicolumn{1}{c|}{\asaq (m=2)}  & & 1.88 & 4.02 & 5.09 & 7.77  & & 0.88 & 2.36 & 2.77 & 5.34  & & 0.81 & 1.83 & 2.46 & 5  \\ \cline{2-2} \cline{4-7} \cline{9-12} \cline{14-17}  
		\multicolumn{1}{|c|}{\multirow{6}{*}{}}  & \multicolumn{1}{c|}{\eri (m=2)}  & & 4.35 & 3.08 & 4.24 & 7.36  & & 0.1 & 0.58 & 2.52 & 4.81  & & 0 & 0.04 & 1.98 & 4.05  \\ \cline{2-2} \cline{4-7} \cline{9-12} \cline{14-17}  
		\multicolumn{1}{|c|}{\multirow{6}{*}{}}  & \multicolumn{1}{c|}{\pl}  & & 3.63 & 3.42 & 3.55 & 5.77  & & 2.52 & 2.72 & 2.75 & 2.95  & & 2.6 & 2.65 & 2.63 & 3.12  \\ \cline{2-2} \cline{4-7} \cline{9-12} \cline{14-17}  
		\multicolumn{1}{|c|}{\multirow{6}{*}{}}  & \multicolumn{1}{c|}{\ml}  & & 9.88 & 9.67 & 9.64 & - \scriptsize{(0)} & & 9.06 & 9.2 & 10.14 & - \scriptsize{(0)} & & 9.06 & 9.19 & 9.73 & - \scriptsize{(0)} \\ \hhline{|=|=|~|====|~|====|~|====|} 
		\multicolumn{2}{|c|}{Global \nj} & & 1.46 & 0.6 & 1.12 & 5.4  & & 0 & 0 & 0.12 & 1.46  & & 0 & 0 & 0.04 & 1.74  \\ \cline{1-9}  \cline{4-7} \cline{9-12} \cline{14-17}  
	\end{tabular}
	\vspace*{2mm}
	\captionsetup{width=1.1\textwidth}
	\caption{ \small\label{tab_mixt_p75} Average Robinson-Foulds distance on mixture data for $p=0.75$. The number in parentheses represents the number of consensus trees that we have been able to reconstruct (\mll(homGMc) may not converge for some 4-uples and in this case we cannot reconstruct the tree). %
		Figure \ref{fig:mixt_p} represents the bar plots of these values. }
\end{table}

\end{document}